\documentclass[12pt,preprint]{aastex}

\usepackage{color}
\usepackage{ascmac}
\usepackage{amsmath}

\slugcomment{Accepted for publication in ApJ: \today}

\shorttitle{Formulation of Non-Steady--State Dust Formation Process}
\shortauthors{Nozawa \& Kozasa}

\begin{document}

\title{FORMULATION OF NON--STEADY--STATE DUST FORMATION PROCESS 
IN ASTROPHYSICAL ENVIRONMENTS}

\author{
Takaya Nozawa\altaffilmark{1} and
Takashi Kozasa\altaffilmark{2}}

\altaffiltext{1}{Kavli Institute for the Physics and Mathematics of 
the Universe (WPI), University of Tokyo, Kashiwa, Chiba 277-8583, Japan; 
takaya.nozawa@ipmu.jp}
\altaffiltext{2}{Department of Cosmosciences, Graduate School 
of Science, Hokkaido University, Sapporo 060-0810, Japan}

\begin{abstract}

The non-steady--state formation of small clusters and the growth of
grains accompanied by chemical reactions are formulated under the 
consideration that the collision of key gas species (key molecule) 
controls the kinetics of dust formation process.
The formula allows us to evaluate the size distribution and 
condensation efficiency of dust formed in astrophysical environments.
We apply the formulation to the formation of C and MgSiO$_3$ grains 
in the ejecta of supernovae, as an example, to investigate how the 
non-steady effect influences the formation process, condensation 
efficiency $f_{{\rm con},\infty}$, and average radius 
$a_{{\rm ave},\infty}$ of newly formed grains in comparison with the 
results calculated with the steady--state nucleation rate.
We show that the steady--state nucleation rate is a good approximation
if the collision timescale of key molecule $\tau_{\rm coll}$ is much 
smaller than the timescale $\tau_{\rm sat}$ with which the 
supersaturation ratio increases; 
otherwise the effect of the non-steady state becomes remarkable,
leading to a lower $f_{{\rm con},\infty}$ and a larger 
$a_{{\rm ave},\infty}$.
Examining the results of calculations, we reveal that the 
steady--state nucleation rate is applicable if the cooling gas 
satisfies 
$\Lambda \equiv \tau_{\rm sat}/\tau_{\rm coll} \ga 30$
during the formation of dust, and find that $f_{{\rm con},\infty}$ and 
$a_{{\rm ave},\infty}$ are uniquely determined by $\Lambda_{\rm on}$ 
at the onset time $t_{\rm on}$ of dust formation.
The approximation formulae for $f_{{\rm con},\infty}$ and 
$a_{{\rm ave},\infty}$ as a function of $\Lambda_{\rm on}$ 
could be useful in estimating the mass and typical size of newly 
formed grains from observed or model-predicted physical properties 
not only in supernova ejecta but also in mass-loss winds from 
evolved stars.

\end{abstract}

\keywords{dust, extinction -- stars: winds, outflows -- 
supernovae: general -- supernovae: individual (SN 2010jl)}

\section{INTRODUCTION}

In astrophysical environments, dust grains condense in the metal-rich 
cooling gas such as the stellar winds from evolved stars and the 
expanding ejecta of novae and supernovae (SNe).
Once newly formed dust grains are injected into the interstellar 
medium (ISM), they cause interstellar extinction and diffuse infrared 
emission, and also serve as catalysts for H$_2$ formation and building 
materials of such planets as we live on.
Hence, the investigation of formation and evolution of dust is
indispensable in disclosing the nature of objects at high redshifts, 
the radiative process and energy balance in the ISM, and the formation 
history of stars and planetary systems.

In particular, the origin of dust has been hotly debated since the 
discoveries of a huge amount of dust grains at redshifts higher than 
$z = 5$ (Gall et al.\ 2011 and references therein).
In an early epoch of the universe, core-collapse SNe arising from
short-lived massive stars are likely to be dominant sources of dust 
(e.g., Dwek et al.\ 2007).
In fact, recent far-infrared to submillimeter observations of young 
supernova remnants,
SN 1987A (Matsuura et al.\ 2011; Laki\'{c}evi\'{c} et al.\ 2012),
Cas A (Sibthorpe et al.\ 2010; Barlow et al.\ 2010), and
Crab (Gomez et al.\ 2012),
have reported the presence of subsolar mass of cool dust formed in 
the ejecta, which seems to be high enough to account for the observed 
amount of dust at high redshifts. 
However, these cool grains have not yet undergone the destruction in 
the hot gas swept up by the reverse and forward shocks, and thus their 
mass does not necessarily represent the amount of dust finally ejected 
by SNe.
What fraction of newly formed grains can survive on their journeys to 
and in the ISM heavily depends on their sizes at the time of formation 
(e.g., Nozawa et al.\ 2006, 2007).
Thus, in order to reveal the roles of SNe as sources of dust in the 
universe, it is essential to understand not only the total mass but 
also the size distribution of dust produced in the ejecta of SNe.

The formation process of dust in the SN ejecta has been studied 
mainly with the classical nucleation theory and its extension
(Kozasa et al.\ 1989, 1991; Todini \& Ferrara 2001; 
Nozawa et al.\ 2003, 2008, 2010, 2011; Bianchi \& Schneider 2007).
In the nucleation theory, the condensation of dust is described by the 
formation of stable seed nuclei (called critical clusters) and their 
growth, where the formation rate of critical clusters is derived by 
assuming the nucleation current to be in a steady state. 
This theory enables us to predict the size distribution and mass of 
condensing grain species, and the results of the dust formation 
calculations have nicely explained the mass of dust formed in SN 1987A 
(Kozasa et al.\ 1991) and the formation and evolution processes of 
dust in Cas A (Nozawa et al.\ 2010).

However, it has been argued that the validity for the application of
classical nucleation theory could not be justified in the rarefied gas 
typical of dust-forming region of astrophysical interest 
(Donn \& Nuth 1985 and references therein);
in much less dense systems, the timescale on which the nucleation 
current achieves a steady state must be longer than the timescales of 
evolutions of the gas density and temperature, which would render the 
application of the steady--state nucleation rate questionable.
On the other hand, Paquette \& Nuth (2011) suggested that the lack of 
the steady--state condition is unlikely to change the radius and
number density of newly formed grains significantly.
They showed that the resulting size distribution of dust is little 
affected even if the steady--state nucleation rate is reduced by a few 
orders of magnitude, although they did not clarify the effect of a 
non-steady state on the formation process of dust.

In this paper, we develop a method managing the dust formation process
without postulating a steady state, which is expected to be more 
appropriate in astrophysical application.
In Section 2, we formulate the non-steady--state dust formation 
process, including chemical reactions at the time of formation of 
clusters and grains.
After describing a simple model of the time evolutions of the 
gas temperature and density in the ejecta of SNe in Section 3,
we present, in Section 4, the results of the calculations for the
formation of C and MgSiO$_3$ grains and discuss the effect of 
the non-steady state and its dependence on the physical conditions
in the ejecta.
In Section 5, we demonstrate that the average radius and condensation 
efficiency can be uniquely determined by the ratio of the 
supersaturation timescale to the collision timescale at the time when
the condensation efficiency rises to $10^{-10}$.
Our conclusions are presented in Section 6.
We also present the detailed derivation of the steady--state 
nucleation rate for the formation of compound grains such as silicates
in Appendix A.

\section{FORMULATION OF NON--STEADY--STATE FORMATION PROCESS OF
CLUSTERS AND DUST GRAINS}

In this section, we formulate the non-steady--state formation of 
clusters and dust grains accompanied by chemical reactions, by means 
of a kinetic approach.
Most of grain species of astrophysical interest, like silicate, have 
no monomer molecule with the same chemical composition as the 
condensate.
This implies that the formation of such compound grains proceeds via 
the chemical reactions involving the relevant gas species, while the 
reaction pathways and their rate constants are not well known.
One of the methods of evading the difficulty in treating the 
formation of compound grains without the detailed knowledge of the 
chemical pathways and reaction constants is to employ the concept 
of a key species (key molecule) that is defined as a gaseous 
reactant with the least collisional frequency among the gaseous 
reactants, as proposed by Kozasa \& Hasegawa (1987).
In this method, two-body reactions between a cluster and the key 
molecule are considered to control the kinetics of the chemical 
reaction; the concept of key molecule has been applied also for the 
growth process of compound grains in circumstellar envelopes as well 
as molecular clouds 
(e.g., Ferrarotti \& Gail 2001; Zhukovska et al.\ 2008).
In what follows, we refer to the cluster composed of $n$--key
molecules as $n$--mer, and assume that clusters are spherical.
We also assume the temperature of clusters to be the same as the gas 
temperature.

First, for simplicity, we shall consider a cooling gas in a closed 
box with the initial concentration of the key molecule $c_{10}$ 
at a time $t = t_0$.
As the gas cools down, the condensation of dust grains proceeds 
through formation and growth of $n$--mer clusters via the attachment 
of the key molecules.
In principle, the time evolution of concentration $c(n,t) = c_n$ 
of $n$--mers can be described by a set of differential equations
\begin{eqnarray}
\frac{dc_n}{dt} = J_n(t) - J_{n+1}(t) 
~~~ {\rm for} ~~ 2 \le n \le n_*,
\end{eqnarray}
where $J_n(t)$ is the net current density from $(n-1)$--mer to 
$n$--mer.
Here we consider that a cluster containing more key molecules 
than $n = n_*$ can be treated as a macroscopic grain (hereafter 
simply called ``grain'').
Given the concentration $c_1$ and the mass $m_1$ of the key molecule, 
the growth rate of grains is given by
\begin{eqnarray}
\frac{da}{dt} = s \Omega_0 \left( \frac{k T}{2 \pi m_1} 
\right)^{\frac{1}{2}} c_1 \left( 1 - \frac{1}{S} \right),
\end{eqnarray}
where $a$ is the grain radius, $s$ is the sticking probability of the 
key molecule onto grains, $\Omega_0$ is the volume of the condensate 
per key molecule, $k$ is the Boltzmann constant, $T$ is the 
temperature of the gas, and $S$ is the supersaturation ratio.

The successive formation of clusters and the growth of grains cause 
the depletion of the key molecules.
The time variation of the concentration $c_1$ of the key molecule is 
determined from the equation of the mass conservation;
\begin{eqnarray}
c_{10} - c_1 = \sum^{n_*-1}_{n=2} n c_n + 
\int^t_{t_0} J_{n_*}(t') \frac{a^3(t, t')}{a^3_0} dt',
\end{eqnarray}
where $a_0 = (3 \Omega_0/4 \pi)^{1/3}$ is the hypothetical radius of 
the condensate per key molecule, and $a(t, t')$ is the radius of a 
grain that reaches $n = n_*$ at $t'$ and is measured at $t$.

In the following subsections, we describe how the current density 
$J_n$ and the supersaturation ratio $S$ can be presented according to
the chemical reaction at the time of formation.

\subsection{Case for a Single--element Grain}

In this subsection, as a reference, we consider the formation of 
clusters whose chemical composition is the same as that of the key 
molecule (hereafter we refer to such a grain as a
``single-element grain'') in order to clarify how the formation 
process of clusters is formulated by means of the kinetic approach.
In this case, the formation of clusters proceeds through attachment 
and detachment of a key molecule as follows;
\begin{eqnarray}
\mathcal{X} + \mathcal{X} \ &\rightleftharpoons& \ \mathcal{X}_2 \\
\mathcal{X}_{n-1} + \mathcal{X} \ &\rightleftharpoons& 
\ \mathcal{X}_{n} ~~~ {\rm for} ~~ 3 \le n \le n_*,
\end{eqnarray}
where $\mathcal{X}$ and $\mathcal{X}_{n}$ represent the key
molecule and the $n$--mer cluster, respectively.
Then, the current density $J_n(t)$ is given by
\begin{eqnarray}
J_n(t) = \alpha_{n-1} c_{n-1} c_1 - \beta_n c_n
~~~ {\rm for} ~~ 2 \le n \le n_*,
\end{eqnarray}
where $\alpha_n$ is the attachment rate coefficient of a monomer to 
an $n$--mer, and $\beta_n$ is the detachment rate coefficient of a 
monomer from an $n$--mer.
In general, $\alpha_n$ has not been measured for the materials of 
astrophysical interest.
Thus, considering that collisions of monomers control the kinetics of
attachment, we evaluate $\alpha_n$ as follows,
\begin{eqnarray}
\alpha_n = \frac{s_n}{1 + \delta_{1n}}\ 4 \pi a_0^2 \ n^{\frac{2}{3}}
\left( \frac{k T}{2 \pi m_{n,1}} \right)^{\frac{1}{2}},
\end{eqnarray}
where $s_n$ is the sticking probability of a monomer onto an $n$--mer, 
$\delta_{1n}$ is the Kronecker's delta, and
$m_{n,1} = n m_1/(n+1)$ is the reduced mass of a monomer and an 
$n$-mer.
The detachment rate coefficient $\beta_n$ ($n \ge 2$) can be 
related to $\alpha_{n-1}$ through the principle of detailed balance;
\begin{eqnarray}
\beta_n = \alpha_{n-1} \frac{\mathring{c}_{n-1}}{\mathring{c}_n} 
\mathring{c}_1,
\end{eqnarray}
where $\mathring{c}_n$ is the concentration of the $n$--mer in the gas 
in thermodynamic equilibrium at a temperature $T$.
Then, the current density $J_n(t)$ is reduced to
\begin{eqnarray}
J_n(t) = \alpha_{n-1} c_1 \left( c_{n-1} - c_n 
\frac{\mathring{c}_{n-1}}{\mathring{c}_n} 
\frac{\mathring{c}_1}{c_1} \right).
\end{eqnarray}
From the law of mass action stemming from the condition that
the sum of chemical potentials of the reactants is equal to that of 
the products in chemical equilibrium (see Landau \& Lifshitz 1976)
\begin{eqnarray}
\frac{\mathring{p}_{n-1}}{\mathring{p}_n} 
\frac{\mathring{p}_1}{p_{\rm s}} =
\exp \left[ \frac{1}{k T} \left( \mathring{g}_n - \mathring{g}_{n-1} 
- \mathring{g}_1 \right) \right],
\end{eqnarray}
where $\mathring{p}_n = \mathring{c}_n k T$ and $\mathring{g}_n$ are, 
respectively, the partial pressure and the chemical potential at a 
standard pressure $p_{\rm s}$ of the $n$--mer, the factor 
$\mathring{c}_{n-1} \mathring{c}_1 / \mathring{c}_n {c_1}$ of the 
second term in the parenthesis on the right-hand side of Equation (9) 
is written as
\begin{eqnarray}
\frac{\mathring{c}_{n-1}}{\mathring{c}_n} 
\frac{\mathring{c}_1}{c_1} =
\exp\left[\frac{1}{k T} \left( \mathring{g}_n - \mathring{g}_{n-1} - 
\mathring{g}_1 \right) - \ln \left( \frac{p_1}{p_{\rm s}} \right) \right].
\end{eqnarray}
By introducing the supersaturation ratio $S$ defined as 
\begin{eqnarray}
\ln S = \ln \left(\frac{p_1}{p_{1 {\rm v}}} \right) 
= - \frac{1}{k T} \left( \mathring{g}_{\rm c} - \mathring{g}_1 \right) 
+ \ln \left(\frac{p_1}{p_{\rm s}} \right),
\end{eqnarray}
where $\mathring{g}_{\rm c}$ and $p_{1 {\rm v}}$ are, respectively, the 
chemical potential at a standard pressure $p_{\rm s}$ and the vapor 
pressure of the bulk condensate, the exponent in Equation (11) is 
represented as
\begin{eqnarray}
\gamma_n = \frac{1}{k T} \left( \mathring{g}_n - \mathring{g}_{n-1} - 
\mathring{g}_{\rm c} \right) - \ln S.
\end{eqnarray}
Then, the current density $J_n(t)$ is expressed as
\begin{eqnarray}
J_n(t) = \alpha_{n-1} c_1 \left[ c_{n-1} - c_n 
\exp(\gamma_n) \right]
\end{eqnarray}
for $2 \le n \le n_*$.

Note that, as is seen from Equations (1) and (14), the current density
$J_{n_*}$ cannot be evaluated without any relation between $c_{n_*}$
and $J_{n_*+1}$.
Thus, in order to close Equations (1) to (3), we introduce a 
closure relation that the current density from $n_*$--mer to 
$(n_*+1)$--mer, $J_{n_*+1}$, is approximated by
\begin{eqnarray}
J_{n_*+1}(t) \simeq \alpha_{n_*} c_1 c_{n_*}
\left[ 1 - \exp(\gamma_{n_*+1}) \right],
\end{eqnarray}
supposing that $c_{n_*} \simeq c_{n_*+1}$ for $n_* \gg 1$.

\subsection{Case for a Multi--element Grain}

In order to derive the formula as generally as possible, we 
consider that an $n$--mer cluster $\mathcal{Z}_n$ containing 
$n$ of the key molecule $\mathcal{X}$ is generated from the 
following chemical reactions;
\begin{eqnarray}
\mathcal{Z}_{n-1} + \left( \mathcal{X} + \nu_1 \mathcal{A}_1 + \dots +
\nu_i \mathcal{A}_i \right)
\rightleftharpoons \ \mathcal{Z}_n + \left( 
\eta_1 \mathcal{B}_1 + \dots + \eta_j \mathcal{B}_j \right)
~~~ \textrm{for} ~~ 3 \le n \le n_*, 
\end{eqnarray}
where $\mathcal{A}_k$ ($k=1$--$i$) and $\mathcal{B}_k$ ($k=1$--$j$)
denote the gaseous reactants and products, respectively, and
the stoichiometric coefficients of the gaseous reactants ($\nu_k$) 
and products ($\eta_k$) are normalized to the key molecule.
In the following, we designate the physical quantities of the 
gaseous reactants $\mathcal{A}_k$ and products $\mathcal{B}_k$ by 
attaching the superscript $A$ and $B$, respectively;
for example, the concentrations (partial pressures) of the gas 
species $\mathcal{A}_k$ are denoted as $c_k^A$ ($p_k^A$).
Below we first formulate the formation of dimer and then describe the
formation of $n$--mer.
In the formulation, we assume that clusters with $n \ge 2$ have the 
same stoichiometric composition as the condensate.

\subsubsection{Formation of dimer ($n = 2$)}

We consider that the dimer formation proceeds through the reaction
\begin{eqnarray}
2(\mathcal{X} + \nu_1 \mathcal{A}_1 + \dots +
\nu_ i \mathcal{A}_i) \rightleftharpoons \ 
\mathcal{Z}_{2} + 2( \eta_1 \mathcal{B}_1 + \dots + 
 \eta_j \mathcal{B}_j).
\end{eqnarray}
Under the consideration that the collision of key molecule controls 
the kinetics of the chemical reaction, the current density $J_2(t)$ 
is written as
\begin{eqnarray}
J_2(t) = \alpha_1 c_1^2 - \beta_2 c_{2} 
\left[ \frac{\prod_{k=1}^{j} (c_k^B)^{\eta_k}}
{\prod_{k=1}^{i} (c_k^A)^{\nu_k}} \right]^2,
\end{eqnarray}
where the forward reaction coefficient $\alpha_1$ is the same as that 
given in Equation (7).
The form of the second term on the right-hand side of Equation (18) 
is based on the principle of detailed balance that the ratio of 
the forward reaction coefficient to the backward reaction coefficient
$K$ is expressed as $K = \mathring{c}_2 \left[ \prod_{k=1}^{j} 
(\mathring{c}_k^B)^{\eta_k} / \mathring{c}_1 \prod_{k=1}^{i} 
(\mathring{c}_k^A)^{\nu_k} \right]^2$
in chemical equilibrium.
Then, the current density $J_2(t)$ is represented as
\begin{eqnarray}
J_2(t) = \alpha_1 c_1 
\left( c_1 - c_2 \frac{c_1}{\mathring{c}_2} \frac{1}{b^2} \right)
\end{eqnarray}
with 
\begin{eqnarray}
b 
= \frac{c_1}{\mathring{c}_1}
\frac{\prod^i_{k=1} \left( c^A_k / \mathring{c}^A_k \right)^{\nu_k}}
     {\prod^j_{k=1} \left( c^B_k / \mathring{c}^B_k \right)^{\eta_k}}
= \frac{p_1}{\mathring{p}_1}
\frac{\prod^i_{k=1} \left( p^A_k / \mathring{p}^A_k \right)^{\nu_k}}
     {\prod^j_{k=1} \left( p^B_k / \mathring{p}^B_k \right)^{\eta_k}},
\end{eqnarray}
where $\mathring{c}^A_k$ and $\mathring{c}^B_k$ ($\mathring{p}^A_k$
and $\mathring{p}^B_k$) are the concentrations (partial gas pressures)
of the $k$--th gaseous reactants and products, respectively, in the 
gas in thermodynamic equilibrium at a temperature $T$.
As is the case for a single-element grain, by applying the law of mass
action and introducing $\omega$ ($\ne 0$)\footnote{ 
Note that the formulation is applicable except for the case of 
$\omega = 0$ such as the endothermic dissociative reaction
C$_2$H$_2$ $+$ C$_2$H$_2$ $\rightleftharpoons$ (C$_2)_2$ $+$ 2H$_2$
for the formation of the dimer (C$_2)_2$.}
defined as
\begin{eqnarray}
\omega = 1 + \sum_{k=1}^i \nu_k - \sum_{k=1}^j \eta_k,
\end{eqnarray}
the factor $c_1 / \mathring{c}_2 b^2$ of the second term in the 
parenthesis on the right-hand side of Equation (19) can be rewritten 
as follows;
since
\begin{eqnarray}
& &
\frac{c_1}{\mathring{c}_2 b^2}
\left[ 
\frac{\prod_{k=1}^i \left( c^A_k / c_1 \right)^{\nu_k}}
     {\prod_{k=1}^j \left( c^B_k / c_1 \right)^{\eta_k}}
\right]^{\frac{1}{\omega}}
= 
\frac{p_1}{\mathring{p}_2} 
\left[
\frac{\mathring{p}_1}{p_1}
\frac{\prod^i_{k=1} \left( \mathring{p}^A_k / p^A_k \right)^{\nu_k}}
     {\prod^j_{k=1} \left( \mathring{p}^B_k / p^B_k \right)^{\eta_k}}
\right]^2
\left[
\left( \frac{p_{\rm s}}{p_1} \right)^{\omega - 1} 
\frac{\prod_{k=1}^i \left( p^A_k / p_{\rm s} \right)^{\nu_k}}
     {\prod_{k=1}^j \left( p^B_k / p_{\rm s} \right)^{\eta_k}}
\right]^{\frac{1}{\omega}}
\nonumber \\
&=&
\frac{p_{\rm s}}{\mathring{p}_2} 
\left( \frac{\mathring{p}_1}{p_{\rm s}} \right)^2
\left[
\frac{\prod^i_{k=1} \left( \mathring{p}^A_k / p_{\rm s} \right)^{\nu_k}}
     {\prod^j_{k=1} \left( \mathring{p}^B_k / p_{\rm s} \right)^{\eta_k}}
\right]^2
\left[
\left( \frac{p_1}{p_{\rm s}} \right)
\frac{\prod^i_{k=1} \left( p^A_k / p_{\rm s} \right)^{\nu_k}}
     {\prod^j_{k=1} \left( p^B_k / p_{\rm s} \right)^{\eta_k}}
\right]^{-2}
\left[
\left( \frac{p_1}{p_{\rm s}} \right)
\frac{\prod_{k=1}^i \left( p^A_k / p_{\rm s} \right)^{\nu_k}}
     {\prod_{k=1}^j \left( p^B_k / p_{\rm s} \right)^{\eta_k}}
\right]^{\frac{1}{\omega}}
\nonumber \\
&=&
\frac{p_{\rm s}}{\mathring{p}_2} 
\left( \frac{\mathring{p}_1}{p_{\rm s}} \right)^2 
\left[
\frac{\prod_{k=1}^i \left( \mathring{p}_k^A / p_{\rm s} \right)^{\nu_k}}
     {\prod_{k=1}^j \left( \mathring{p}_k^B / p_{\rm s} \right)^{\eta_k}} 
\right]^2
\left[
\left( \frac{p_1}{p_{\rm s}} \right)
\frac{\prod_{k=1}^i \left( p^A_k / p_{\rm s} \right)^{\nu_k}}
     {\prod_{k=1}^j \left( p^B_k / p_{\rm s} \right)^{\eta_k}}
\right]^{\frac{1}{\omega} - 2},
\end{eqnarray}
we have
\begin{eqnarray}
\frac{p_{\rm s}}{\mathring{p}_2} 
\left( \frac{\mathring{p}_1}{p_{\rm s}} 
\right)^2 \left[
\frac{\prod_{k=1}^i \left( \mathring{p}_k^A / p_{\rm s} \right)^{\nu_k}}
     {\prod_{k=1}^j \left( \mathring{p}_k^B / p_{\rm s} \right)^{\eta_k}} 
\right]^2 
=
\frac{c_1 \Pi}{\mathring{c}_2 b^2}
\left( \frac{p_1}{p_{\rm s}} \Xi \right)^{2 - \frac{1}{\omega}},
\end{eqnarray}
where
\begin{eqnarray}
\Pi &=& \left[ 
\frac{\prod_{k=1}^i \left( c^A_k / c_1 \right)^{\nu_k}}
     {\prod_{k=1}^j \left( c^B_k / c_1 \right)^{\eta_k}}
\right]^{\frac{1}{\omega}}
\\
\Xi &=& \frac{\prod_{k=1}^{i} \left( p_k^A/p_{\rm s} \right)^{\nu_k}}
      {\prod_{k=1}^{j} \left( p_k^B/p_{\rm s} \right)^{\eta_k}}.
\end{eqnarray}
Then, by applying the law of mass action
\begin{eqnarray}
\frac{p_{\rm s}}{\mathring{p}_2} 
\left( \frac{\mathring{p}_1}{p_{\rm s}} \right)^2 \left[
\frac{\prod_{k=1}^i \left( \mathring{p}_k^A / p_{\rm s} \right)^{\nu_k}}
     {\prod_{k=1}^j \left( \mathring{p}_k^B / p_{\rm s} \right)^{\eta_k}} 
\right]^2 
= \exp \left[ \frac{1}{k T} 
\left( \mathring{g}_2 - 2 {\it \Delta} \mathring{g}_{\rm gas} 
\right) \right],
\end{eqnarray}
with
\begin{eqnarray}
{\it \Delta} \mathring{g}_{\rm gas} = 
\mathring{g}_1 + \sum_{k=1}^{i} \nu_k \mathring{g}_k^A 
- \sum_{k=1}^{j} \eta_k \mathring{g}_k^B,
\end{eqnarray}
where $\mathring{g}_k^A$ and $\mathring{g}_k^B$ are the chemical 
potentials of $k$--th gaseous reactants and products at a standard 
pressure $p_{\rm s}$, respectively, the factor in the second term in 
the parenthesis on the right-hand side of Equation (19) is reduced to
\begin{eqnarray}
\frac{c_1}{\mathring{c}_2 b^2} =
\frac{1}{\Pi} \exp \left\{
\frac{1}{k T} \left( \mathring{g}_2 - 2 {\it \Delta} 
\mathring{g}_{\rm gas} \right) \right. 
- \left.
\left( 2 - \frac{1}{\omega} \right) \left[ \ln \left( 
\frac{p_1}{p_{\rm s}} \right) + \ln \Xi \right] \right\}.
\end{eqnarray}
The exponent in Equation (28) is written as
\begin{eqnarray}
\gamma_2 = \frac{1}{kT} \left[ \mathring{g}_2 - 
\left( 2 - \frac{1}{\omega} \right) \mathring{g}_{\rm c} 
- \frac{1}{\omega} {\it \Delta} \mathring{g}_{\rm gas} \right]
- \left( 2- \frac{1}{\omega} \right) \ln S,
\end{eqnarray}
where the supersaturation ratio $S$ is defined as
\begin{eqnarray}
\ln S = - \frac{1}{k T}
\left( \mathring{g}_{\rm c} - {\it \Delta} \mathring{g}_{\rm gas} 
\right) + \ln \left( \frac{p_1}{p_{\rm s}} \right) + \ln \Xi,
\end{eqnarray}
and Equation (19) is finally reduced to
\begin{eqnarray}
J_2(t) = \alpha_1 c_1 \left[ c_1 - c_2 \frac{1}{\Pi} 
\exp(\gamma_2) \right].
\end{eqnarray}
%

\subsubsection{Formation of $n$--mer ($n \ge 3$)}

For the chemical reaction (16) for the formation of $n$--mers 
($3 \le n \le n_*$), the current density $J_n(t)$ is given by
\begin{eqnarray}
J_n(t) = \alpha_{n-1} c_{n-1} c_1 - \beta_n c_n 
\frac{\prod_{k=1}^{j} (c_k^B)^{\eta_k}}
{\prod_{k=1}^{i} (c_k^A)^{\nu_k}}
\end{eqnarray}
with $\alpha_n$ defined by Equation (7), and the principle of detailed 
balance leads to the equation
\begin{eqnarray}
J_n(t) = \alpha_{n-1} c_1 \left( c_{n-1} - c_n 
\frac{\mathring{c}_{n-1}}{\mathring{c}_n} \frac{1}{b}
\right).
\end{eqnarray}
From the law of mass action
\begin{eqnarray}
\frac{\mathring{p}_{n-1}}{\mathring{p}_n} 
\frac{\mathring{p}_1}{p_{\rm s}}
\frac{\prod_{k=1}^i \left( \mathring{p}_k^A / p_{\rm s} \right)^{\nu_k}}
     {\prod_{k=1}^j \left( \mathring{p}_k^B / p_{\rm s} \right)^{\eta_k}} 
= \exp \left[
\frac{1}{k T} \left( \mathring{g}_n - \mathring{g}_{n-1} 
- {\it \Delta} \mathring{g}_{\rm gas} \right) \right],
\end{eqnarray}
the factor $\mathring{c}_{n-1}/\mathring{c}_n b$ in Equation (33) 
can be reduced to
\begin{eqnarray}
\frac{\mathring{c}_{n-1}}{\mathring{c}_n} \frac{1}{b} 
= \exp \left[
\frac{1}{k T} \left( \mathring{g}_n - \mathring{g}_{n-1} - 
{\it \Delta} \mathring{g}_{\rm gas} \right)
- \ln \left( \frac{p_1}{p_{\rm s}} \right)
- \ln \Xi \right].
\end{eqnarray}
With $\ln S$ defined by Equation (30), the exponent in Equation (35) 
is rewritten as
\begin{eqnarray}
\gamma_n = \frac{1}{k T} \left(
\mathring{g}_n - \mathring{g}_{n-1} - \mathring{g}_{\rm c}
\right) -\ln S,
\end{eqnarray}
and consequently, the current density $J_n(t)$ for $3 \le n \le n_*$ is
given by
\begin{eqnarray}
J_n(t) = \alpha_{n-1} c_1 \left[ c_{n-1} - c_n 
\exp(\gamma_n) \right].
\end{eqnarray}

It should be mentioned here that the chemical reaction at the time of 
formation is included in the current density only through the factor 
$\Pi$ given by Equation (24), $\omega$ in Equation (21), and the 
supersaturation ratio $S$ defined by Equation (30).
The formation process of multi-element grains formulated by introducing 
the key molecule can be treated as the natural extension of the 
formation process of single-element grains;
in fact, Equations (31) and (37) can be reduced to Equation (14) by
substituting $\Pi = \omega = 1$ calculated for 
$\nu_k^A = \nu_k^B = 0$.

Note that in principle the current density $J_n$ can be evaluated once 
the chemical potentials of $n$--mers are given.
However, the chemical potential has been available only for tiny 
clusters ($n \la 5$) of very few materials of astrophysical interest
(e.g., Goumans \& Bromley 2012).
Therefore, the so-called capillary approximation, which is the 
practice for estimating the chemical potential of an $n$--mer in
terms of the chemical potential of a monomer in the bulk 
condensate and the surface energy (Abraham 1974; Blander \& Katz 1972),
is generally adopted for evaluating the current density as well as 
the steady--state nucleation rate.
For example, $\mathring{g}_n$ is expressed as
\begin{eqnarray}
\mathring{g}_n = 4 \pi a_0^2 \sigma (n-1)^{\frac{2}{3}} 
+ (n-1) \mathring{g}_{\rm c} + \mathring{g}_1
\end{eqnarray}
for a single-element grain (e.g., Yasuda \& Kozasa 2012), where 
$\sigma$ is the surface tension of bulk condensate.
By substituting $\mathring{g}_n$ into Equation (13), $\gamma_n$ for 
a single-element grain is written as
\begin{eqnarray}
\gamma_n = \mu \left[ (n-1)^{\frac{2}{3}} -
(n-2)^{\frac{2}{3}} \right] -\ln S,
\end{eqnarray}
where $\mu = 4 \pi a_0^2 \sigma / k T$.
Inspecting the change of the chemical potential $\gamma_2$ for the 
formation of a dimer from the reactants given by Equation (29) for a 
multi-element grain and $\gamma'_n$ (Equation (A5)) for the formulation 
of the steady--state nucleation rate in comparison with the corresponding
one given by Equation (13) for a single-element grain, we can see
that the factor $1/\omega$ in Equation (29) represents the contribution 
of the key molecule to the change of chemical potential.
Thus, the chemical potential of an $n$--mer with $n \ge 2$ for a 
multi-element grain can be defined as 
\begin{eqnarray}
\mathring{g}_n = 
4 \pi a_0^2 \sigma \left( n - \frac{1}{\omega} \right)^{\frac{2}{3}} 
+ \left(n - \frac{1}{\omega} \right) \mathring{g}_c + 
\frac{1}{\omega} {\it \Delta} \mathring{g}_{\rm gas},
\end{eqnarray}
so as to be consistent with the formula for a single-element grain.
Then, $\gamma_n$ is evaluated by
\begin{eqnarray}
\gamma_2 = \mu \left( 2 - \frac{1}{\omega} \right)^{\frac{2}{3}}
- \left( 2 - \frac{1}{\omega} \right) \ln S
~~~ {\rm for} ~~ n = 2,
\end{eqnarray}
and
\begin{eqnarray}
\gamma_n = \mu \left[ \left( n - \frac{1}{\omega} \right)^{\frac{2}{3}}
- \left( n - 1 - \frac{1}{\omega} \right)^{\frac{2}{3}}
\right] - \ln S ~~~ {\rm for} ~~ 3 \le n \le n_*.
\end{eqnarray}
%

\subsection{Formulation of Cluster and Dust Formation in a Cooling
Gas Flow}

Equations (1) and (3) describe, respectively, the time evolution of 
concentrations of $n$--mer clusters and the conservation of the key 
molecule in a fixed volume.
However, the formation of dust takes place generally in a cooling gas 
flow such as the stellar winds from evolved stars and the expanding 
ejecta of SNe and novae.
Therefore, it is more useful to formulate the formation process of 
dust in a frame comoving with the gas.

Let us consider a specific volume $V(t)$ comoving with the gas.
The nominal concentration of the key molecule $\tilde{c}_1(t)$ is 
defined as the concentration without the depletion due to the 
formation of clusters and dust grains.
Thus, $\tilde{c}_1(t)$ at a time $t$ holds the relation 
$c_{10} V(t_0) = \tilde{c}_1(t) V(t)$.
The current from the $(n-1)$--mer to the $n$--mer in $V(t)$ is 
given as 
$J_n(t) V(t)$, and the time variation of the number of $n$--mers
in $V(t)$, $N_n(t) = c_n(t) V(t)$, is expressed as
\begin{eqnarray}
\frac{d N_n}{dt} = V(t) \left( J_n - J_{n+1} \right) 
~~~{\rm for} ~~ 2 \le n \le n_*.
\end{eqnarray}
Then, Equation (43), being divided by $\tilde{c}_1 V$, is reduced to
\begin{eqnarray}
\frac{d Y_n}{dt} = I_n - I_{n+1} ~~~{\rm for} ~~ 2 \le n \le n_*,
\end{eqnarray}
where $Y_n = c_n / \tilde{c}_1$ represents the normalized concentration 
of $n$--mers, and the normalized current density from the
$(n-1)$--mer to the $n$--mer $I_n = J_n / \tilde{c}_1$ is given by
\begin{equation}
I_n = \tau_{n-1}^{-1} \times
\begin{cases}
\left[ Y_{n-1} - Y_n \Pi^{-1} \exp \left( \gamma_n \right) \right] 
~~~ {\rm for} ~~~ n = 2 \\
\left[ Y_{n-1} - Y_n \exp \left( \gamma_n \right) \right]
~~~~~ {\rm for} ~~~ 3 \le n \le n_* \\
Y_{n-1} \left[ 1 - \exp \left( \gamma_n \right) \right]
~~~~~ {\rm for} ~~~ n = n_* + 1
\end{cases}
\end{equation}
with $\tau_{n-1}^{-1} = \alpha_{n-1} c_{1}$.

The equation of the mass conservation for the key molecule given by
\begin{eqnarray}
\tilde{c}_1 V - c_1 V = \sum^{n_*-1}_{n=2} n c_n V +
\int^t_{t_0} V(t') J_{n_*}(t') \frac{a^3(t, t')}{a_0^3} dt',
\end{eqnarray}
being divided by $\tilde{c}_1 V$, is also rewritten as
\begin{eqnarray}
1 - Y_1 = \sum^{n_*-1}_{n=2} n Y_n + K_3.
\end{eqnarray}
By introducing
\begin{eqnarray}
K_i(t) = \int^t_{t_0} I_*(t') \frac{a^i(t, t')}{a_0^i} dt'
 ~~~{\rm for} ~~ i = 0{\rm -}3
\end{eqnarray}
and $I_* = I_{n_*}$,
$K_3$ is calculated by solving the following simultaneous 
differential equations
\begin{eqnarray}
\frac{dK_i}{dt} &=& I_*(t) n_*^{\frac{i}{3}} 
+ \frac{i}{a_0} \left( \frac{da}{dt} \right) K_{i-1}
~~~{\rm for} ~~ i = 1{\rm -}3  \nonumber \\
&=& I_*(t) \hspace{3.8cm} {\rm for} ~~ i = 0,
\end{eqnarray}
and $Y_1$ is calculated from Equation (47).
The concentrations of gaseous reactants and products except for the 
key molecule taking into account the depletion due to the formation 
of clusters and the growth of grains are evaluated, respectively, by
\begin{eqnarray}
Y_k^A = \frac{c_k^A}{\tilde{c}_1} 
= \frac{\tilde{c}_k^A}{\tilde{c}_1} - \nu_k^A (1 - Y_1),
\end{eqnarray}
and
\begin{eqnarray}
Y_k^B = \frac{c_k^B}{\tilde{c}_1} 
= \frac{\tilde{c}_k^B}{\tilde{c}_1} + \eta_k^B (1 - Y_1),
\end{eqnarray}
where $\tilde{c}_k^A$ and $\tilde{c}_k^B$ are the nominal concentrations
of $k$--th gaseous reactant and product, respectively.
These equations can be solved, given the initial abundances of gaseous
reactants and products as well as clusters ($Y_n$) at $t = t_0$ together
with the time evolutions of gas temperature and density.

Note that $Y_1 = c_1 / \tilde{c}_1$ represents the number fraction of 
the key molecules left in the gas phase (so-called depletion efficiency), 
$K_0$ the number density of dust grains 
($K_0 = n_{\rm dust} / \tilde{c}_1$), and 
$K_3$ the number fraction of the key molecules locked in dust grains.
Hence, the condensation efficiency $f_{\rm con}(t)$ and 
volume-equivalent average radius $a_{\rm ave}(t)$ are calculated by
\begin{eqnarray}
f_{\rm con}(t) = K_3(t) ~~~{\rm and} ~~~
a_{\rm ave} = 
a_0 \left[ \frac{K_3(t)}{K_0(t)} \right]^{\frac{1}{3}},
\end{eqnarray}
respectively.
In addition, since the grains nucleated in a time interval between 
$t$ and $t + dt$ have the radii between $a$ and $a + da$, the size 
distribution function $f(a)$ of newly formed grains is calculated by
\begin{eqnarray}
f(a) da = \tilde{c}_1(t) I_*(t) dt.
\end{eqnarray}

\section{APPLICATION TO DUST FORMATION IN THE EJECTA OF SNe}

We apply the derived formula to the formation of dust grains in the 
ejecta of SNe.
The aim is to reveal how the non-steady--state effect and the physical 
conditions of the dust-forming regions affect
the condensation efficiency, average grain radius, and
size distribution.
Also, in the following sections, we discuss the applicability of
the steady--state nucleation rate.

\subsection{Grain Species and Elemental Composition of the Gas}

In the ejecta of SNe, the macroscopic mixing of elements is likely to 
be caused by the Rayleigh--Taylor instability.
Although Cherchneff \& Dwek (2009) have claimed that hydrogen atoms 
mixed with heavy elements play a critical role in the formation of 
precursor molecules of dust grains,
the microscopic mixing of hydrogen penetrating into the inner 
layer is absolutely impossible within the timescale of a few 
years, because molecular diffusion length is much smaller than the 
typical sizes of the gas shell and clumps in the ejecta
(Deneault et al.\ 2003; Clayton 2013).
Thus, we suppose the onion-like composition as an elemental 
composition in the inner ejecta of SNe, and consider the formation of 
C and MgSiO$_3$ grains individually, which are expected to form in 
the carbon-rich layer and the oxygen-rich layer of SNe, respectively.
The chemical reactions at the time of formation of C and MgSiO$_3$ 
grains and clusters, along with the physical constants necessary for 
the calculations, are taken from Table 2 of Nozawa et al.\ (2003).

It has been believed that carbon reacts with oxygen to produce CO 
molecules, and that carbon atoms tied up in CO cannot be available for 
the formation of C grains because CO is stable against dissociation.
However, in the ejecta of SNe, some of CO molecules might be destroyed 
by the collisions with energetic electrons and charge transfer
reactions with the ionized inert gas (Liu \& Dalgarno 1994, 1996;
Clayton et al.\ 1999, 2001; Clayton 2013).
Here, we do not consider the formation and destruction processes of 
CO molecules, since we treat the initial abundance of carbon atoms 
available for dust formation as a parameter.

For MgSiO$_3$ grains, we assume that the key molecule is SiO 
molecule, which has been considered to be a precursor for the 
formation of silicate grains in SNe (e.g., Kozasa et al. 1989; 
Kotak et al.\ 2009).
The initial abundance of SiO is also treated as a parameter.
The number ratios of Mg and O atoms to SiO molecules are taken to be 
$c_{{\rm Mg},0}/c_{{\rm SiO},0} = 2$ and 
$c_{{\rm O},0} /c_{{\rm SiO},0} = 20$.
These abundance ratios are of typical in the oxygen-rich layer of 
solar-metallicity SNe if almost all Si atoms are bound to SiO 
molecules (see e.g., Figure 1 of Nozawa et al.\ 2010).

\subsection{Evolution of the Gas Density and Temperature}

The number density and size distribution of newly formed dust depend 
on the time evolution of the density and temperature of the gas.
In the ejecta of SNe, the gas expands homologously after $\sim$1 day, 
and the nominal concentration of a gas species decreases as
\begin{eqnarray}
\tilde{c}(t) = c_{0} \left( \frac{t}{t_0} \right)^{-3},
\end{eqnarray}
where $c_{0}$ is the concentration at a time $t = t_0$.

On the other hand, the temperature of the gas in the ejecta is
determined by the balance between the energy input due to the decay 
of radioactive elements and the energy output due to expansion and 
radiative cooling.
In this study, as in some previous works (e.g., Kozasa et al.\ 
1989), we assume the time evolution of the gas temperature as
\begin{eqnarray}
T(t) = T_0 \left( \frac{t}{t_0}\right)^{-3 (\gamma-1)},
\end{eqnarray}
where $T_0$ is the gas temperature at $t_0$, and $\gamma$ is a 
constant parameter.
In the calculations of dust formation, we employ the capillary 
approximation expressed by Equations (38)--(42) for evaluating the 
chemical potentials $\mathring{g}_n$ and $\gamma_n$.

As the gas cools down, it shifts from unsaturated states ($\ln S < 0$)
to supersaturated ones ($\ln S > 0$) in which the formation of 
dust grains occurs.
Thus, we take $t_0$ as a time when $\ln S = 0$, and determine the
equilibrium temperature $T_0$ from the equation
\begin{eqnarray}
\ln S = \frac{A}{T_0} - B + \ln \left( \frac{c_{10} k T_0}{p_{\rm s}} 
\right) + \ln \Xi = 0
\end{eqnarray}
for a given initial concentration of the key molecule $c_{10}$ 
(and given abundance ratios of reactants and products).
Throughout this paper, the chemical potential for the formation of 
a bulk condensate from reactants per key molecule is approximated as
$(\mathring{g}_{\rm c} - {\it \Delta} \mathring{g}_{\rm gas}) / k T 
= - A/T + B$ with the numerical values $A$ and $B$ from Table 2 of 
Nozawa et al.\ (2003).
Figure 1 shows the equilibrium temperature $T_0$ for C and 
MgSiO$_3$ grains as a function of $c_{10}$.
For MgSiO$_3$, $c_{\rm Mg}/c_{\rm SiO} = 2$ and
$c_{\rm O}/c_{\rm SiO} = 20$ are adopted as mentioned above, but 
$T_0$ is insensitive to the changes in the abundance ratios.
For both the grain species, $T_0$ is higher for higher $c_{10}$, 
and $T_0 \simeq 2000$ K for C grains and $T_0 \simeq 1500$ K for 
MgSiO$_3$ grains with $c_{10} = 10^8$ cm$^{-3}$.

As mentioned in Section 2, the condensation of dust is achieved 
through the formation of clusters and their growth.
Thus, it is convenient to define the timescales characterizing these 
processes along with given temporal evolutions of the gas density and 
temperature.
The time evolution of current density and thus the formation of 
clusters are regulated mainly by the time evolution of 
$\ln S$ in $\gamma_n$ (see Equations (41), (42), and (45)).
Then, assuming that the depletion of the key molecule is negligible 
at the earlier stage of dust formation, we introduce the timescale 
of supersaturation $\tau_{\rm sat}$ with which the supersaturation 
ratio $S$ increases as follows,
\begin{eqnarray}
\tau_{\rm sat}^{-1} \equiv \frac{d \ln S}{dt}
= \frac{3 ( \gamma - 1)}{t} \left[ \frac{A}{T} 
- \frac{\gamma \omega}{\gamma - 1}  \right]
\sim \frac{A}{T} \tau_{\rm cool}^{-1},
\end{eqnarray}
where $\tau_{\rm cool} = t/3 (\gamma-1)$ is the timescale of gas 
cooling for the current model.
The second term on the right-hand side in Equation (57) generates
from the time differentiation of the term $\ln [ (p_1/p_{\rm s}) \Xi]$
in Equation (30) for the time evolutions of the gas density and 
temperature in Equation (54) and (55).
On the other hand, the growth of dust grains proceeds through the 
collision (attachment) of the key species onto their surfaces, and
the collision timescale is defined as
\begin{eqnarray}
\tau_{\rm coll}^{-1} \equiv s 4 \pi a_0^2 \tilde{c}_{1}
\left( \frac{k T}{2 \pi m_1} \right)^{\frac{1}{2}}.
\end{eqnarray}
%

\subsection{Calculations of Dust Formation}

In what follows, we adopt $n_* =100$ as the minimum number of the key 
molecule that is regarded as a grain.
This number corresponds to the minimum radius of grains $a_* = 5.9$ 
\AA~for C grains and $a_* = 10.8$ \AA~for MgSiO$_3$ grains.
The effect of changing $n_*$ on the results of calculations is 
examined in Appendix B.
We take a sticking probability $s_n = 1$ onto all sizes of clusters 
and grains.
Among the three free parameters $c_{10}$, $\gamma$, and $t_0$ in the 
models described above, we adopt $\gamma = 1.25$ and $t_0 = 300$ 
days as the standard values representing the gas cooling rate and the 
equilibrium time at which $\ln S = 0$ in the SN ejecta, although 
the cases for different $\gamma$ and $t_0$ are examined as well.
The calculations are performed up to the time long enough so that
the current density $I_*$ and the growth rate of grains can become 
negligibly small.

The formation of dust grains from the gas phase takes place when 
$\ln S > 0$.\footnote{The steady--state nucleation rate in Equation 
  (59) can be applied only for $S > 1$; see Appendix A.}
On the other hand, our non-steady--state calculations with the initial
conditions of $\ln S < 0$ and $c_n = 0$ ($n \ge 2$) have confirmed 
the following;
the formation of small ($n \la 10$) clusters is possible even if 
$\ln S < 0$, although their abundances are very small.
The abundances of small clusters at $\ln S \la 0$ are the same as 
the steady--state values as a result of too high backward reaction rates 
($\exp(\gamma_n) \gg 1$, see Equation (45)).
Therefore, the steady--state abundances of small clusters at 
$\ln S = 0$ are taken as the initial values for the simulations
starting from $t = t_0$.

We also compare the results obtained from a set of formulae as 
described in Section 2 (hereafter referred to as the non-steady 
model) with those calculated from the revised steady--state 
nucleation rate (hereafter the steady model) given by
\begin{eqnarray}
J_{\rm s} = 
s_{\rm crit} \Omega_0 
\left( \frac{2 \sigma}{\pi m_1} \right)^{\frac{1}{2}} 
\ c_1^2 \ \Pi \ \exp \left[ - \frac{4}{27} \frac{\mu^3}{(\ln S)^2}
\right],
\end{eqnarray}
for which the detailed derivation is presented in Appendix A.
In the steady model which does not involve the formation of clusters, 
the calculations of dust formation are performed by replacing $I_*$ 
with $I_{\rm s} = J_{\rm s}/\tilde{c}_1$ in Equation (47) without 
the first term on the right-hand side
and by replacing $n_*$ with $n_{\rm crit}$ given in Equation (A9).
In what follows, we refer to $I_n$ and $I_{\rm s}$ as current densities
and steady--state current density, respectively, for convenience.

\section{RESULTS OF DUST FORMATION CALCULATIONS}

Given the parameters $\gamma$ and $t_0$ representing the cooling and
dynamical times of the gas, the concentration $c_{10}$ controls the
behavior of the formation processes of clusters and grains.
We first present the results of the calculations in the case that
the initial concentration of the key molecule is high enough that 
the assumption of a steady state is considered to be a good 
approximation.
Then, we demonstrate the results for the low-density case.
We also explore the dependence of the results on $t_0$ and $\gamma$
in Section 4.3.

\subsection{High Density Case}

Figure 2 illustrates the formation process of C grains as a function 
of time ($x = t/t_0$) for $c_{10} = 10^8$ cm$^{-3}$,
$t_0 = 300$ days, and $\gamma = 1.25$;
Figure 2(a) depicts the evolutions of abundances of $n$--mer clusters 
$Y_n$ ($n \ge 2$) as well as of the key molecules $Y_1$, and
Figure 2(b) the time evolutions of the current density of $n_*$--mer 
$I_*$, average grain radius $a_{\rm ave}$, and condensation efficiency 
$f_{\rm con}$.
Note that, in the figure, the time evolution of the average radius
is plotted after the time at which the condensation efficiency 
reaches $10^{-10}$, and hereafter the time is referred to as the onset 
time $x_{\rm on}$ of dust formation.\footnote{
  The threshold value $10^{-10}$ is arbitrary, but, given that 
  $f_{\rm con}$ rises up quickly with time, it does not affect the 
  conclusion of this paper as long as the value is less than 
  $\sim$$10^{-5}$.}
Figure 2(c) presents the time evolutions of the supersaturation ratio
$S$ and the critical size $n_{\rm crit}$ that is defined as the 
size satisfying the condition $\gamma_n = 0$, and Figure 2(d) the time 
evolutions of current densities $I_n$ for the formation of given 
$n$--mers.
The results for MgSiO$_3$ grains are provided in Figure 3.

For both the grain species, the non-steady--state formation process 
of dust is described as follows.
As can be seen from (c) and (d) of Figures 2 and 3, the increase in 
$\ln S$, induced by the decrease in gas temperature with time, leads 
to the formation of clusters with larger $n$ progressively.
Once $\ln S$ reaches $\simeq$2 at which $n_{\rm crit} \simeq 100$, 
$I_*$ becomes high enough that some amount of grains with $n \ge 100$ 
start to form.
A further increase in $\ln S$ enhances $I_n$, producing a much greater 
number of clusters and grains.
Note that the onset time of dust formation is later for C grains 
($x_{\rm on} \simeq 1.08$) than for MgSiO$_3$ grains 
($x_{\rm on} \simeq 1.03$), which stems partly from the longer 
timescale of supersaturation and partly from the larger surface
tension:
$\tau_{\rm sat} \simeq$ 10 days and $\sigma = 1400$ erg cm$^{-3}$ 
for C grains 
($\tau_{\rm sat} \simeq$ 2.5 days and $\sigma = 400$ erg cm$^{-3}$  
for MgSiO$_3$ grains).
Since newly formed grains grow efficiently to cause the consumption 
of the key molecule, the supersaturation ratio $S$ reaches a maximum 
and then decreases.
The critical size $n_{\rm crit}^{\rm smax}$ at $S = S_{\rm max}$ is 
$\sim$20 for C grains ($\sim$10 for MgSiO$_3$ grains).
The current densities $I_n$ for the formation of clusters with 
$n < n_{\rm crit}^{\rm smax}$ cease almost abruptly just before $S$ 
becomes $S_{\rm max}$, whereas those for $n > n_{\rm crit}^{\rm smax}$
--mer reach almost the same maximum value at $S \simeq S_{\rm max}$ 
and then decrease quickly.
Accordingly, $I_*$ has a sharp peak;
the gas temperature at the peak of $I_*$ is 1850 K (1490 K) for C 
(MgSiO$_3$) grains, being lower than its equilibrium temperature 
$T_0 = 1990$ K (1530 K).
After then, dust grains continue to grow until $f_{\rm con} \simeq 1$
by consuming almost all of the key molecules.

It should be noticed here that the current densities for the formation 
of clusters with $n \ga n_{\rm crit}^{\rm smax}$ around 
$S = S_{\rm max}$, being almost independent of $n$, reach a 
steady--state value in this high-density case for which the condition 
of $\tau_{\rm coll} \ll \tau_{\rm sat}$ is satisfied (see Section 5).
In addition, the steady--state value excellently matches the 
steady--state current density $I_{\rm s}$, as is seen from (b) and (d) 
in Figures 2 and 3, where we overplot the results obtained from the 
steady model (dotted lines).

The behavior of the formation process of clusters and grains in this 
high-density case can be qualitatively understood by inspecting the 
time evolution of $\mathrm{e}^{\gamma_n}$ regulating the backward 
reactions in Equation (45).
The factor $\mathrm{e}^{\gamma_n}$ is a decreasing function of $n$, 
and in a supersaturated gas, it becomes below unity for $n$ larger 
than a critical size $n_{\rm crit}$ approximately given as
\begin{eqnarray}
n_{\rm crit} - \frac{1}{\omega}
\simeq \left( \frac{2 \mu}{3 \ln S} \right)^3
\end{eqnarray}
for $n_{\rm crit} \gg 1$.
Note that this expression for $n_{\rm crit}$ is equivalent to that 
for the steady--state current density as defined by Equation (A9).
At the initial phase of $\ln S \la 2$, $n_{\rm crit}$ is very large 
($n_{\rm crit} > 100$) (see Figures 2(c) and 3(c)), so the backward 
reactions are dominant ($\mathrm{e}^{\gamma_{n}} \gg 1$) for any 
size of $n$--mers.
In this case, the abundances of $n$--mers at each time approximately 
take the values, $Y_n \simeq Y_{n-1} \mathrm{e}^{- \gamma_n}$, and 
the current to larger $n$--mers is considerably smaller 
($I_n/I_{n-1} \ll 1$).
On the other hand, as $\ln S$ increases with time, $n_{\rm crit}$ 
falls below $n_* = 100$, reaching down to 10--20.
In this phase, the backward reactions are suppressed 
($\mathrm{e}^{\gamma_{n}} < 1$) for $ n \ga n_{\rm crit}$, so the 
$n$--mers can grow exclusively through the collisions with the key 
molecules.
Since the collision timescale is extremely small 
($\tau_{\rm sat} / \tau_{\rm coll} \gg 1$) in this high-density case, 
the reactions for $n \ga n_{\rm crit}$ proceed instantaneously, which 
makes the current densities $I_n$ being in a steady state 
(i.e., $I_n \simeq I_{n-1}$ for $n \ga n_{\rm crit}$).
Furthermore, the agreement of $I_*$ with $I_{\rm s}$ can be interpreted
as follows;
even if $\ln S$ is high enough, the backward reactions remain 
predominant for $n \la n_{\rm crit}$, where the relation 
$Y_n \simeq Y_{n-1} \mathrm{e}^{- \gamma_n}$ holds.
Thus, the abundance of $n_{\rm crit}$--mers can be approximately 
estimated as
\begin{eqnarray}
Y_{n_{\rm crit}} 
\simeq Y_{1} \exp \left(- \sum_{n=2}^{n_{\rm crit}} \gamma_n \right) 
\simeq Y_{1} \exp \left[- \mu \left( n_{\rm crit} - \frac{1}{\omega} 
\right)^{\frac{2}{3}} + 
\left( n_{\rm crit} - \frac{1}{\omega} \right) \ln S \right].
\end{eqnarray}
Then, the current of $n_*$--mers is found to be on orders of 
\begin{eqnarray}
I_* \simeq I_{n_{\rm crit}} 
\sim \tau_{n_{\rm crit} - 1}^{-1} Y_{n_{\rm crit}} 
\sim (n_{\rm crit} -1)^{\frac{2}{3}} \tau_{\rm coll}^{-1} Y_1
\exp \left[- \frac{4 \mu^3}{ 27 (\ln S)^2} \right].
\end{eqnarray}
The exponential term in Equation (62), which dominates the time 
evolution of the current density, has the same form as the term in 
the steady--state current density $I_{\rm s}$.
Hence, the current density for the formation of $n_*$--mers
$I_*$ is 
essentially equal to the steady--state current density $I_{\rm s}$.
This allows us to conclude that the application of the steady--state 
current density could be valid as long as the consumption of the key 
molecules due to formation and growth of clusters and grains causes the
supersaturation ratio $S$ to decrease in the course of time evolution 
of gas density and temperature, as is demonstrated in the case of a 
high initial gas density.

Figures 2(e) and 3(e) show the final size distributions of grains 
and clusters.
The size distribution is lognormal--like with 
$a_{{\rm ave},\infty} = 0.07$ $\mu$m for C grains and 
$a_{{\rm ave},\infty} = 0.08$ $\mu$m for MgSiO$_3$ grains.
Since the size distributions of grains follow the time evolution of 
$I_*$ and $I_{\rm s}$ is equal to $I_*$, we can see that the size 
distribution in the non-steady model is the same as those in the 
steady model.
In this high-density case with the final condensation efficiency 
$f_{{\rm con},\infty} \simeq 1$, the abundance of the key molecules 
locked in the clusters is extremely small ($\sum n Y_n < 10^{-7}$). 
Our calculations show that, when the initial concentration is as high 
as $c_{10} \ga 10^7$ cm$^{-3}$, all carbon atoms and SiO molecules 
are ultimately locked in grains with the size distributions identical 
to those in the steady models.

\subsection{Low Density Case}

Figures 4 and 5 show the formation processes of C and MgSiO$_3$ grains, 
respectively, for $c_{10} = 10^5$ cm$^{-3}$, $t_0 = 300$ days, and 
$\gamma = 1.25$.
Even for such a low initial gas density, the formation of $n$--mers
progresses as $\ln S$ increases, as is the same as the high-density 
case.
However, even if grains with $n\ge 100$ are produced, they cannot 
grow efficiently through attachment of the key molecules because the
collision timescale is considerably longer
($\tau_{\rm sat}/\tau_{\rm coll} \sim$ 1--10).
Thus, despite the fact that the formation and growth of clusters and 
grains significantly consume the key molecules, $\ln S$ continues to 
increase (Figures 4(c) and 5(c)), and $I_*$ gradually decreases after 
passing the peak in contrast to the high-density case (Figures 4(d) 
and 5(d)), which results in the formation of many small grains with 
$n \la 1000$.
Finally, the depletion due to formation of clusters and grains and/or 
the dilution due to the expansion of the gas makes the concentration 
of the key molecules too low to advance the further growth, and the
abundances of clusters approach to the constant values 
(see Figures 4(a) and 5(a)).

Figures 4(b) and 5(b) compare the time evolutions of $I_*$ (and 
$I_{\rm s}$), $a_{\rm ave}$, and $f_{\rm con}$ between the non-steady 
and the steady models.
It can be seen that the non-steady current density $I_*$ rises up at 
a time later than the steady--state current density $I_{\rm s}$, 
corresponding to the later onset time of dust formation, and that its 
peak value is much smaller than that in the steady model.
This is at odds with the high-density case, indicating that the steady 
model is no longer appropriate for this low-density case with 
$\tau_{\rm sat} / \tau_{\rm coll} \la 10$ during the evolution.
For clusters with $n \la n_{\rm crit}$, as mentioned in the previous 
subsection, $Y_n$ evolves as 
$Y_n \simeq Y_{n-1} \mathrm{e}^{- \gamma_n}$, and
$I_{n_{\rm crit}}$ at each time is on order of $I_{\rm s}$ 
(c.f., Equation (62)).
However, the collision timescale is too long for the current densities 
$I_n$ to establish a steady state at $n \ga n_{\rm crit}$, so $I_*$ 
remains much lower than $I_{n_{\rm crit}} \simeq I_{\rm s}$ 
(Figures 4(d) and 5(d)).

There also appear differences in the final average radius and 
condensation efficiency of dust grains between the non-steady and 
steady models. 
The final condensation efficiency in the non-steady model is 
$f_{{\rm con},\infty} \simeq 0.3$ for C grains 
($f_{{\rm con},\infty} \simeq 0.01$ for MgSiO$_3$ grains), 
which is lower than $f_{{\rm con},\infty} = 1$ in the steady model for 
both the grain species.
In both the models, the key molecules are little left in the gas phase 
($f_{{\rm dep},\infty} < 10^{-5}$) in the end, indicating that 70 \% 
(99 \%) of them are finally bound to clusters for the non-steady model.
On the other hand, the final average radius of dust grains
is larger for the non-steady model ($a_{{\rm ave},\infty} = 0.0007$ 
$\mu$m for C grains and 0.0011 $\mu$m for MgSiO$_3$ grains) 
than for the steady model ($a_{{\rm ave},\infty} = 0.0004$ $\mu$m for 
C grains and 0.0005 $\mu$m for MgSiO$_3$ grains).

The discrepancies in the average grain radius and condensation
efficiency between the two models simply reflect the difference in 
the minimum size considered as grains.
As can be seen from Figures 4(e) and 5(e), the final size distribution 
in the steady model is quite similar to the combined size distribution 
of grains and clusters in the non-steady model.
This is because $n_{\rm crit}$ and $\tau_{\rm coll}$ are the same 
in both the models, so the number of $n_{\rm crit}$--mer formed
at a given time and the growth rate are essentially identical.
However, in the steady model, clusters that meet $n \ge n_{\rm crit}$ 
are taken as bulk grains, and $n_{\rm crit}$ is normally less than 10,
even down to $\simeq$1--2 in the low density case as shown in Figures 
4 and 5. 
Thus, the steady model, which regards small $n$--mers as grains, 
leads to a smaller average grain radius and a higher condensation 
efficiency.
Although we cannot have a clear number of constituent atoms to 
distinguish between small clusters and grains, it must be 
unreasonable to consider clusters with $n \le 10$ to hold the 
properties of bulk grains.

The results of calculations for this low initial density clearly 
attest that the application of the steady--state current density
overestimates the condensation efficiency and underestimates the 
average grain radius for dust formation in less dense and/or rapidly 
cooling environments.
Also, it may be useful to point out here that the application of the 
steady--state current density with a given cut-off value of critical 
size (e.g., Bianchi \& Schneider 2007) in low-density/rapidly cooling
environments, arguing the inadequacy of the extension to smaller 
critical sizes, leads to the considerable depression of the 
condensation efficiency and enhancement of the average grain radius, 
and cannot reproduce the combined size distribution of clusters and 
grains.

\subsection{Dependence on $t_0$ and $\gamma$}

In the subsections 4.1 and 4.2, we have demonstrated how the initial 
concentration of the key molecule affects the formation process
and properties of newly formed grains.
In this subsection, we investigate the dependence of other free
parameters $t_0$ and $\gamma$ on the formation process, average 
radius, and size distribution of dust grains.

Figure 6(a) plots the formation process of C grains for $t_0 =$ 100, 
300, and 600 days with $c_{10} = 10^7$ cm$^{-3}$ and $\gamma =$ 1.25. 
The results of the calculations show that a larger $t_0$ leads to a 
smaller peak of $I_*$ as well as a little earlier onset time of
dust formation $x_{\rm on}$.
This is explained as follows:
as seen from Equation (57), the timescale of supersaturation in 
terms of $x$, $(d \ln S / dx)^{-1}$ is independent of $t_0$, 
while the timescale of grain growth, 
$(d \ln a / dx)^{-1} \propto \tau_{\rm coll} / t_0$, is inversely 
proportional to $t_0$.
This means that the increase in $t_0$ makes grain growth more active
but has little impact on the number of clusters at a given time $x$.
Therefore, for a larger $t_0$, dust grains capture the key molecules
more efficiently through their growth, which causes a faster rise of 
$f_{\rm con}$ and a faster drop of $I_*$ (a smaller peak of $I_*$)
and results in the grain size distribution weighted toward a larger 
radius (see Figure 6(b)).
Thus, the increase in $t_0$ enhances the effect of grain growth 
relative to formation of clusters and act to produce dust grains with 
large average radii.

Figure 7 gives the results of the calculations for the formation 
of MgSiO$_3$, adopting $\gamma =$ 1.25, 1.4, and 1.6 for
$c_{10} = 10^7$ cm$^{-3}$ and $t_0$ = 300 days.
For a larger $\gamma$, which corresponds to a more rapid cooling of 
the gas, the onset time of dust formation is earlier, and $I_*$ has 
a higher peak.
Again considering the timescale in terms of $x$, the timescale of 
supersaturation, 
$(d \ln S / dx)^{-1} \propto x^{-3 \gamma+4}/(\gamma-1)$ decreases 
with increasing $\gamma$, whereas the timescale of grain growth,
$(d \ln a / dx)^{-1} \propto x^{3 (\gamma+1) / 2}$, increases.
Hence, for a larger $\gamma$, a more rapid increase in $\ln S$ as
well as a more rapid decrease in $n_{\rm crit}$ leads to the formation
of a larger number of clusters with $n \ga n_{\rm crit}$ at an earlier 
time before grain growth efficiently consumes the key molecules.
Here it should be noted that the collision timescale $\tau_{\rm coll}$
in the models considered here is still short enough so that the 
consumption of the key molecules due to grain growth makes $\ln S$ 
decrease during the evolution.
Consequently, in the model with larger $\gamma$, $I_*$ increases more 
rapidly and has a higher and narrower peak at an earlier time, and the 
average grain radius as well as the peak radius of size distribution 
becomes smaller, as seen from Figures 7(a) and 7(b).
In conclusion, the increase in $\gamma$ makes the formation of 
clusters more active relative to grain growth and induces the 
formation of dust grains with small average radii even if 
$\tau_{\rm sat}/\tau_{\rm coll}$ is not much larger than unity.

\section{THE SCALING RELATIONS FOR AVERAGE GRAIN RADIUS AND
CONDENSATION EFFICIENCY}

The objects to be clarified in the study of dust formation in
astrophysical environments are not only the chemical composition
of dust grains but also their amount and size distribution.
In the ejecta of SNe, the knowledge on size distribution of newly 
formed dust is crucial for unraveling what amount and size of dust 
grains are finally ejected from SNe to the ISM, because the destruction 
efficiency of dust by the reverse shock heavily depends on the size 
distribution (e.g., Nozawa et al.\ 2007).
Our results show that the size distribution of dust formed for given 
time evolutions of gas density and temperature is lognormal--like 
as long as $\tau_{\rm sat}/\tau_{\rm coll} \gg 1$ during the formation 
of dust. 
Therefore, the average grain radius can be taken as a representative 
measurement of the size distribution.
The condensation efficiency is also a fundamental quantity in 
estimating the mass of newly formed dust.
In this section, we explore how the average grain radius and 
condensation efficiency can be constrained from a physical condition 
at the time of dust formation, and derive the scaling relations for
the average radius and condensation efficiency, referring to the 
results of the calculations presented in the previous section.
In addition, we clarify in what conditions the steady--state nucleation
rate is applicable.
Then, we shall present some examples of the application of the
scaling relations for the formation of dust in SNe.

\subsection{The Physical Quantity Characterizing Dust Formation Process
and the Scaling Relations}

The results in Section 4 demonstrate that the formation process of 
dust is determined by the competition between the formation of 
clusters and the growth of grains.
Although the onset time of dust formation (and the condensation time
at which the current density $I_*$ reaches the maximum), average grain 
radius, and condensation efficiency depend on $c_{10}$, $\gamma$, and 
$t_0$ in a complicated manner, we have shown that the behavior of 
formation processes of dust grains can be qualitatively interpreted in
terms of $\tau_{\rm sat}$ and $\tau_{\rm coll}$ during the formation 
of dust.
It has been shown in the studies based on the steady--state nucleation 
rate that the average radius and number density of dust grains formed in
a cooling gas undergoing macroscopic motion can be scaled by a 
non-dimensional quantity $\Lambda = \tau_{\rm sat}/\tau_{\rm coll}$
at the condensation time when the nucleation rate reaches the maximum 
(e.g., Hasegawa \& Kozasa 1988).

The formation time of dust is the most direct information that can be 
obtained from observations of SNe.
Hence, it is useful to relate the average grain radius and condensation 
efficiency to the gas density and temperature at the time of dust 
formation, in order to get the information on physical conditions in 
the ejecta from the observations and vice versa.
One of the best indicators for the formation time of dust would be 
the condensation time, $t_{\rm c}$, defined as the time when $I_*$ 
has a peak.
However, the depletion of the key molecules due to the formation and 
growth of clusters and grains at $t_{\rm c}$ is considerably large 
($Y_1 \simeq$ 0.1--0.2 at $t_{\rm c}$), which has significant effects 
on the relevant physical quantities.
Thus, we adopt, as the time of dust formation, the onset time of 
dust formation, $t_{\rm on}$, defined as the time at which 
$f_{\rm con}$ reaches $10^{-10}$.
Then, the non-dimensional physical quantity $\Lambda_{\rm on}$ 
characterizing the formation process of dust grains is given as 
\begin{eqnarray}
\Lambda_{\rm on} 
&\equiv& \frac{ \tau_{\rm sat}(t_{\rm on}) }{ \tau_{\rm coll} (t_{\rm on}) }
\sim \frac{t_{\rm on}}{3 (\gamma -1)} \frac{T_{\rm on}}{A}
\times s 4 \pi a_0^2 \tilde{c}_{\rm on}
\left( \frac{ k T_{\rm on} }{ 2 \pi m_1} \right)^{\frac{1}{2}} 
\nonumber \\
&\sim& \frac{C}{ \gamma - 1 }
\left( \frac{s}{1.0} \right)
\left( \frac{\tilde{c}_{\rm on}}{10^8 ~{\rm cm}^{-3}} \right)
\left( \frac{T_{\rm on}}{2,000 ~{\rm K}} \right)^{\frac{3}{2}}
\left( \frac{t_{\rm on}}{300~{\rm days}} \right),
\end{eqnarray}
where $\tilde{c}_{\rm on} = \tilde{c}_1(t_{\rm on})$ and 
$T_{\rm on} = T(t_{\rm on})$, and $C = 1.94 \times 10^3$ 
($1.15 \times 10^3$) for C grains (MgSiO$_3$ grains).
In Equation (63), we employ the approximation
$\tau_{\rm sat}^{-1} \simeq (A/T) \tau_{\rm cool}^{-1}$ 
(see Equation (57)), although we have calculated $\Lambda_{\rm on}$ 
without using this approximation in the following figures.

Figure 8 presents the final average grain radius $a_{{\rm ave},\infty}$
and condensation efficiency $f_{{\rm con},\infty}$ as a function of 
$\Lambda_{\rm on}$ calculated for $\gamma =$ 1.1, 1.3, 1.5, and 1.7
by covering a wide range of $c_{10}$ and $t_0$. 
The figures show that, as $\Lambda_{\rm on}$ increases, 
$a_{{\rm ave},\infty}$ and $f_{{\rm con},\infty}$ increase, and
$f_{{\rm con},\infty} = 1$ at $\Lambda_{\rm on} \ga$ 20--30 for 
both C and MgSiO$_3$ grains;
for a larger $\Lambda_{\rm on}$, grain growth becomes more dominant 
over the formation of clusters, and as a result larger grains are 
formed to lock up all of the key molecules.
The remarkable consequence of Figure 8 is that $a_{{\rm ave},\infty}$ 
and $f_{{\rm con},\infty}$ for different $\gamma$ (especially for
$\gamma \ga 1.2$) are, respectively, plotted almost completely 
as a single curve for both C grains (Figure 8(a)) and MgSiO$_3$ 
grains (Figure 8(b)).
This means that the average grain radius and condensation efficiency 
can be uniquely determined by one parameter $\Lambda_{\rm on}$, 
except for C grains formed in extremely slowly cooling gas with low 
densities corresponding to the case of $\gamma = 1.1$ with 
$\Lambda_{\rm on} \la 10$.
In the figures, we also plot $a_{{\rm ave},\infty}$ and 
$f_{{\rm con},\infty}$ from the steady model for $\gamma = 1.3$.
They deviate from those from the non-steady model at 
$\Lambda_{\rm on} \la 30$, where the steady model predicts too small 
$a_{{\rm ave},\infty}$ to be regarded as bulk grains with keeping 
$f_{{\rm con},\infty} = 1$.
In other words, the steady--state nucleation rate is applicable
only if $\Lambda_{\rm on} \ga 30$.

Then, we derive the approximation formulae 
describing the dependence of $a_{{\rm ave},\infty}$ and
$f_{{\rm con},\infty}$ on $\Lambda_{\rm on}$ for the non-steady model,
which are, respectively, given by
\begin{eqnarray}
\log \left( \frac{a_{{\rm ave},\infty} }{ a_* } -1 \right) =
\epsilon_1 + \epsilon_2 \log \Lambda_{\rm on}
\end{eqnarray}
and
\begin{eqnarray}
\log f_{{\rm con},\infty} = 
\chi_1 \left[ \tanh \left( \chi_2 \log \Lambda_{\rm on} + \chi_3 
\right) -1 \right],
\end{eqnarray}
where the fitting parameters $\epsilon_1$, $\epsilon_2$, and 
$\chi_k$ ($k =$ 1--3) are given in Table 1.
In Figure 9, $a_{{\rm ave},\infty}$ and $f_{{\rm con},\infty}$ 
calculated by the above fitting formulae are compared with the results 
of simulations for $\gamma = 1.25$.
Equation (64) reproduces the calculated average radii with the 
error less than 5 \% for $\Lambda_{\rm on} \le 10^6$.

It should be emphasized here that the scaling relations given above
are independent of the initial conditions of the calculations and the
time evolution of the gas density.
In fact, we have performed the dust formation calculations for
expanding gas flows with a constant velocity by changing $\gamma$, 
and confirmed that the resulting $a_{{\rm ave},\infty}$ and 
$f_{{\rm con},\infty}$ entirely coincide with those shown in Figure 8.
Therefore, the average grain radius and condensation efficiency
of a given grain species can be universally described by the 
corresponding non-dimensional physical quantity $\Lambda _{\rm on}$.

\subsection{Application of the Scaling Relations to Dust Formation 
in SNe}

Equation (64) allows us to estimate the typical size of 
newly formed grains once we know the density of the gas in the ejecta 
and the onset time of dust formation (or the formation time of dust).
For example, from Figures 1 and 2 in Nozawa et al.\ (2003), the 
concentration of carbon atoms in the carbon-rich He layer of Type 
II--P SNe is found to be $\tilde{c}_{\rm on} \simeq$ $10^8$--$10^9$ 
cm$^{-3}$ at $t_{\rm on} = 330$ days.
For the reasonable values of $\gamma$ ($\simeq$1.5--1.7) and  
$T_{\rm on}$ ($\simeq$2,000 K), Equation (64) presents 
$a_{{\rm ave},\infty} =$0.03--0.3 $\mu$m ($\Lambda_{\rm on} =$ 
(0.3--8$) \times 10^4$, see Figure 9(a)), which is consistent with 
the average grain radii given in Figure 7 of Nozawa et al.\ (2003).
In Type IIb SNe with much less massive hydrogen envelopes, the 
density of the gas at $t_{\rm on} = 330$ days is by a factor of 
100--300 lower than in Type II--P 
(see Figure 2 of Nozawa et al.\ (2010)), and Equation (64) leads to 
$a_{{\rm ave},\infty} \sim$ 0.001 $\mu$m 
($\Lambda_{\rm on} \sim 100$).
This radius also agrees with that obtained in Nozawa et al.\ (2010).
These simple analyses suggest that our previous calculations by a
theory of the non-steady--state nucleation and grain growth applying
the steady--state nucleation rate with a relaxation time toward the
steady--state rate (see Nozawa et al.\ 2003) were performed under 
the condition that the steady--state approximation is appropriate.

Observationally, there have been few studies that reported a typical 
size of dust formed in SNe.
Recently, Maeda et al.\ (2013) clearly detected the formation of C 
grains in the luminous Type IIn SN 2010jl around day 550 after the 
explosion by the optical through the near-infrared observation.
They suggested that the typical radius of the dust grains is less 
than 0.1 $\mu$m (more probably $\la$0.01 $\mu$m) to account for
the wavelength-dependence of obscuration of hydrogen emission lines.
These C grains are likely to have formed not in the ejecta but in 
relatively dense clumps in the shocked circumstellar shell, but it 
would be interesting to compare with our results.
Based on the simple argument of optical depth, Maeda et al.\ (2013) 
showed that the gas density in the interclump medium must be
$c_{\rm gas} \la 10^9$ cm$^{-3}$.
Then, assuming the abundance of carbon atoms to be 
$c_1/c_{\rm gas} = 10^{-4}$, and with typical values of $\gamma$ and 
$T_{\rm on}$, $\Lambda_{\rm on} \la 8 \times 10^2 (D / 100)$, where 
$D$ is the density contrast between the dense clumps and the 
interclump medium. 
Adopting $D =$ 100--1000, Equation (64) yields 
$a_{{\rm ave},\infty} \le$ 0.01--0.08 $\mu$m, which is consistent 
with the grain size estimated from the observation.
This emphasizes that Equations (64) and (65) could provide powerful 
constraints on the properties of newly formed grains through the gas 
density and condensation time extracted from observations as well as
those predicted from theoretical models.

\section{Summary}

We have developed a new formulation describing the non-steady--state 
formation of small clusters and grains in a self-consistent manner,
taking into account chemical reactions at the time of dust formation 
and assuming that the temperatures of small clusters are the same as
that of the gas.
Given the chemical potentials of small clusters, the formula can be 
applied to investigate the formation process of dust grains in 
rarefied astrophysical environments, where the steady--state 
nucleation rate is not applicable.
It should be pointed out here that the formation process of dust is
formulated in the present study under the assumption that the 
smallest cluster (dimer) has the same chemical composition as the 
grains.
However, the formulation can be extended for the case that chemical
compositions of small clusters are different from the grains, given 
the chemical reaction paths and chemical potentials.
Also, the formula can be extended and applied to explore the effects
of the difference in temperatures of small clusters and the gas
(Kozasa et al.\ 1996; Yasuda \& Kozasa 2012) as well as the
temperature fluctuation (Keith \& Lazzati 2011) and the shape of
small clusters (Fallest et al.\ 2011) on the formation process of
dust grains.
These subjects will be studied in the future works.

Applying the new formulation with the capillary approximation for
evaluating the chemical potentials of small grains, we have 
investigated the formation processes of C and MgSiO$_3$ grains over 
a wide range of physical conditions expected in the ejecta of SNe.
The results of the calculations have shown that the behavior of 
non-steady--state formation process of small clusters and grains
can be qualitatively interpreted in terms of the temporal evolutions
of the collision timescale of key molecule $\tau_{\rm coll}$ and
the supersaturation timescale of the gas $\tau_{\rm sat}$ during the 
formation of dust;
in the condition that $\tau_{\rm coll} \ll \tau_{\rm sat}$, the
formation process of dust grains can be completely reproduced by the
steady--state nucleation rate, and grains form with the condensation 
efficiency $f_{\rm con} \simeq 1$, otherwise the formation of
clusters and grains proceeds in a non-steady state, and the resulting
condensation efficiency is $f_{\rm con} < 1$ with the efficiency of
growth of clusters and grains being depressed considerably.

Analyzing the results of the model calculations, we found that the 
condensation efficiency and average radius of newly formed grains 
can be fully described by one non-dimensional quantity
$\Lambda_{\rm on}$, the ratio of the supersaturation timescale to the 
collision timescale at the onset time of dust formation, although the 
time evolutions of gas temperature and density considerably influence
the formation process as well as the average grain radius and
condensation efficiency.
Also, we have revealed that the steady--state nucleation rate is
applicable under the condition of $\Lambda_{on} \ga 30$, irrespective 
of grain species;
otherwise the application of the steady--state nucleation rate results 
in the formation of a large number of unreasonably small grains 
with the condensation efficiency considerably higher that calculated 
by the non-steady rate. 
Furthermore, we have derived the scaling relations for the
average radius and condensation efficiency of C and MgSiO$_3$ grains
as a function of $\Lambda_{\rm on}$.
The approximation formulae depend neither on the time evolution of the
gas density and temperature nor on the initial condition, and thus 
could serve as a universal relation to predict the mass and average 
size of newly formed grains from the observations and/or the model 
calculations of explosions of supernovae and novae as well as
mass-loss winds from stars.

\acknowledgments

We are grateful to the anonymous referee for critical comments
that improved the manuscript.
This research has been supported by World Premier International 
Research Center Initiative (WPI Initiative), MEXT, Japan, and by the 
Grant-in-Aid for Scientific Research of the Japan Society for the 
Promotion of Science (20340038, 22684004, and 23224004).

\newpage

\appendix

\section{A Revised Formula of Steady-state Nucleation Rate}

The rate of the steady--state homogeneous nucleation accompanied by
chemical reactions of relevant gaseous molecules has been developed 
by introducing the concept of the key species (Kozasa \& Hasegawa 1987)
and has been applied for the formation of dust not only in the ejecta 
of SNe but also in the solar nebula (Kozasa \& Hasegawa 1988) and in
circumstellar envelopes of asymptotic giant branch stars 
(e.g., Kozasa \& Sogawa 1997, 1998; Chigai et al.\ 1999).
Yamamoto et al.\ (2001) tried to generalize the formulation for the 
steady--state nucleation rate and derived an analytic formula in which 
the effect of chemical reactions at the time of formation is included 
as a correction factor in a form of functions of partial pressures of 
reactants except for the key species and products as well as the 
standard gas pressure.
However, the formula is incomplete because the nucleation rate depends 
on the standard gas pressure explicitly in the case that the number 
of product molecules is different from that of reactant molecules 
except for the key species.
Thus, in this Appendix, we reformulate the analytic formula for the
rate of the steady--state nucleation proceeding through the chemical 
reactions at the time of formation. 

Here we consider the formation of dust grains through the chemical 
reactions (16) and (17). 
In a steady state, the current density $J_n$ from $(n-1)$--mer to 
$n$--mer is independent of $n$, being identical to the steady--state 
nucleation rate $J_{\rm s}$.
Thus, Equations (19) and (33) lead to the following relations, 
respectively,
\begin{eqnarray}
\frac{J_{\rm s}}{\alpha_1 c_1^2} 
= 1 - \frac{c_2}{\mathring{c}_2} \frac{1}{b^2}
~~~~~~~~~~~
\end{eqnarray}
and 
\begin{eqnarray}
\frac{J_{\rm s}}{\alpha_{n-1} c_1 \mathring{c}_{n-1}}
= \frac{c_{n-1}}{\mathring{c}_{n-1}} -
\frac{c_n}{\mathring{c}_n} \frac{1}{b} 
~~~ {\rm for} ~~ n \ge 3,
\end{eqnarray}
where $b$ is defined by Equation (20).
By summing up Equation (A1) and Equation (A2) multiplied by 
$1/b^{n-1}$ successively, the nucleation rate $J_{\rm s}$ can be
derived from the equation
\begin{eqnarray}
J_{\rm s} \left( \frac{1}{\alpha_1 c_1^2} +
\sum_{i=2}^{n} \frac{1}{\alpha_i c_1 \mathring{c}_i b^i} \right)
= 1 - \frac{c_n}{\mathring{c}_n} \frac{1}{b^n}.
\end{eqnarray}
By applying Equations (28) and (35), the factor 
$1/\mathring{c}_n b^n$
on the second term in the right-hand side of Equation (A3) is written 
as
\begin{eqnarray}
\frac{1}{\mathring{c}_n b^n} = \frac{1}{c_1 \Pi} \
\exp \left\{ \frac{1}{k T} \left( \mathring{g}_n - 
n {\it \Delta} \mathring{g}_{\rm gas} \right)
- \left( n - \frac{1}{\omega} \right) \left[
\ln \left( \frac{p_1}{p_{\rm s}} \right) +\ln \Xi \right] \right\} 
\equiv \frac{1}{c_1 \Pi} \exp\left(\gamma'\right) 
\end{eqnarray}
with
\begin{eqnarray}
\gamma'_n = \frac{1}{k T} \left[
\mathring{g}_n - \left( n - \frac{1}{\omega} \right) 
\mathring{g}_{\rm c}
- \frac{1}{\omega} {\it \Delta} \mathring{g}_{\rm gas} \right]
- \left( n - \frac{1}{\omega} \right) \ln S,
\end{eqnarray}
\noindent
where $S$ is the supersaturation ratio defined by Equation (30). 
Since $\gamma'_n \simeq -n \ln S$ and  
$\mathring{c}_n b^n \propto S^n$ for $n \gg 1$, 
the second term on the right-hand side of Equation (A3) approaches 
to zero for $S >1$ as $n \rightarrow \infty$.
Then, the steady state nucleation rate $J_{\rm s}$ is given by
\begin{eqnarray}
\frac{1}{J_{\rm s}} = \frac{1}{\alpha_1 c_1^2} +
\sum_{i=2}^{\infty} \frac{1}{\alpha_i c_1 \mathring{c}_i b^i}=
\frac{1}{\alpha_1 c_1^2} +
\sum_{i=2}^{\infty} \frac{1}{\alpha_i c_1^2 \Pi} \exp(\gamma'_i).
\end{eqnarray}

As long as $1 / \alpha c_1^2 \ll 1$, the summation in Equation (A6) 
can be replaced with the integration, and the nucleation rate 
$J_{\rm s}$ can be approximately calculated as
\begin{eqnarray}
\frac{1}{J_{\rm s}} \simeq \frac{1}{c_1^2 \Pi}
\int_{2}^{\infty} \frac{1}{\alpha_i} \exp(\gamma'_i) di.
\end{eqnarray}
Under the capillary approximation given by Equation (40), 
$\gamma'_n$ is expressed as
\begin{eqnarray}
\gamma'_n = \mu \left( n -\frac{1}{\omega} \right)^{\frac{2}{3}} 
- \left( n - \frac{1}{\omega} \right) \ln S,
\end{eqnarray}
and have a maximum at $n_{\rm crit}$ given by
\begin{eqnarray}
\left( n_{\rm crit} -\frac{1}{\omega} \right)^{\frac{1}{3}}=
\frac{2}{3} \frac{\mu}{\ln S}.
\end{eqnarray}
Thus, the integration of Equation (A7) with the saddle-point method 
results in the steady state nucleation rate given by
\begin{eqnarray}
J_{\rm s} = s_{n_{\rm crit}} 
\Omega_0 \left( \frac{2 \sigma}{\pi m_{n_{\rm crit},1}} 
\right)^{\frac{1}{2}} \frac{n_{\rm crit}^{\frac{2}{3}}}{\left( 
n_{\rm crit} - 1 / \omega \right)^{\frac{2}{3}}}
\ c_1^2 \ \Pi \ \exp \left[ - \frac{4}{27} \frac{\mu^3}{(\ln S)^2}
\right],
\end{eqnarray}
and, for $n_{\rm crit} \gg 1$, 
\begin{eqnarray}
J_{\rm s} = s_{n_{\rm crit}} 
\Omega_0 \left( \frac{2 \sigma}{\pi m_1} \right)^{\frac{1}{2}} 
\ c_1^2 \ \Pi \ \exp \left[ - \frac{4}{27} \frac{\mu^3}{(\ln S)^2}
\right].
\end{eqnarray}
This revised formula of the steady state nucleation rate includes
a correction factor $\Pi$ defined as
\begin{eqnarray}
\Pi 
= \left[ \frac{1}{c_1^{\omega - 1}}
\frac{\prod_{k=1}^i \left( c^A_k \right)^{\nu_k}}
     {\prod_{k=1}^j \left( c^B_k \right)^{\eta_k}}
\right]^{\frac{1}{\omega}}
= \left[ 
\frac{\prod_{k=1}^i \left( c^A_k / c_1 \right)^{\nu_k}}
     {\prod_{k=1}^j \left( c^B_k / c_1 \right)^{\eta_k}}
\right]^{\frac{1}{\omega}}.
\end{eqnarray}
It should be emphasized that the factor $\Pi$ is expressed as
functions of concentration ratios of gaseous reactants/products to 
the key molecule and does not include the standard pressure 
$p_{\rm s}$ explicitly.

In most of the past works for the formation of dust in the ejecta of 
SNe (e.g., Nozawa et al.\ 2003; Todini \& Ferrara 2001), the factor 
$\Pi$ has been set to be unity.
Since $\Pi > 1$ for the formation of a multi-element grain with no 
gaseous product such as in the ejecta of SNe (see Nozawa et al.\ 2003), 
the revised formula provides a nucleation rate larger than the previous 
one;
for the elemental abundances of the gas
($c_{{\rm Mg},0}/c_{{\rm SiO},0} = 2$ and 
$c_{{\rm O},0} /c_{{\rm SiO},0} = 20$) for the formation of 
MgSiO$_3$ used in the text, the revised formula presents the nucleation 
rate only by a factor of $\Pi = 5.3$ higher than the old one. 
This enhancement cannot be significant because Paquette \& Nuth (2011) 
showed that the resulting size distribution of dust is little 
affected even if the steady state nucleation rate is changed by a few 
orders of magnitude.

\section{Dependence on $n_*$}

Here we describe how the change in $n_*$ in Equation (1) affects the 
formation process of dust as well as the average radius and 
condensation efficiency of newly formed grains.
First of all, as demonstrated in Section 4.1 and discussed in 
Section 5, the change of $n_*$ does not influence the results of 
calculations as long as the formation of dust grains proceeds under 
the condition of $\Lambda = \tau_{\rm sat}/\tau_{\rm coll} \gg 1$ 
and $n_* \gg n_{\rm crit}^{\rm smax}$, because the current densities 
$I_n$ for $n > n_{\rm crit}^{\rm smax}$ achieve a steady--state
value, being independent of $n$.
Thus, here we examine the effect of change in $n_*$ for the model
in which the non-steady--state effect for the formation process of 
grains is significant. 

Figure 10(a) presents the time evolutions of the current $I_*$, 
average grain radius $a_{\rm ave}$, and condensation efficiency 
$f_{\rm con}$ of C grains for $n_* = 100$, 300, and 1000 with 
$c_{10} = 10^6$ cm$^{-3}$, $t_0 = 300$ days, and $\gamma = 1.25$.
The figure shows that, as $n_*$ increases, the onset time of dust 
formation is slightly delayed, and the peak value of $I_*$ decreases.
In addition, a larger $n_*$ leads to a smaller $f_{\rm con}$ and 
a larger $a_{\rm ave}$:
$f_{\rm con} = 1.0$ and $a_{\rm ave} = 0.0016$ $\mu$m for $n_* = 100$, 
while $f_{\rm con} = 0.9$ and $a_{\rm ave} = 0.0019$ $\mu$m 
for $n_* = 1000$.
Thus, the formation process as well as the resulting properties of 
dust grains seems to depend on the adopted value of $n_*$.

However, we recognize that the dependence of $n_*$ is superficial.
As is seen from the final size distributions displayed in Figure 10(b), 
where the values for $n_* = 300$ and 1000 are shifted down, 
respectively, by factors of 0.1 and 0.01 for clarity, the combined 
size distributions of clusters and grains are identical with each 
other, regardless of $n_*$.
It should be noted that this demonstrates that the change in $n_*$ 
does not influence the formation processes of clusters and grains 
themselves.
The independence of the combined size distribution on $n_*$ can be
understood as follows; 
the evolutions of $I_n$ and $Y_n$ are the same for 
$n \le n_{*,{\rm min}} = 100$ in all the three cases.
On the other hand, for $n > n_{*,{\rm min}}$, the current density 
$I_n \simeq \alpha_{n-1} c_1 Y_{n-1}$ under the condition that 
$\ln S \gg (2 \mu) / (3 n^{1/3})$, as can be seen from Equations (42) 
and (45).
Thus, the combined size distribution can be expected to be independent 
of $n_{*,{\rm min}}$ since, in the model considered here, $\ln S$ is 
not depressed very much by the depletion of the key molecules due to 
grain growth.
The increase in $n_*$ only enhances the minimal size of grains, 
which results in the reduced number of atoms finally locked in 
grains (that is, condensation efficiency) and the increased 
average radius of grains for the given combined size distribution of 
clusters and grains.

In conclusion, the change in $n_*$ does not vary the formation process
of dust grains but affects the resulting condensation efficiency 
and average radius of grains.
The value of $n_*$ is important only for dust formation at low gas 
densities where the average radius of newly formed grains is very 
small ($a_{\rm ave} \la 0.005$ $\mu$m), as is exemplified above.
It should be kept in mind that, even if $n_*$ is enhanced from 
$n_* = 100$ to 1000, the minimal size of grains is only increased 
by a factor of $10^{1/3} = 2.15$.
We also note that small grains formed in the SN ejecta would be 
destroyed quickly by the passage of the reverse shock before being 
injected into the ISM (Nozawa et al.\ 2007, 2010; 
Silvia et al.\ 2010, 2012).
Thus, when we consider the final mass of dust ejected from SNe, the 
uncertainty in the mass of small grains, resulting from the 
uncertainty in $n_*$, could not be important.


\clearpage
\begin{figure}
\epsscale{0.8}
\plotone{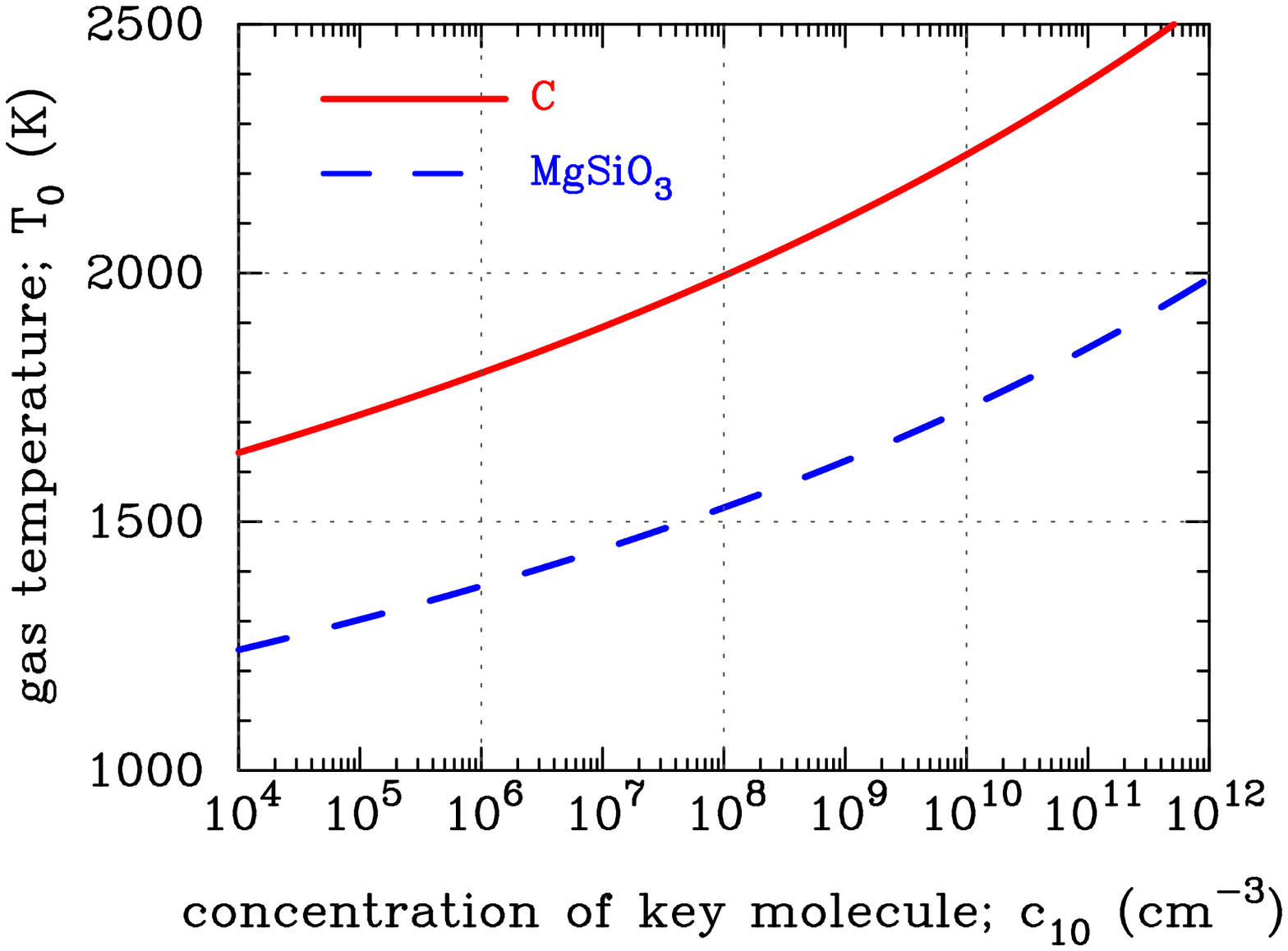}
\caption{
Gas temperature $T_0$ versus the concentration of the key molecule 
$c_{10}$ at $\ln S = 0$ for C (solid) and MgSiO$_3$ grains (dashed).
The key molecule for MgSiO$_3$ grains are assumed to be SiO molecules, 
and the number ratios of Mg and O atoms to SiO molecules are taken 
to be $c_{{\rm Mg},0}/c_{{\rm SiO},0} = 2$ and 
$c_{{\rm O},0} /c_{{\rm SiO},0} = 20$, respectively.
\label{fig1}}
\end{figure}

\clearpage
\begin{figure}
\epsscale{1.05}
\plottwo{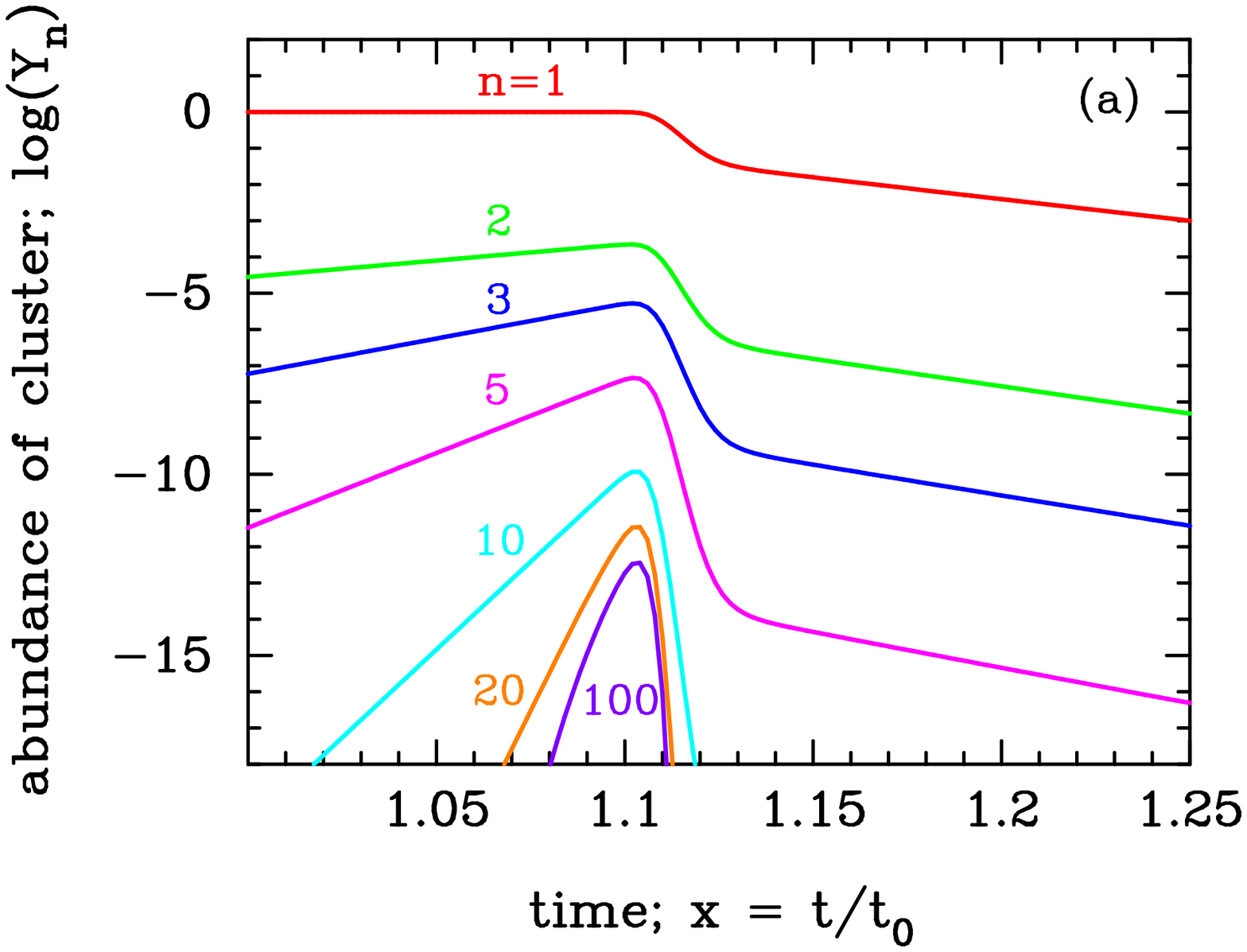}{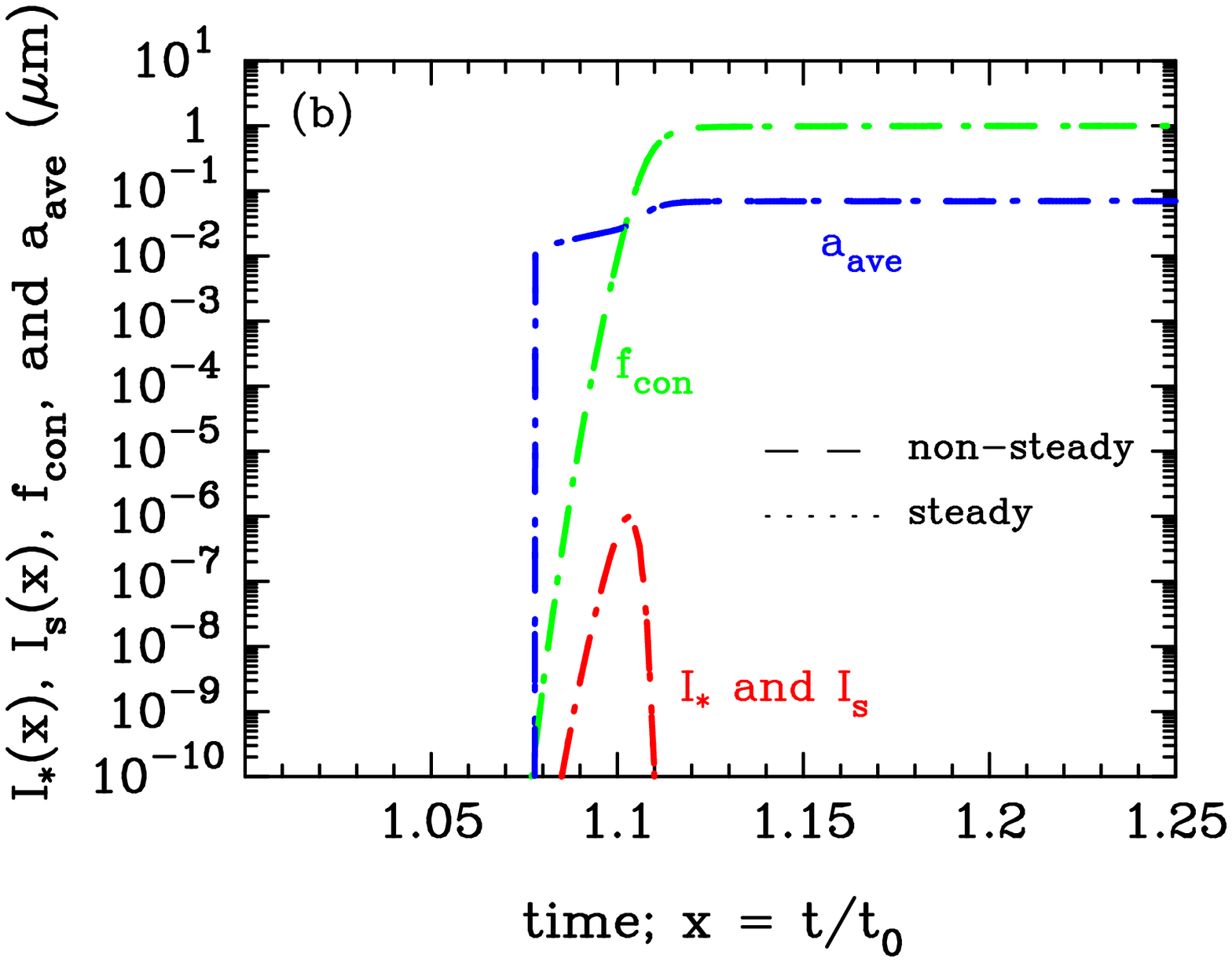}
\vspace{0.5 cm}
\plottwo{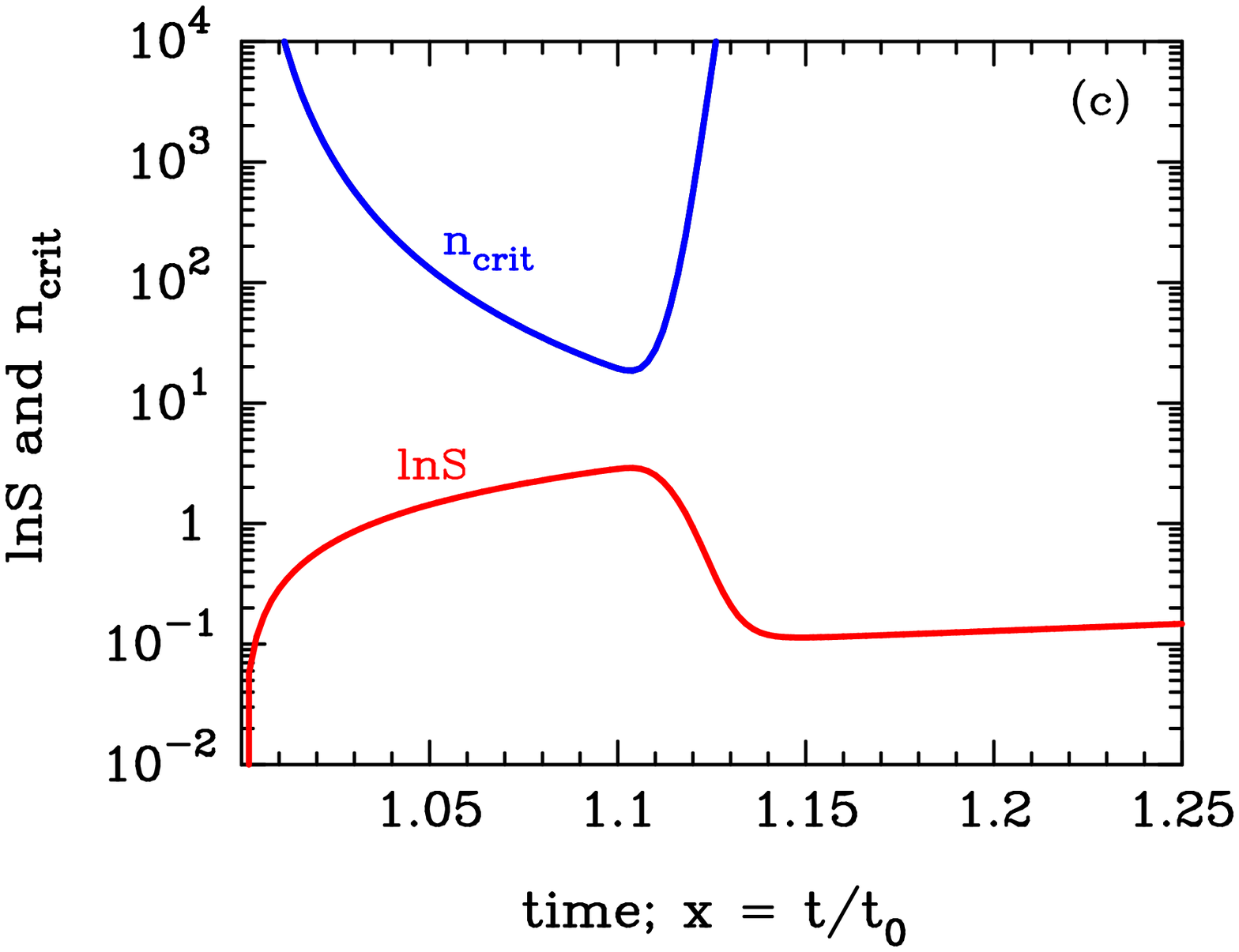}{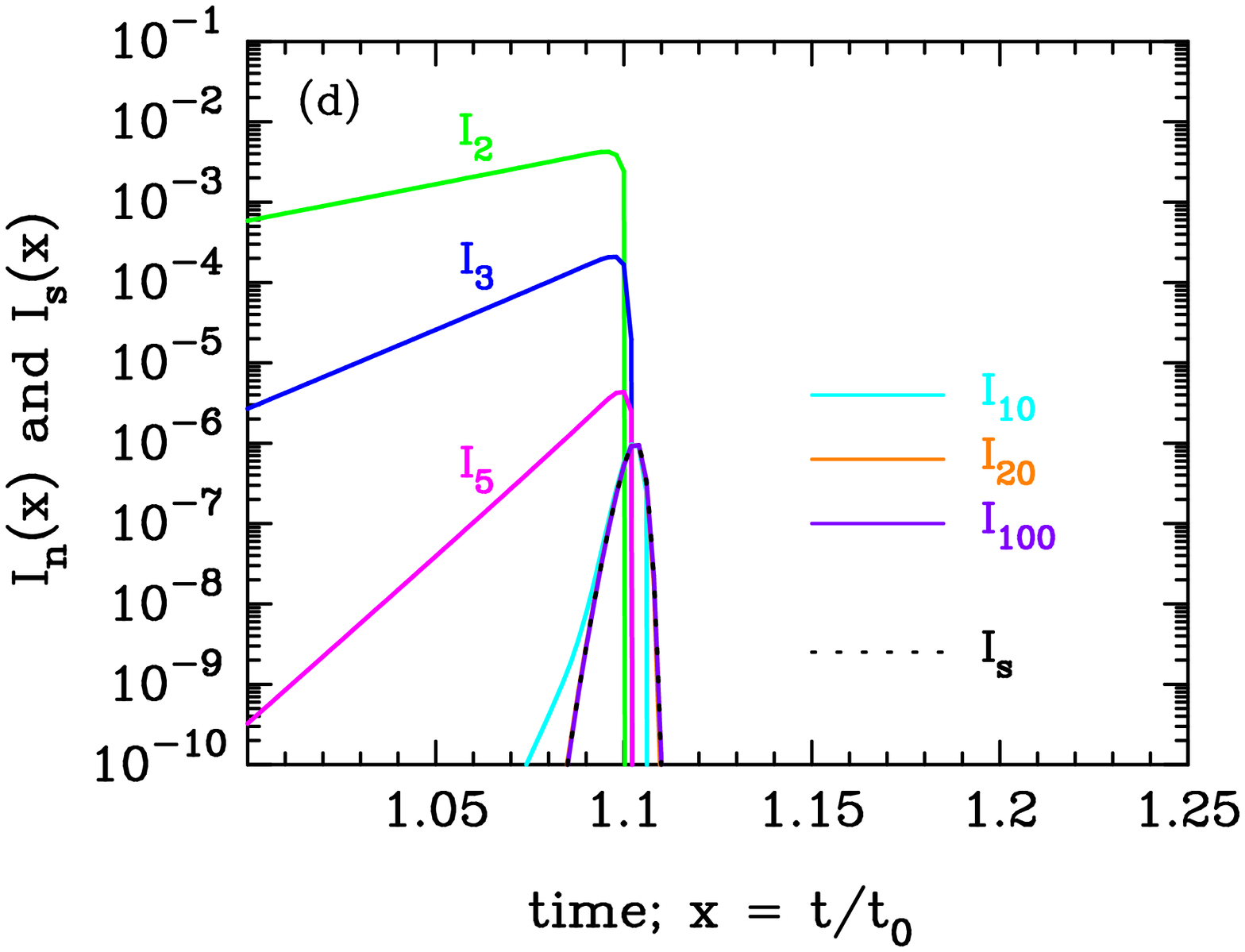}
\vspace{0.5 cm}
\epsscale{0.46}
\plotone{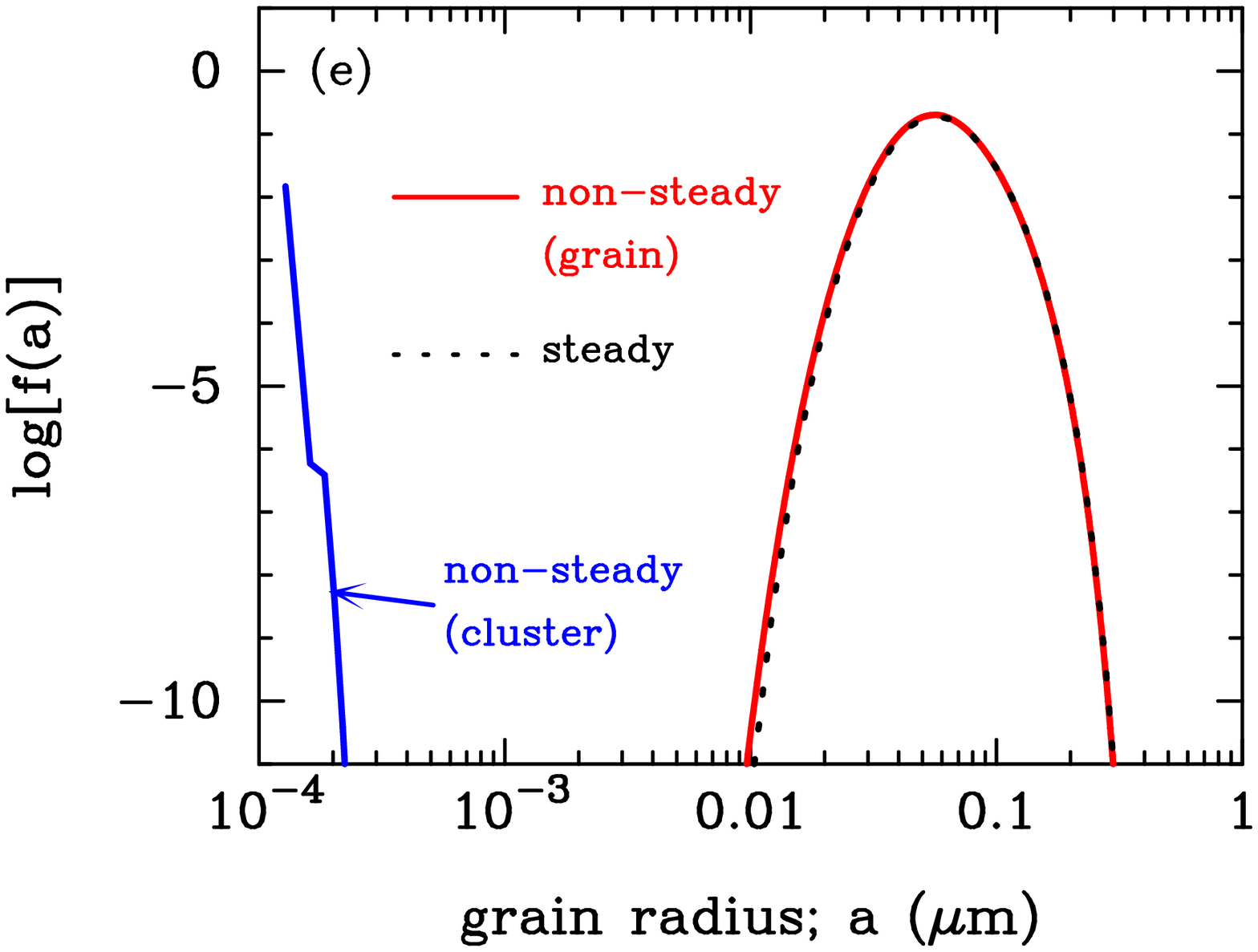}
\caption{
Formation process of C grains as a function of time ($x = t / t_0$) and 
the resulting size distribution for 
$c_{10} = 10^8$ cm$^{-3}$, $t_0 = 300$ days, and $\gamma = 1.25$;
(a) abundances of the key molecule ($n = 1$) and $n$--mer clusters with 
    $n =$ 2--100,
(b) current density of $n_*$--mer ($I_*$), average grain radii 
    ($a_{\rm ave}$) in units of $\mu$m, and condensation efficiency 
    ($f_{\rm con}$),
(c) supersaturation ratio ($S$) and critical size ($n_{\rm crit}$),
(d) current densities of $n$--mers ($I_n$) and the steady--state 
    current densies ($I_{\rm s}$, dotted line),
(e) final size distribution spectra of clusters (blue) and 
grains (red).
For comparison, the results from the steady model are shown by dotted 
lines in (b) and (e).
In (d), $I_{20}$, $I_{100}$, and $I_{\rm s}$ are overlapped.
\label{fig2}}
\end{figure}

\clearpage
\begin{figure}
\epsscale{1.05}
\plottwo{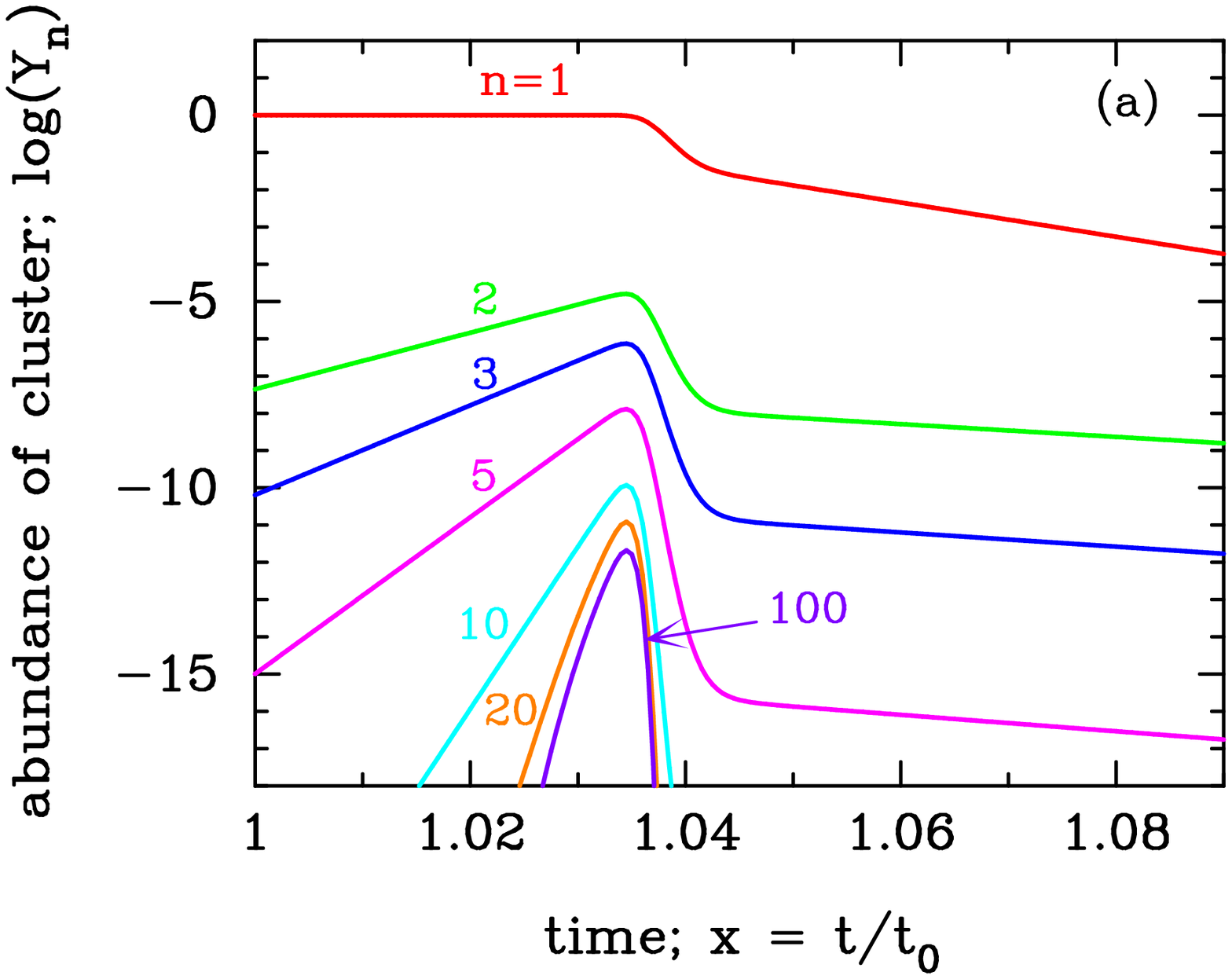}{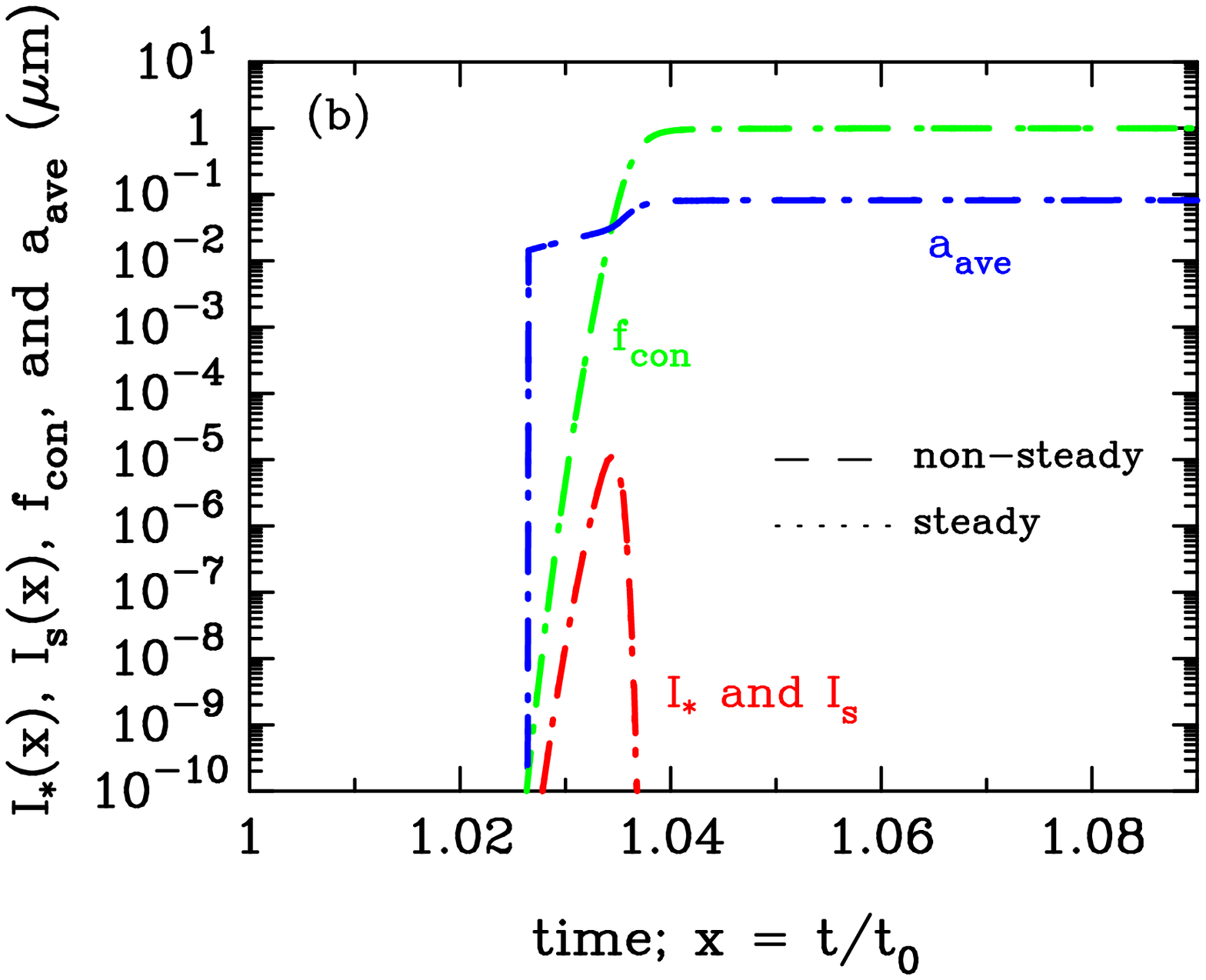}
\vspace{0.5 cm}
\plottwo{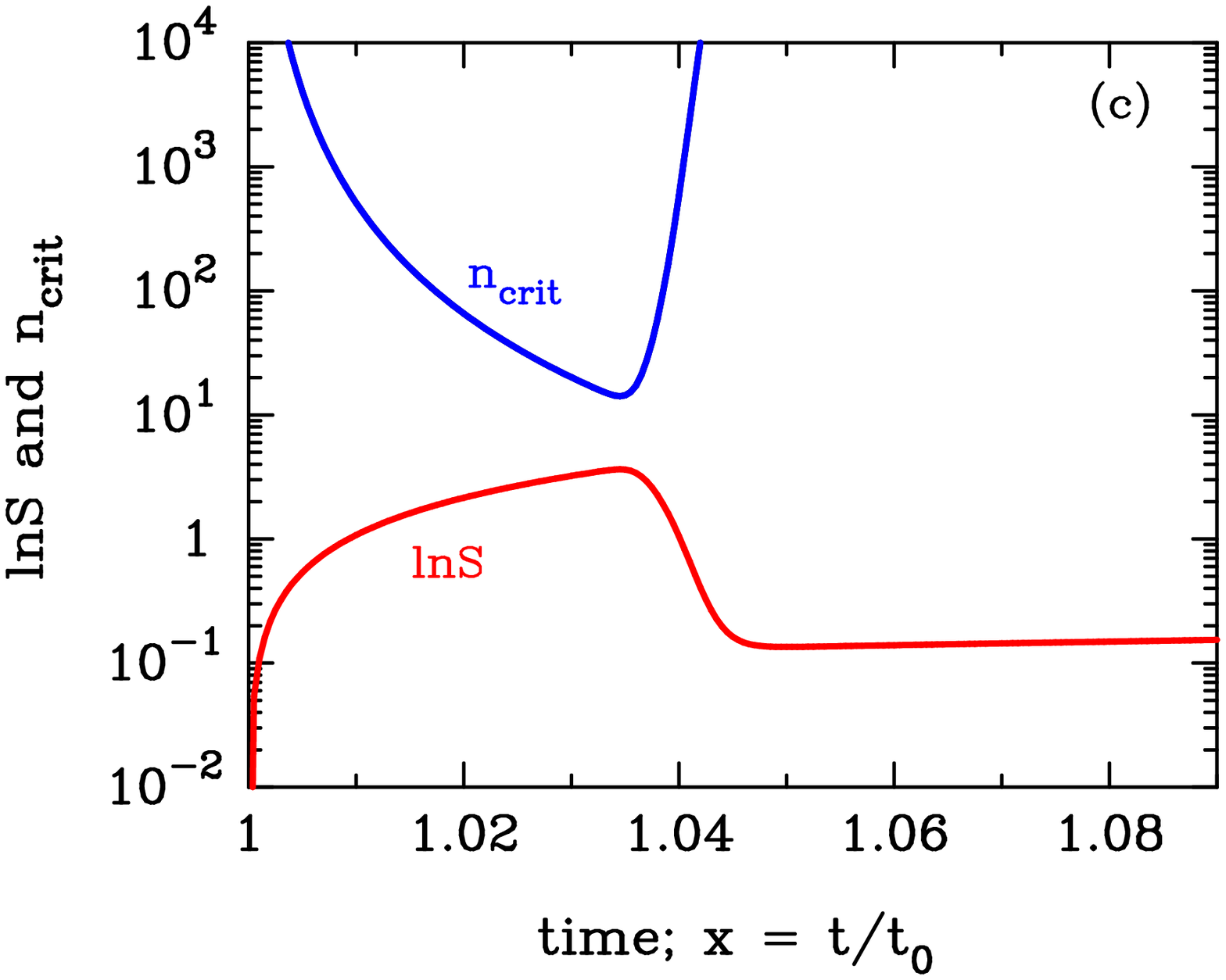}{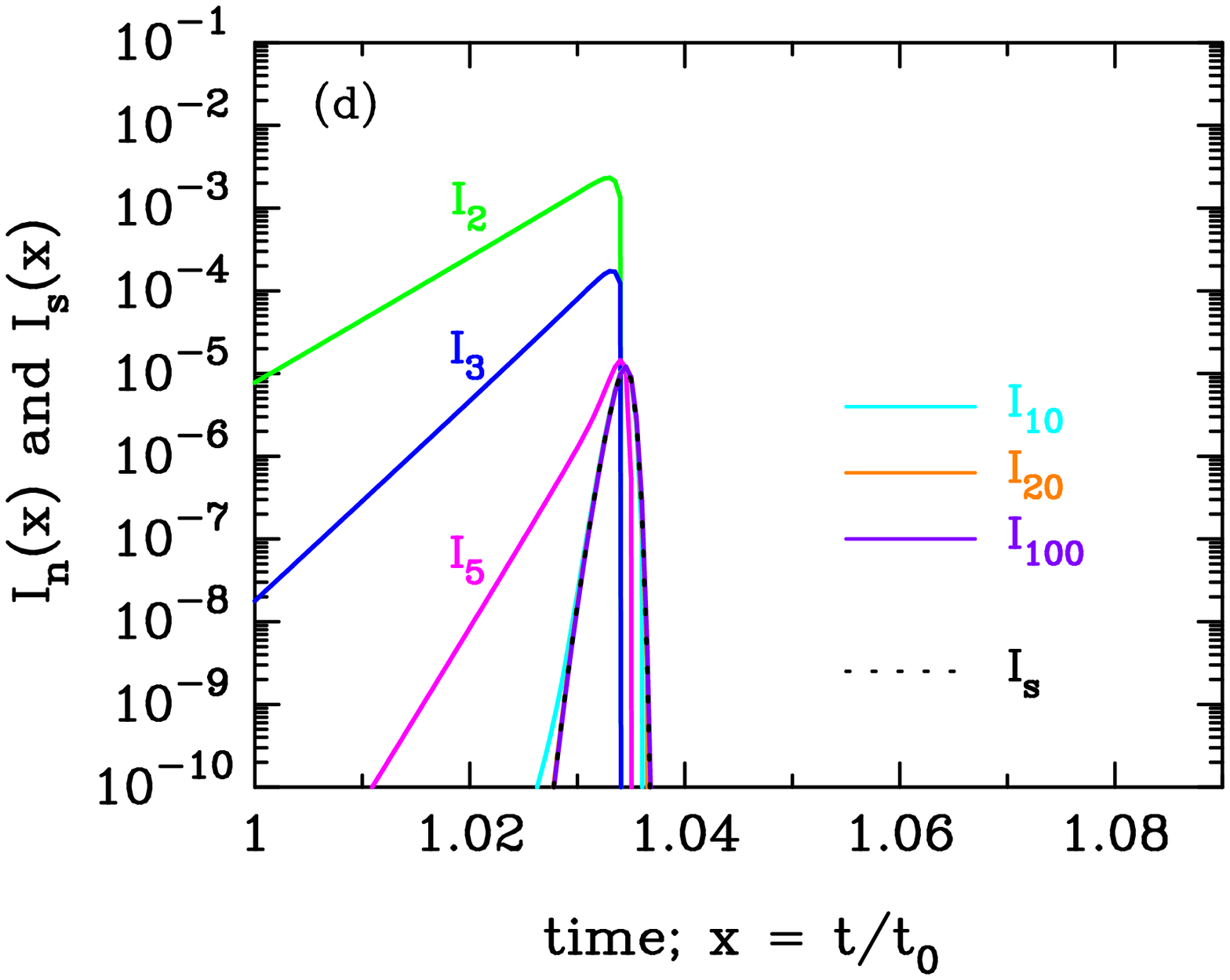}
\vspace{0.5 cm}
\epsscale{0.46}
\plotone{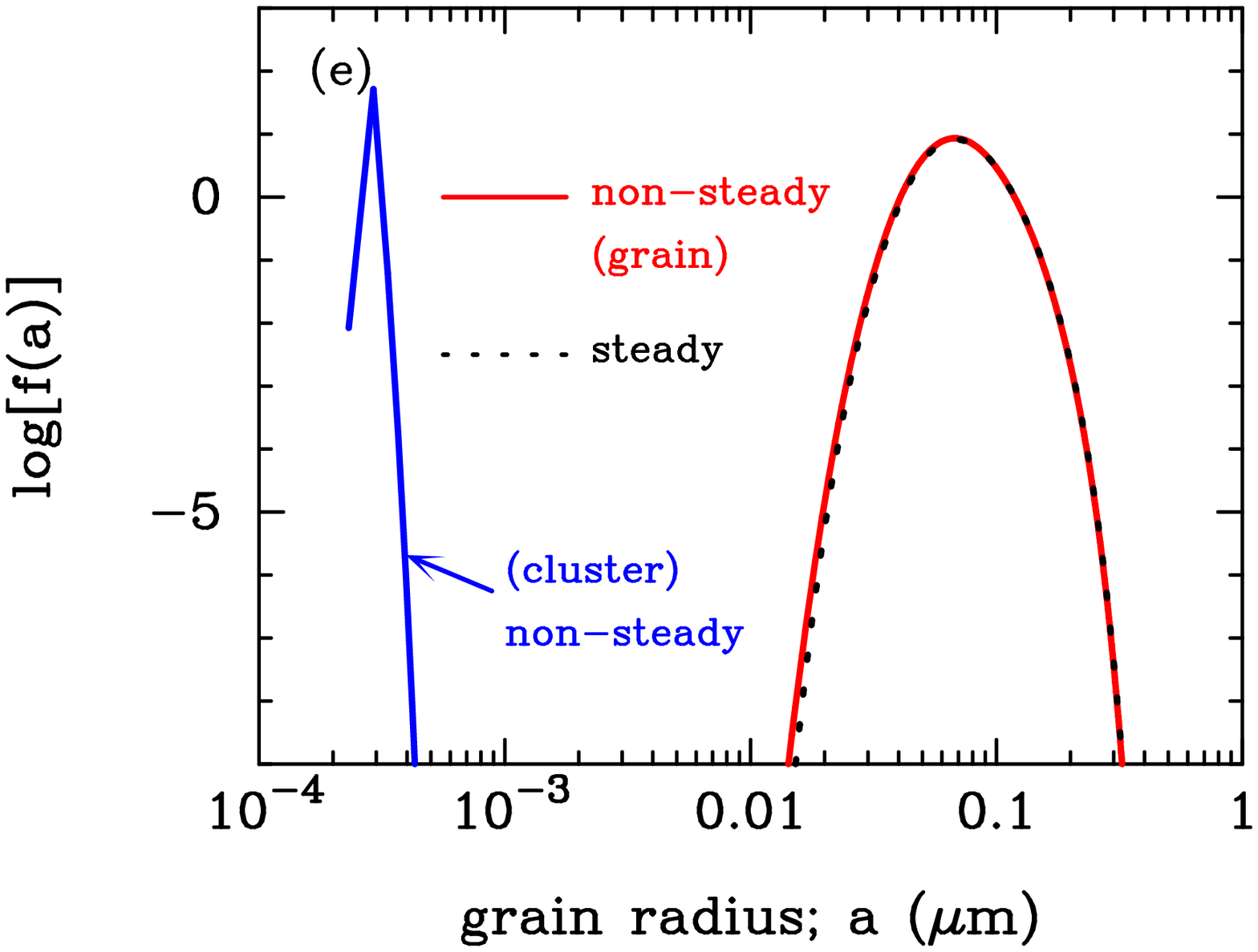}
\caption{
Same as Figure 2 but for MgSiO$_3$ grains.
\label{fig3}}
\end{figure}

\clearpage
\begin{figure}
\epsscale{1.05}
\plottwo{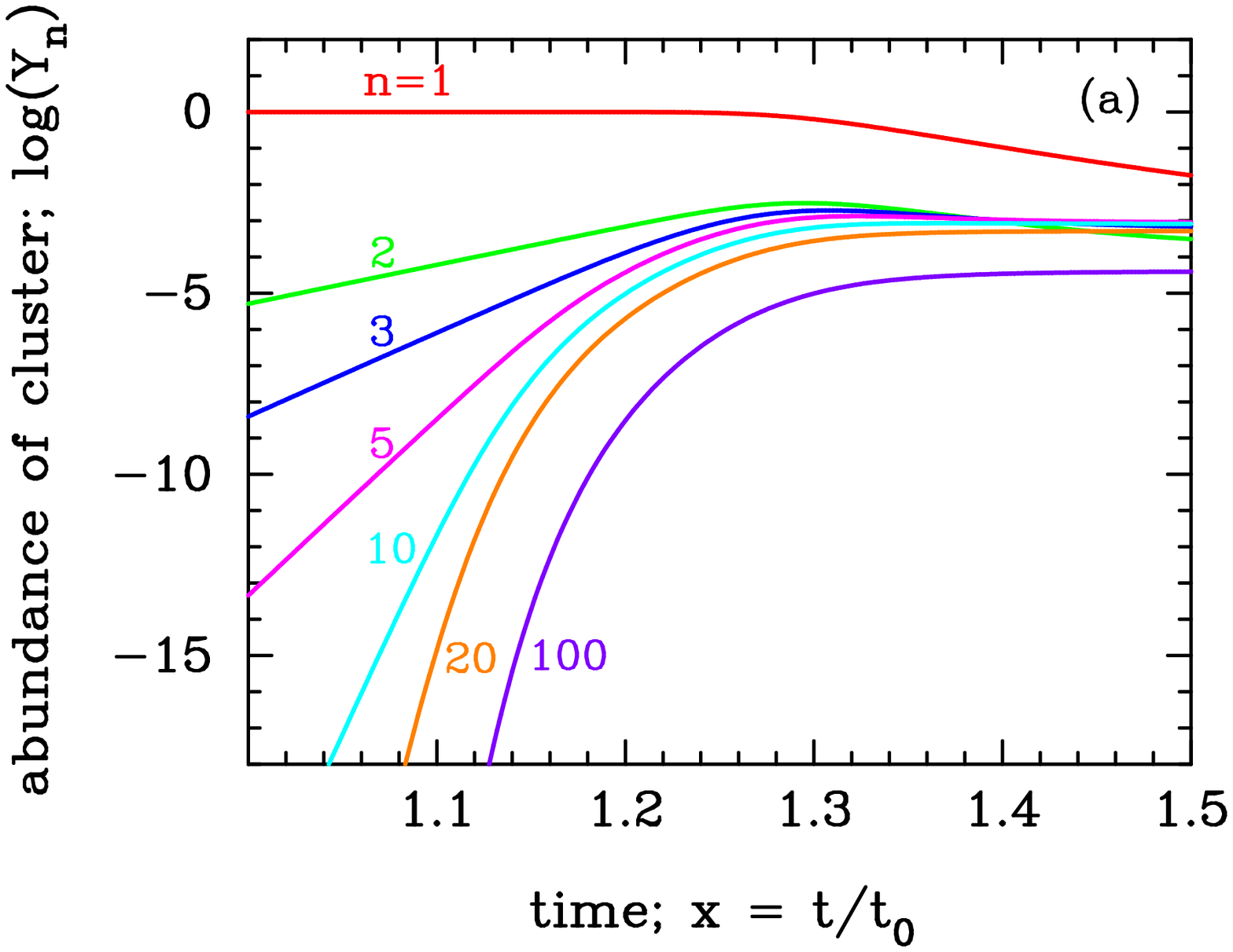}{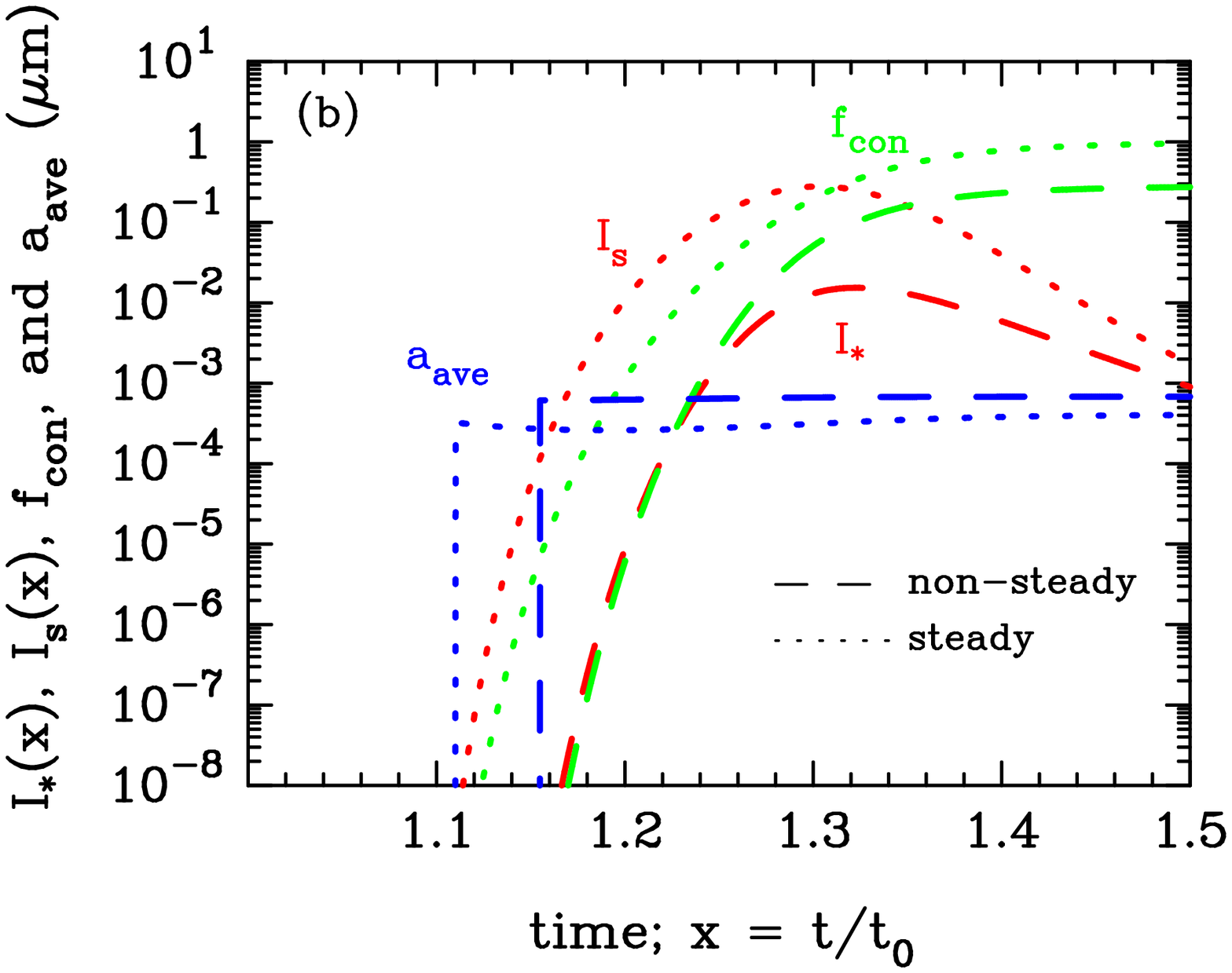}
\vspace{0.5 cm}
\plottwo{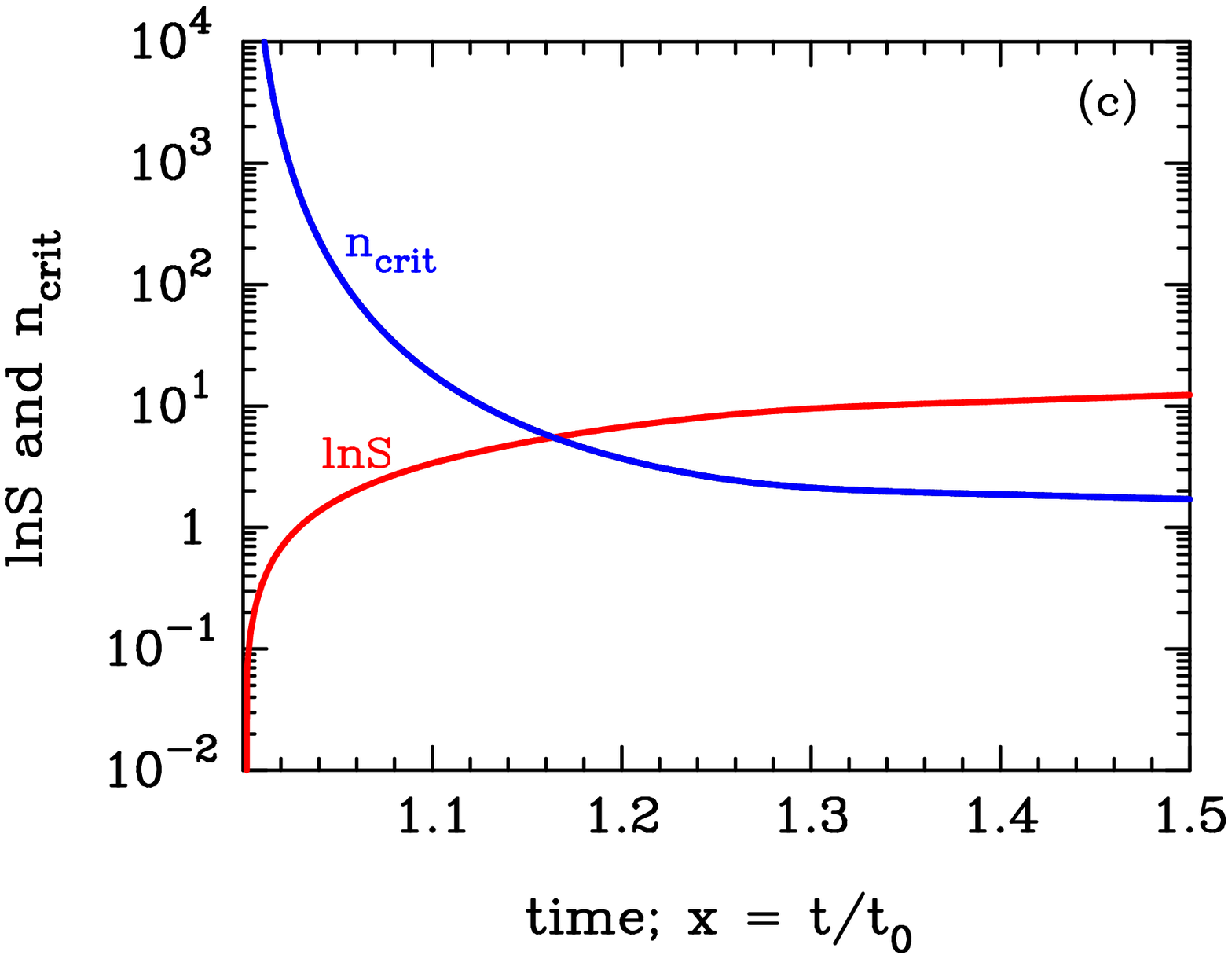}{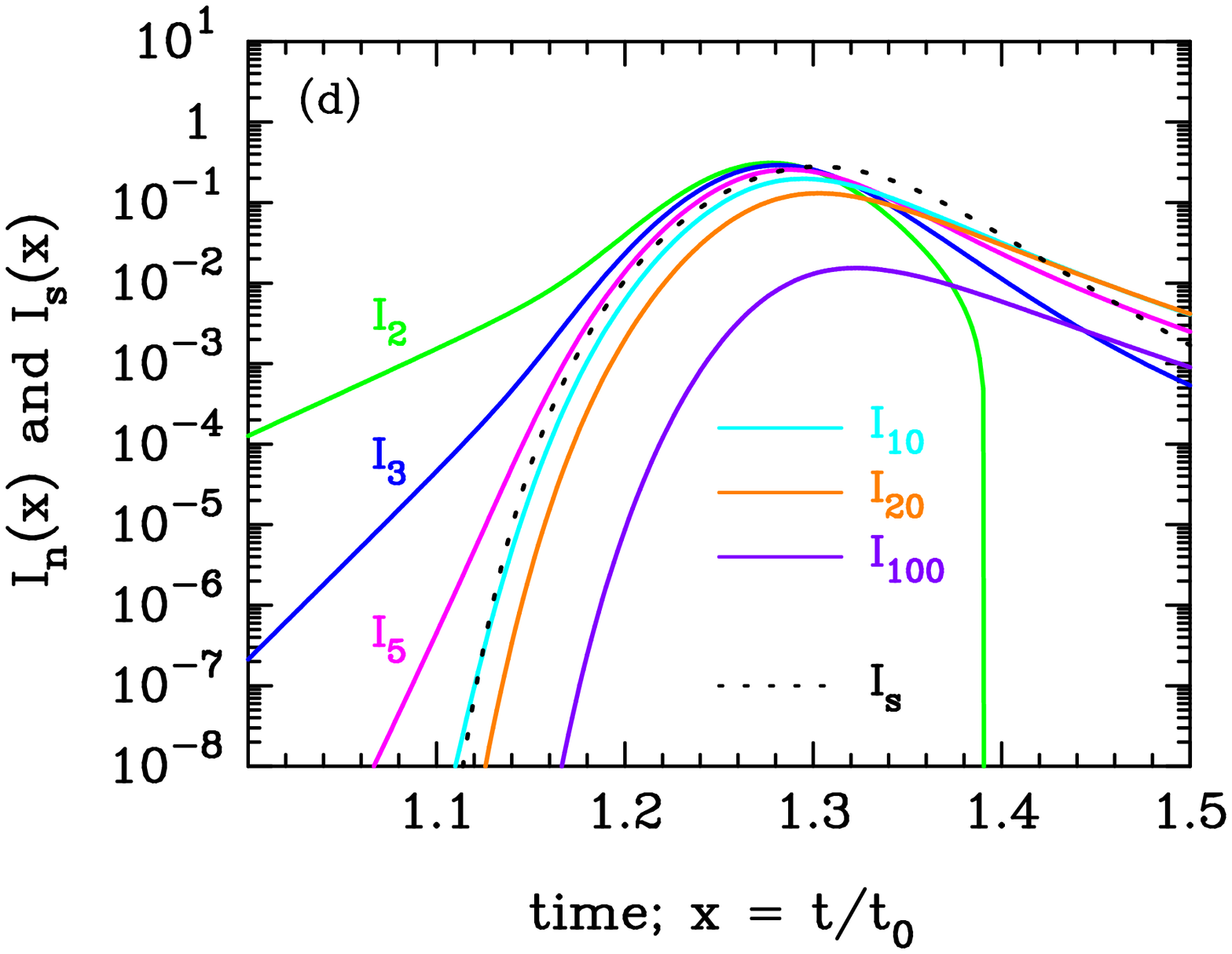}
\vspace{0.5 cm}
\epsscale{0.46}
\plotone{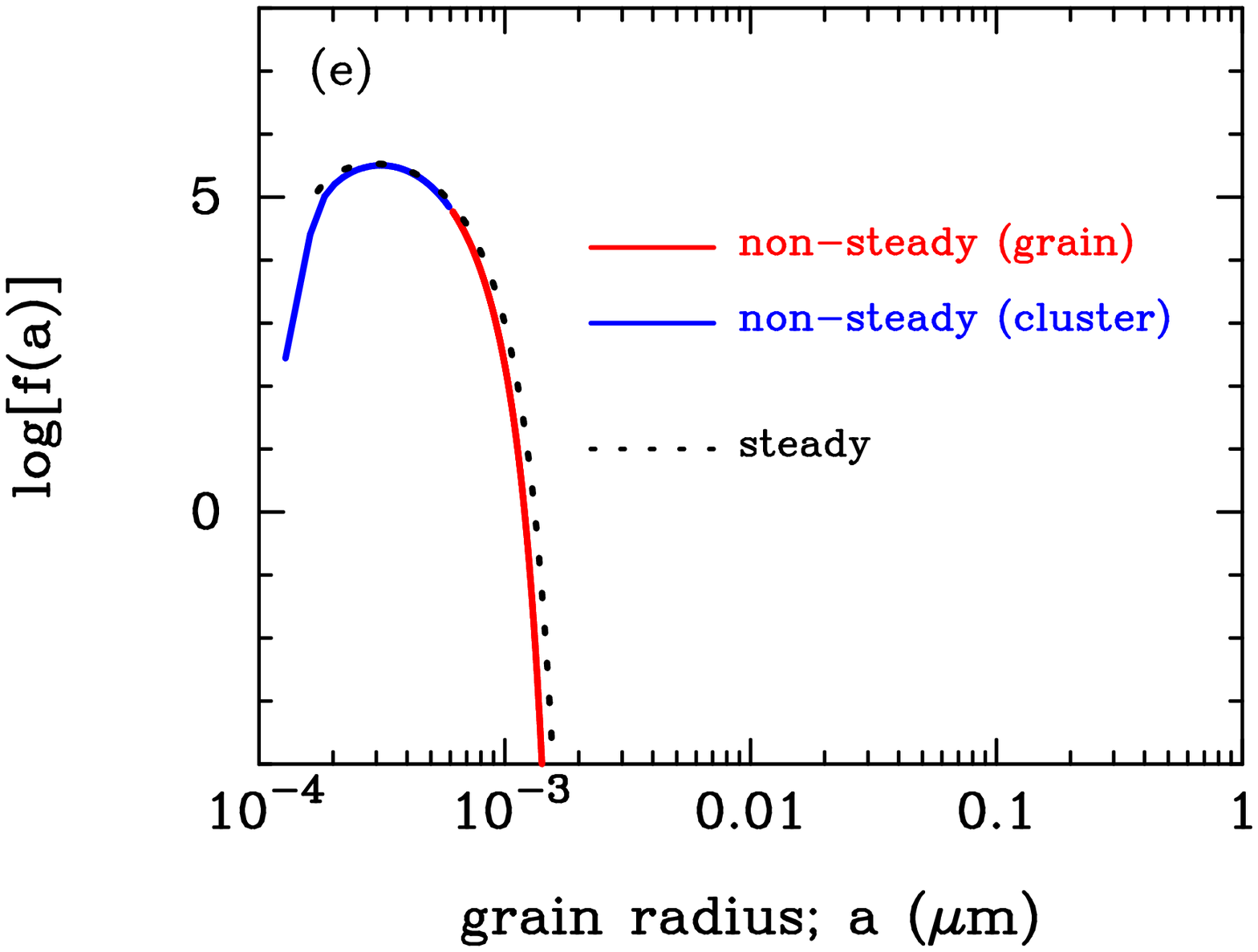}
\caption{
Same as Figure 2 but for 
$c_{10} = 10^5$ cm$^{-3}$, $t_0 = 300$ days, and $\gamma = 1.25$.
\label{fig4}}
\end{figure}

\clearpage
\begin{figure}
\epsscale{1.05}
\plottwo{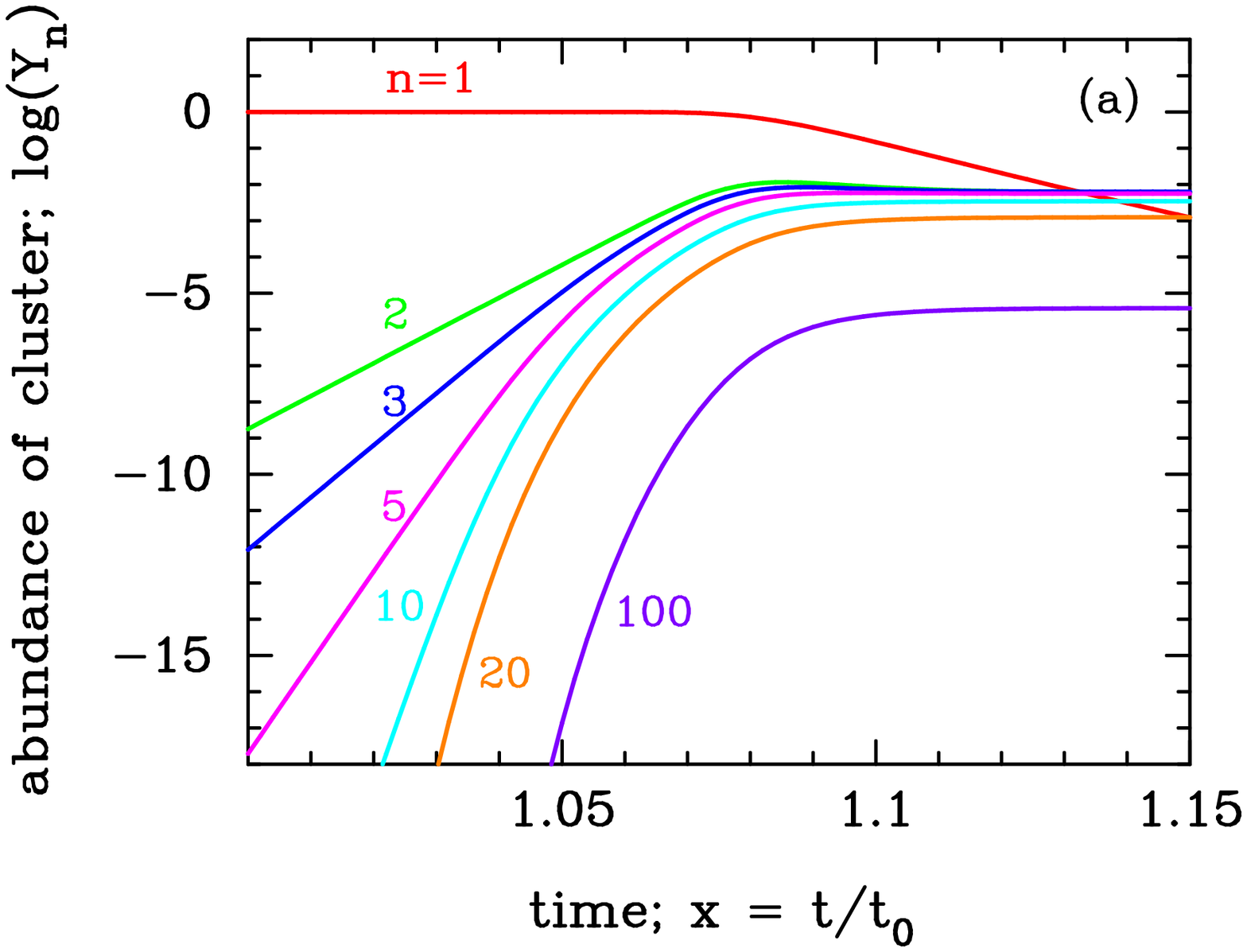}{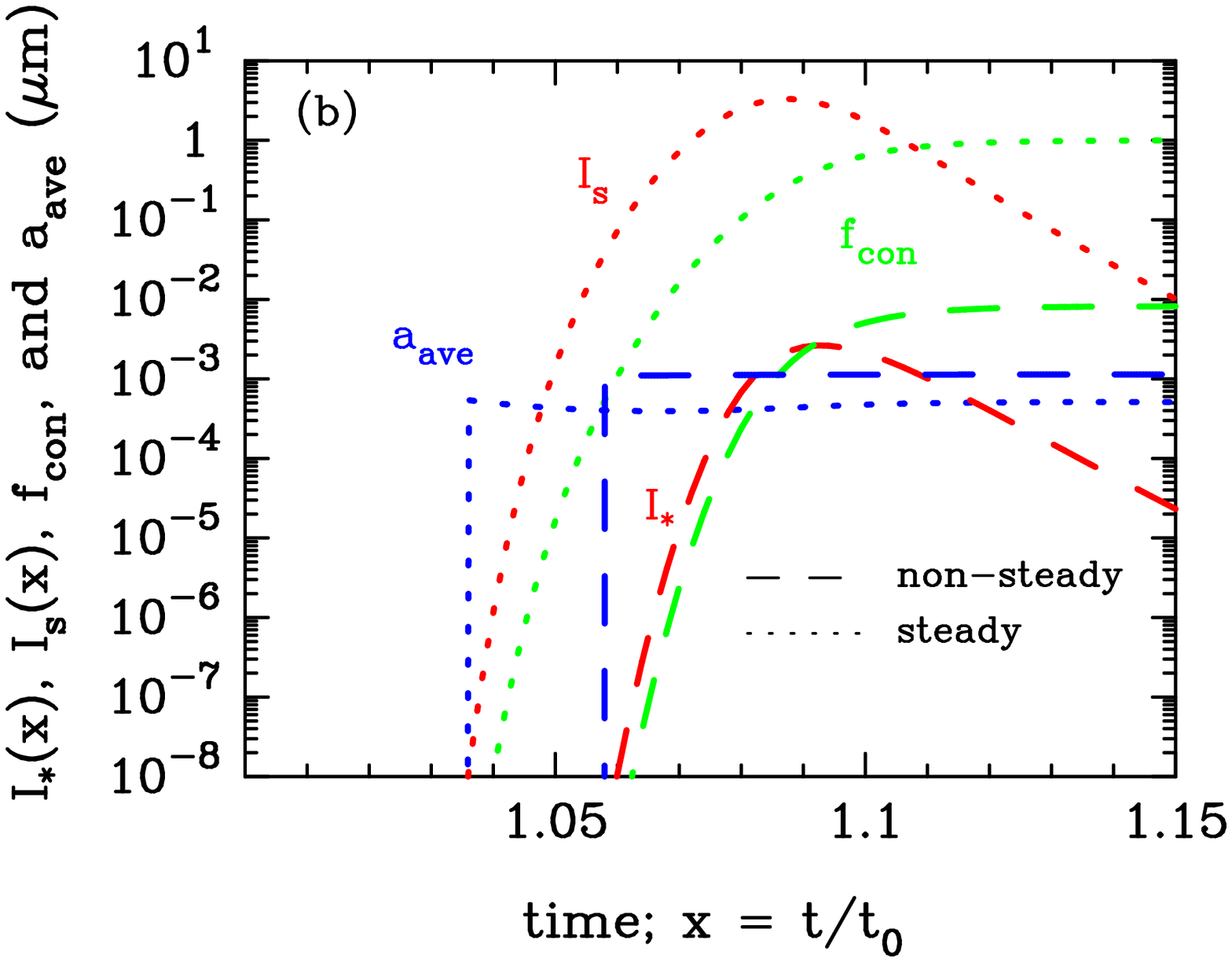}
\vspace{0.5 cm}
\plottwo{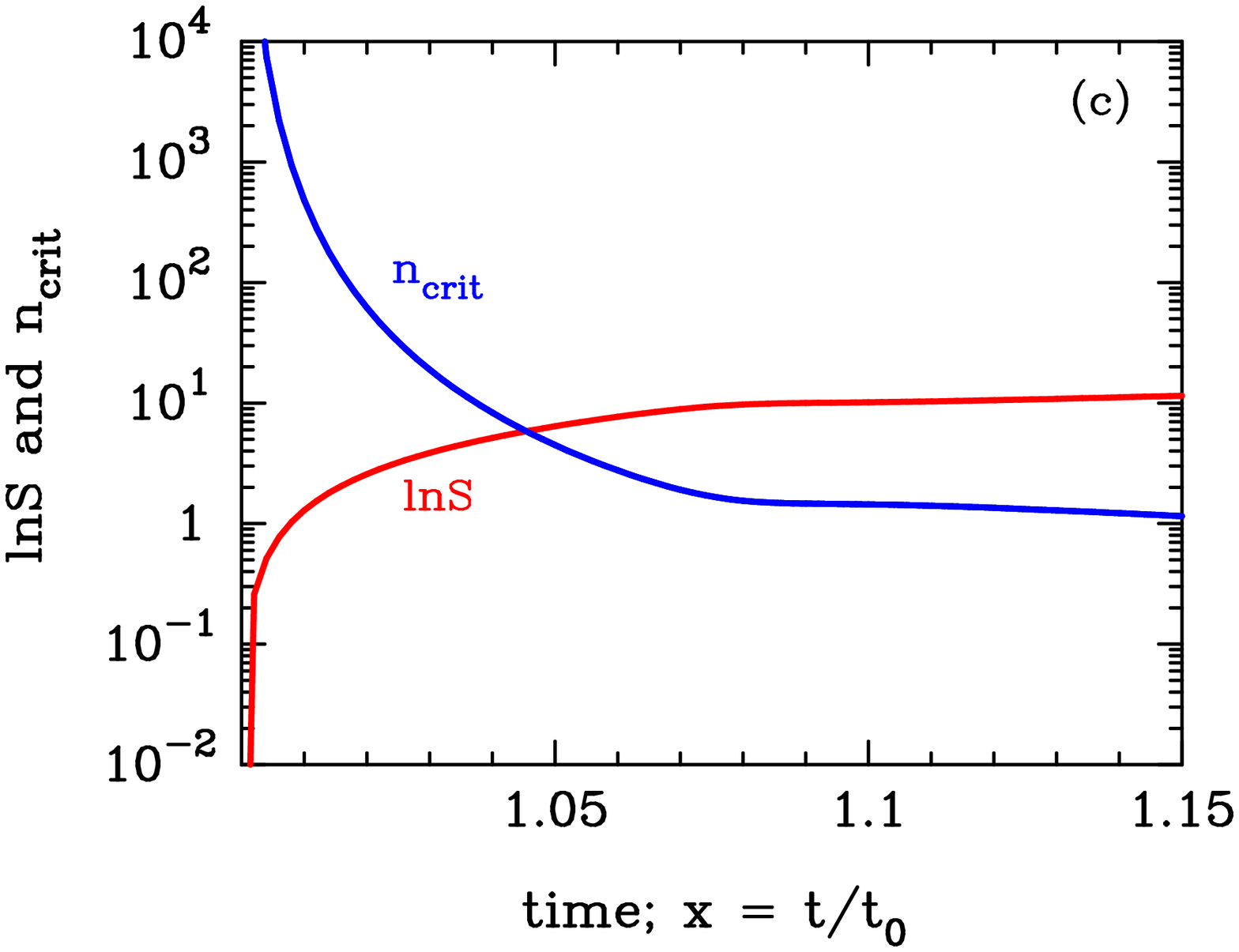}{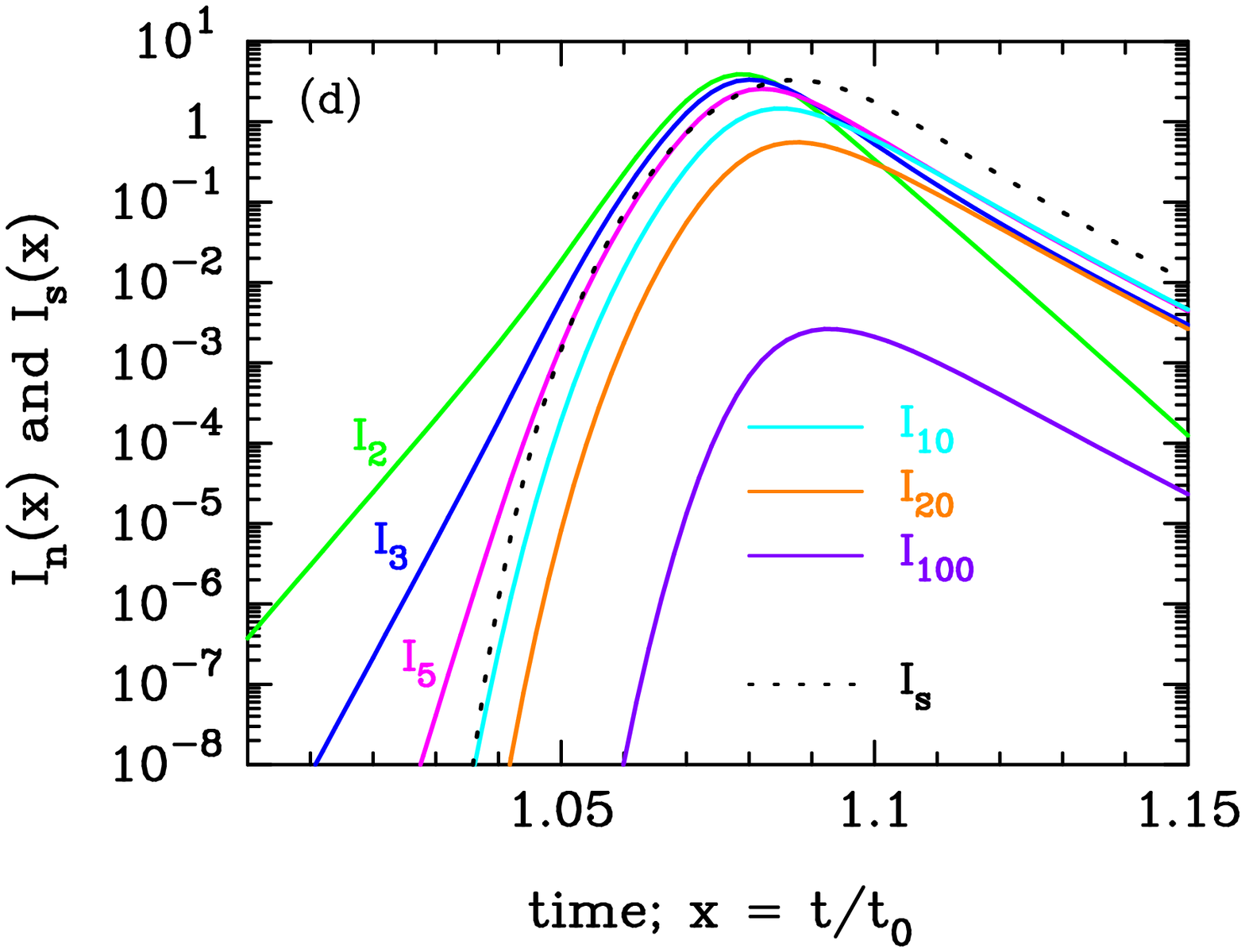}
\vspace{0.5 cm}
\epsscale{0.46}
\plotone{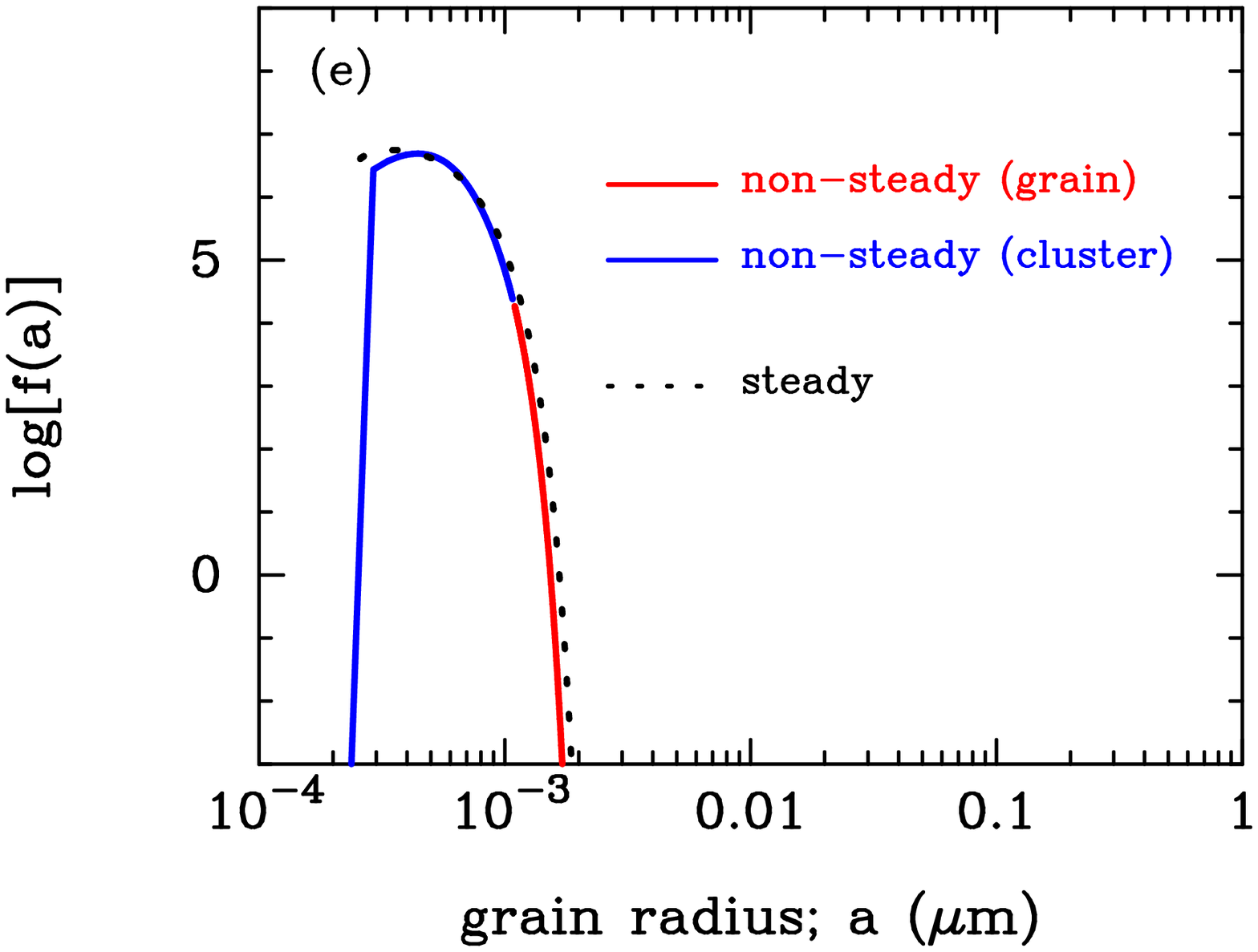}
\caption{
Same as Figure 4 but for MgSiO$_3$ grains.
\label{fig5}}
\end{figure}

\clearpage
\begin{figure}
\epsscale{0.6}
\plotone{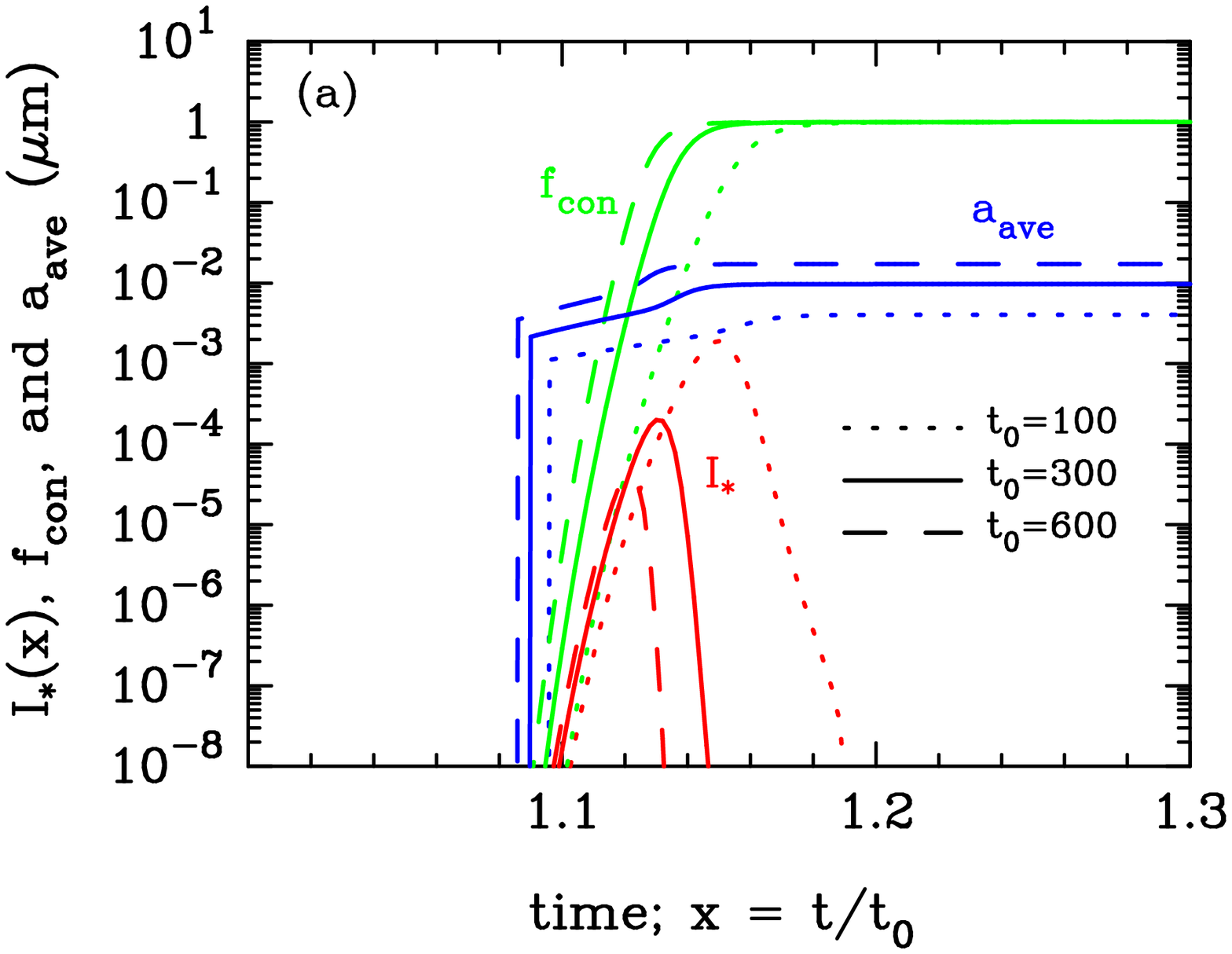}
\vspace{0.8 cm}
\plotone{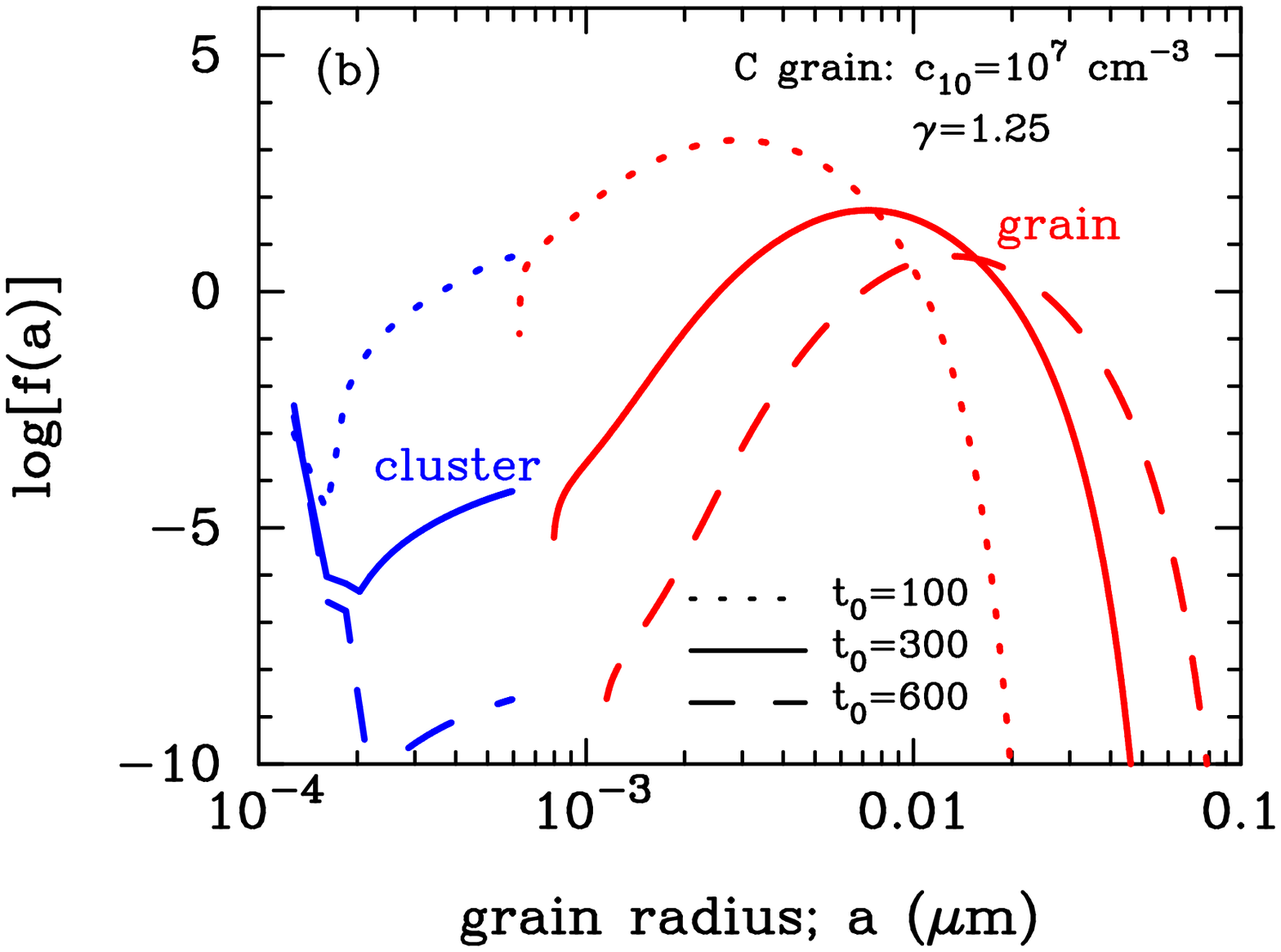}
\caption{
(a) Time evolutions of $I_*$, $a_{\rm ave}$, and 
$f_{\rm con}$ and (b) the final size distribution spectra 
of newly formed 
C grains for $t_0 = 100$ days (dotted), 300 days (solid), 
and 600 days (dashed).
The other parameters are set to be $c_{10} = 10^7$ cm$^{-3}$
and $\gamma = 1.25$.
\label{fig6}}
\end{figure}

\clearpage
\begin{figure}
\epsscale{0.6}
\plotone{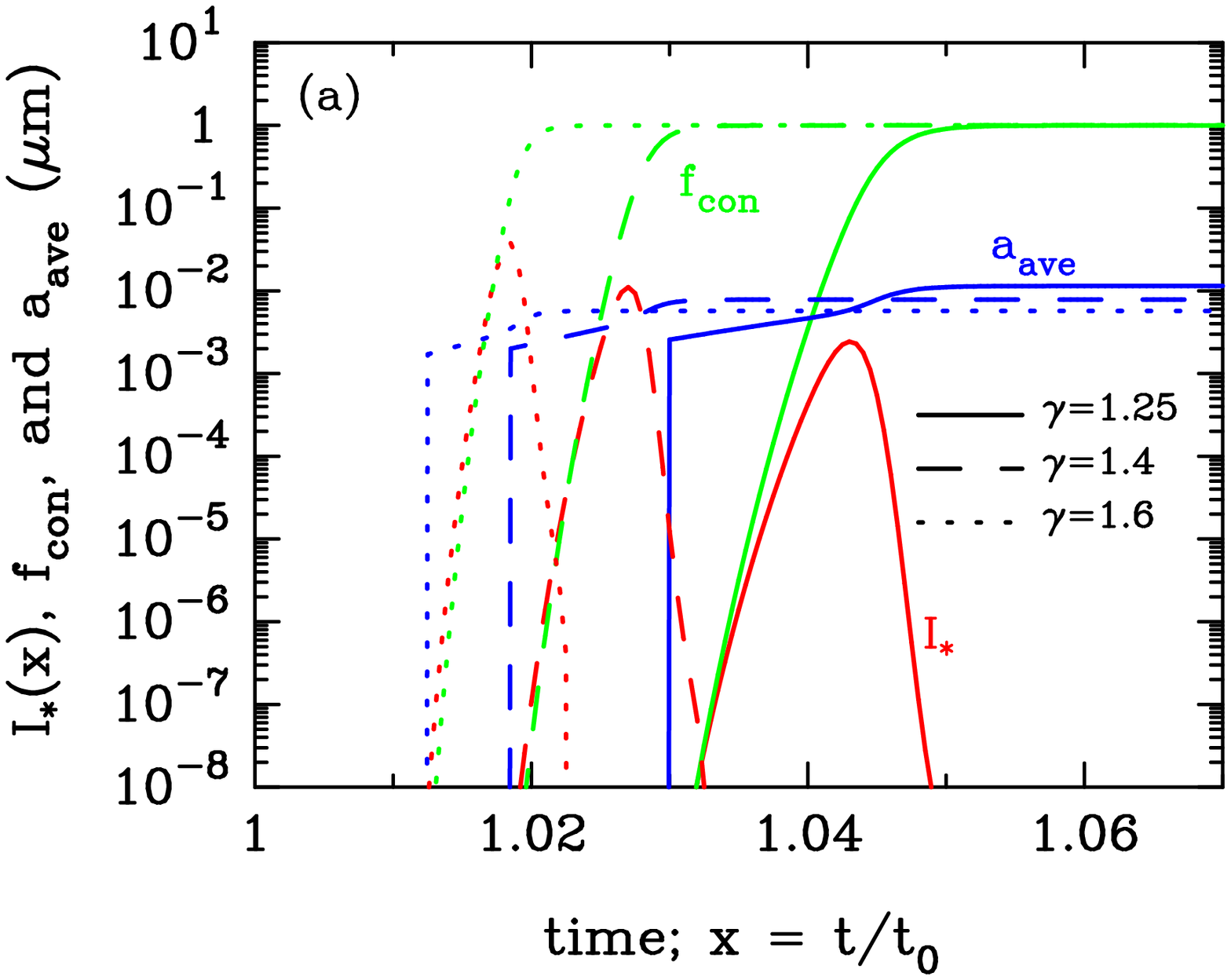}
\vspace{0.8 cm}
\plotone{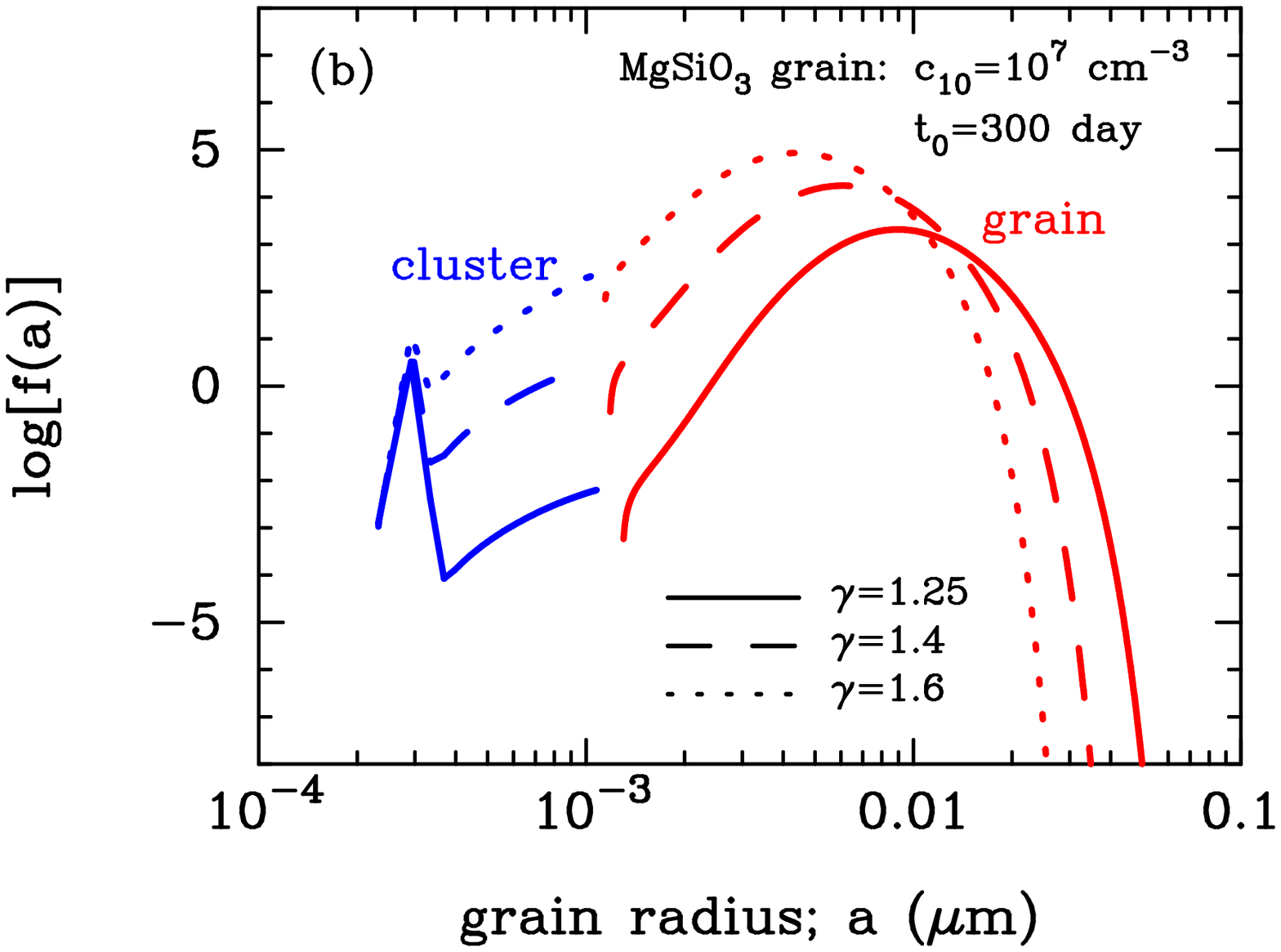}
\caption{
(a) Time evolutions of $I_*$, $a_{\rm ave}$, and 
$f_{\rm con}$, and (b) the final size distribution spectra 
of newly formed 
MgSiO$_3$ grains for $\gamma = 1.25$ (solid), 1.4 (dashed), and 1.7 
(dotted).
The other parameters are set to be $c_{10} = 10^7$ cm$^{-3}$ and
$t_0 = 300$ days.
\label{fig7}}
\end{figure}

\clearpage
\begin{figure}
\epsscale{0.6}
\plotone{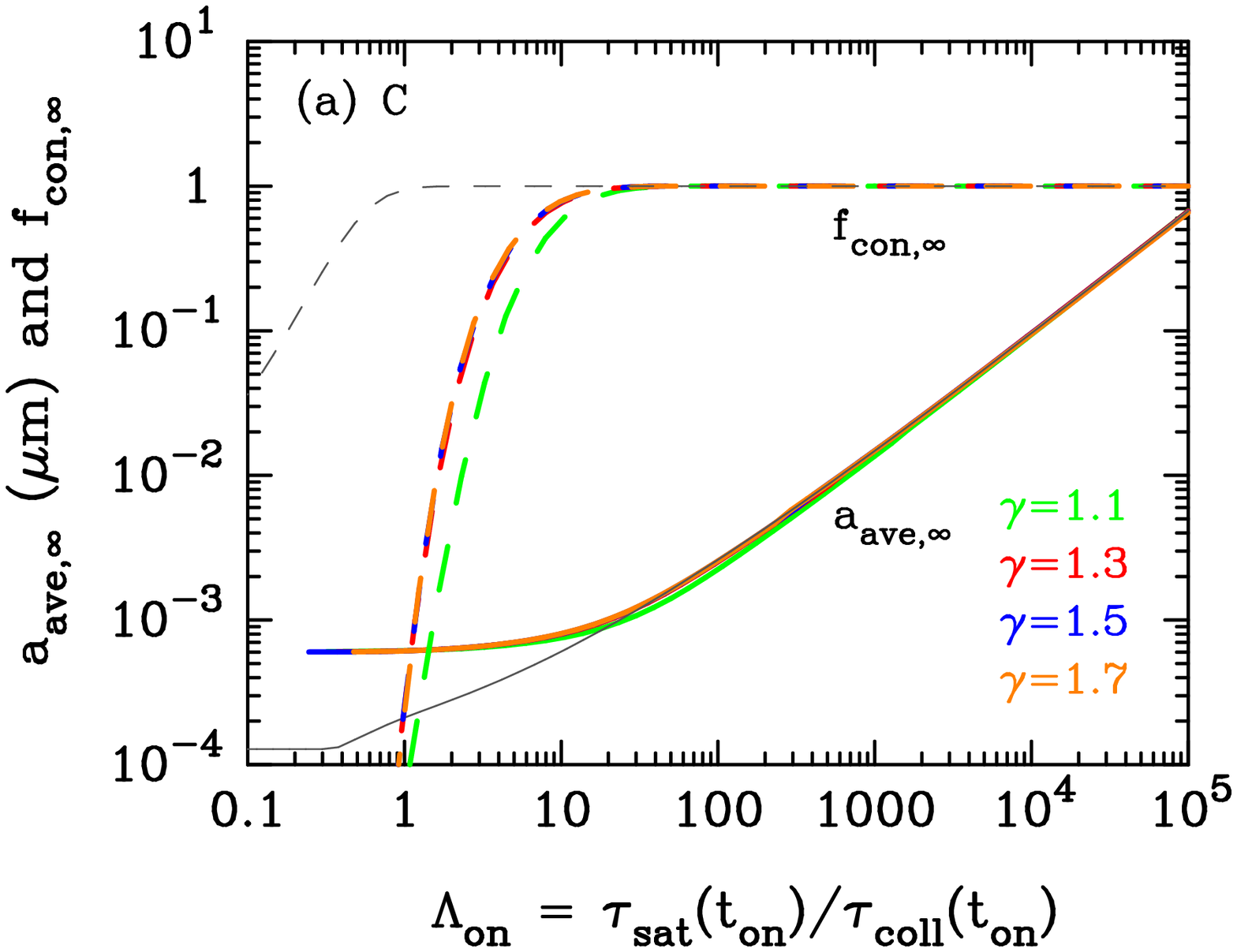}
\vspace{0.8 cm}
\plotone{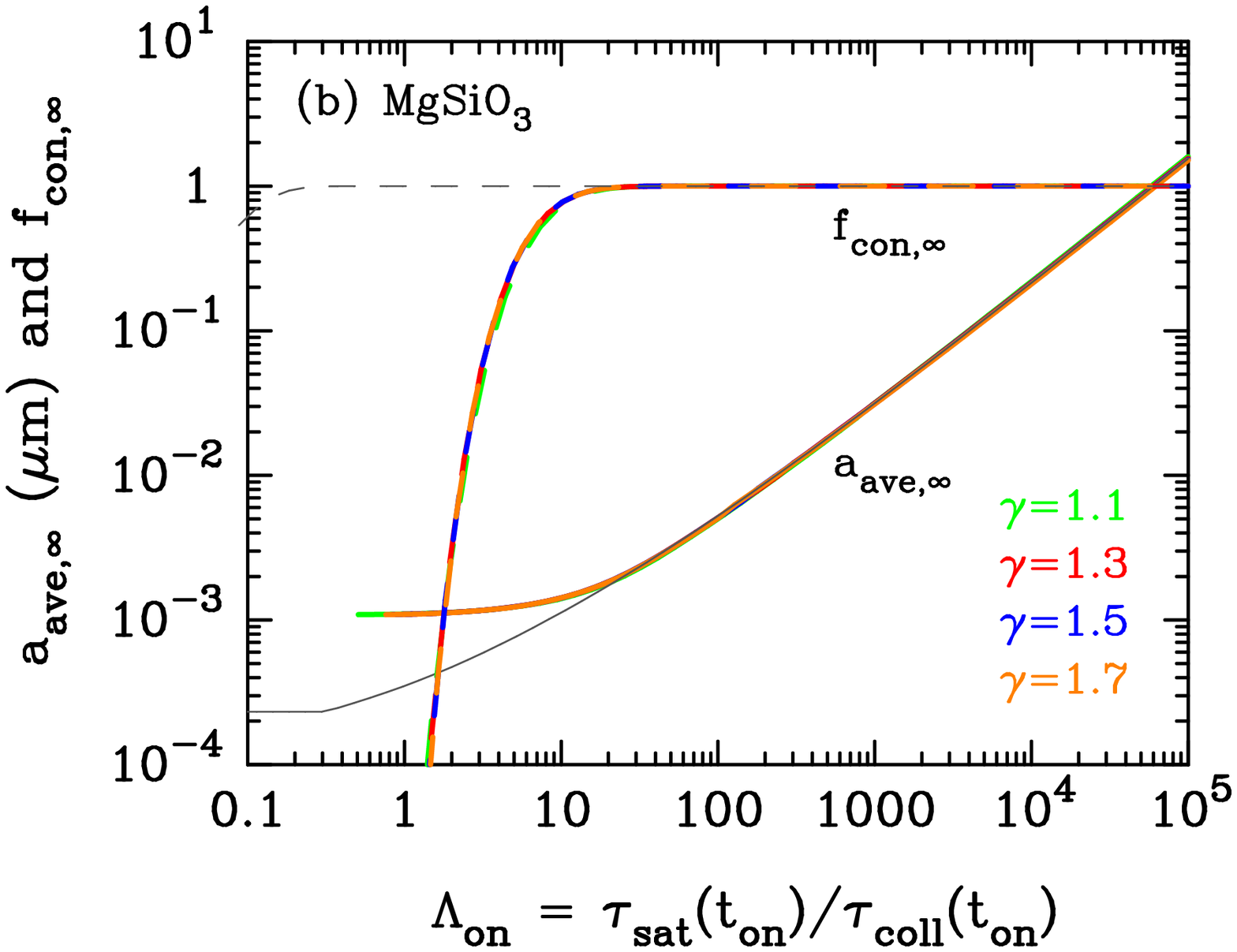}
\caption{
Final condensation efficiencies $f_{{\rm con},\infty}$ and 
average grain radii $a_{{\rm ave},\infty}$ as a 
function of $\Lambda_{\rm on}$ for $\gamma =$ 1.1, 1.3, 1.5, and 1.7;
(a) for C grains and (b) for MgSiO$_3$ grains.
The thin gray lines are the results from the steady model 
with $\gamma = 1.3$.
\label{fig8}}
\end{figure}

\clearpage
\begin{figure}
\epsscale{0.6}
\plotone{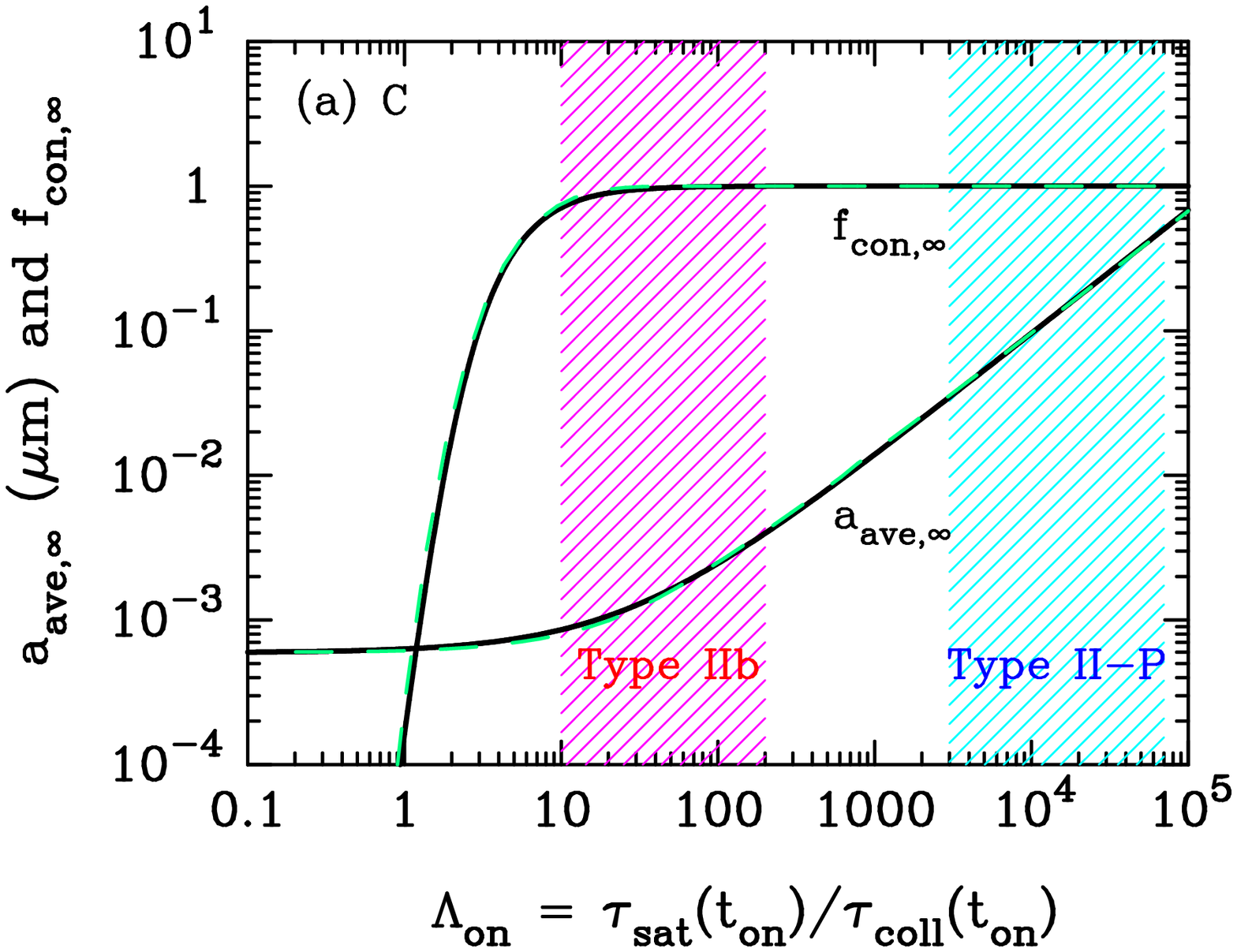}
\vspace{0.8 cm}
\plotone{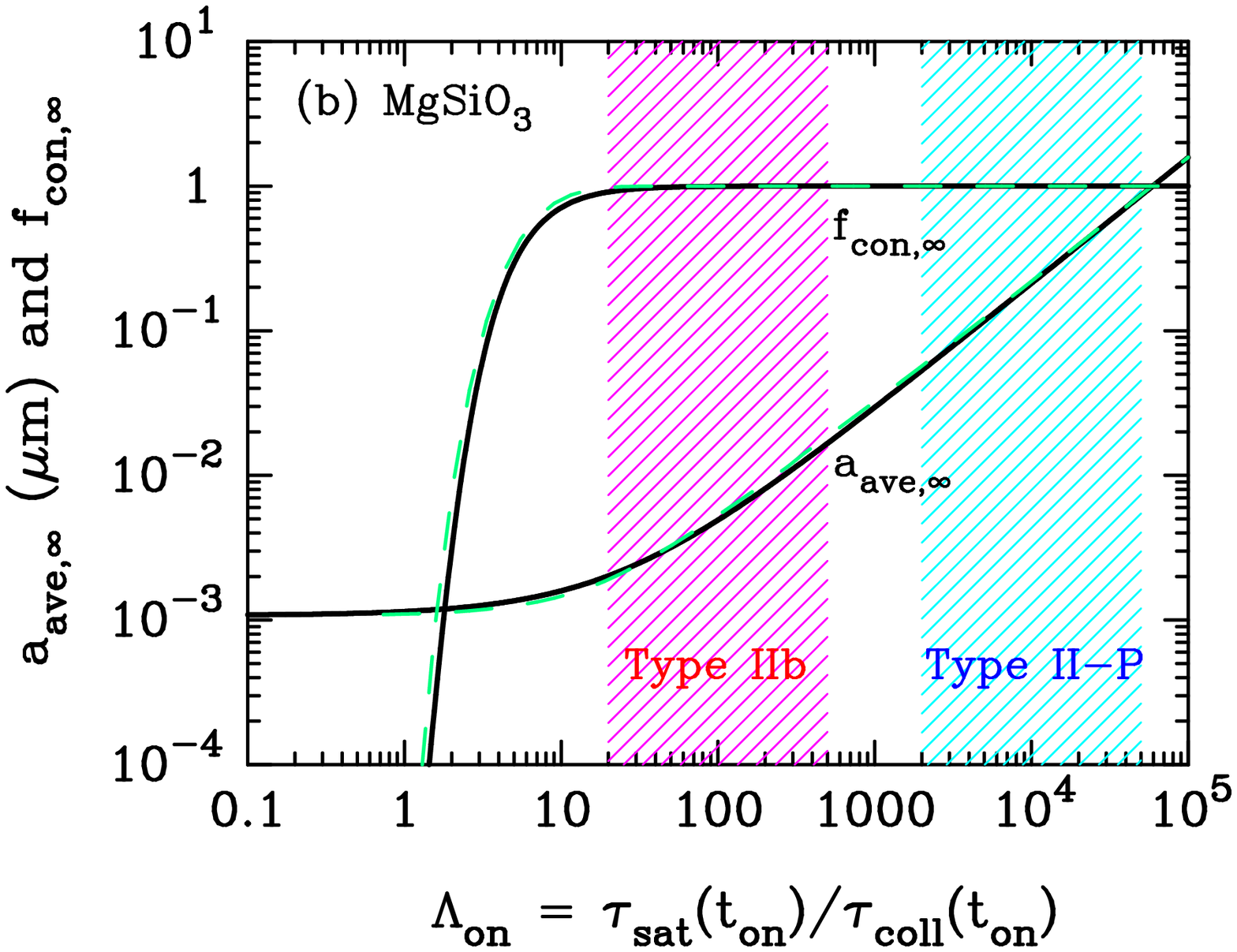}
\caption{
Final average grain radii $a_{{\rm ave},\infty}$ and condensation 
efficiencies $f_{{\rm con},\infty}$ calculated by the fitting formulae,
respectively, Equations (64) and (65) as a function of $\Lambda_{\rm on}$
(solid lines), and the results of the simulations for $\gamma =$ 1.25 
(dashed lines); (a) for C grains and (b) for MgSiO$_3$ grains.
The hatched regions display the typical ranges of $\Lambda_{\rm on}$
for dust formation (a) in the carbon-rich He layer and (b) in the 
oxygen-rich layer, for Type II--P SNe (cyan, Nozawa et al.\ 2003) and 
Type IIb SNe (magenta, Nozawa et al.\ 2010).
\label{fig9}}
\end{figure}

\clearpage
\begin{figure}
\epsscale{0.6}
\plotone{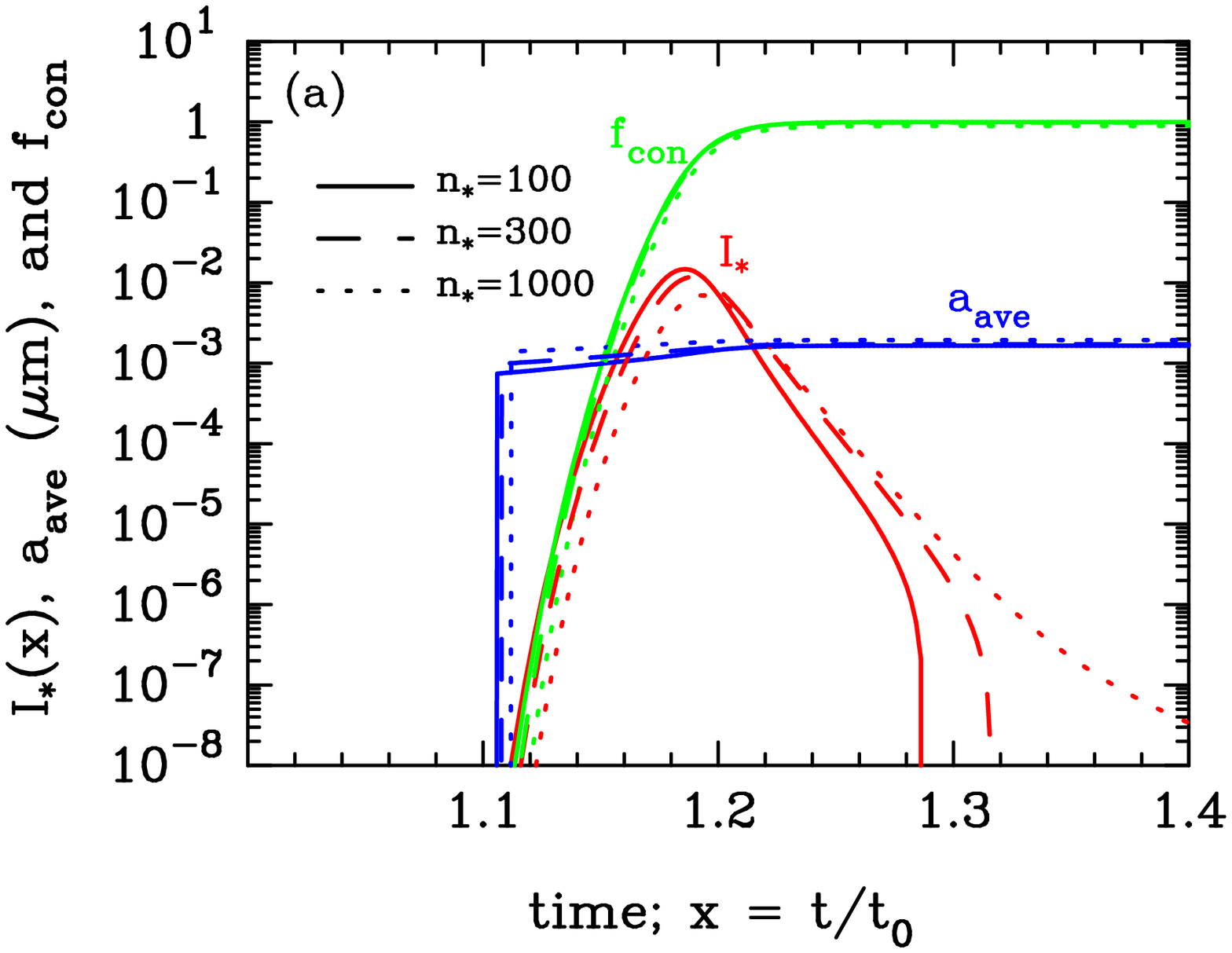}
\vspace{0.8 cm}
\plotone{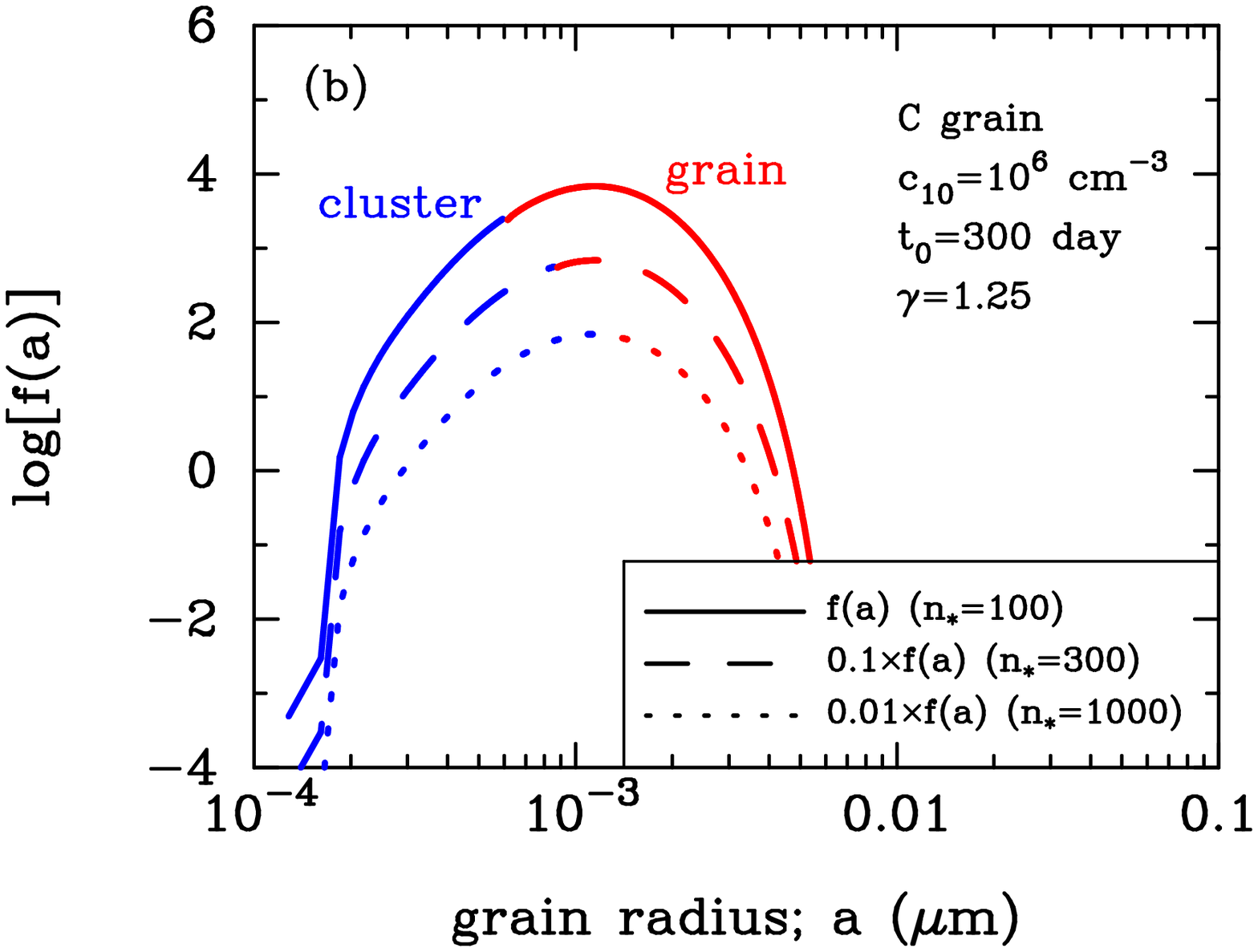}
\caption{
(a) Time evolutions of $I_*$, $a_{\rm ave}$, and 
$f_{\rm con}$, and (b) the final size distribution spectra 
of newly formed 
C clusters (blue) and grains (red) for $n_* = 100$ (solid), 300 
(dashed), and 1000 (dotted).
The other parameters are set to be $c_{10} = 10^6$ cm$^{-3}$,
$t_0 = 300$ days, and $\gamma = 1.25$.
For clarity, the size distributions for $n_* = 300$ and 1000 are 
reduced by a factor of 0.1 and 0.01, respectively.
\label{fig11}}
\end{figure}

\clearpage

\begin{deluxetable}{lcc}
\tablewidth{0pt}
\tablecaption{Fitting parameters for Equations (64) and (65)}
\tablehead{ 
 & \colhead{C} & 
\colhead{MgSiO$_3$}
}
\startdata
$a_*$ ($\mu$m)  &  $5.946 \times 10^{-4}$
                &  $1.076 \times 10^{-3}$ \\
$\epsilon_1$    &  $-1.21$  & $-1.19$ \\
$\epsilon_2$    &  0.854  & 0.871 \\
\hline
\hline \\
$\chi_1$    &  4.15  &  6.38 \\
$\chi_2$    &  1.91  &  2.17 \\
$\chi_3$    &  0.088  &  0.049 \\
\enddata
\tablecomments{
The minimal radius of bulk grain $a_*$ is for
the number of constituent monomer $n_* = 100$.
}
\end{deluxetable}


\newpage

\begin{thebibliography}{}

\bibitem[Abraham(1974)]{abr74}
    Abraham, F. F. 1974, Homogeneous Nucleation Theory, Academic Press,
    INC, New York
\bibitem[Barlow et al.(2010)]{bar10}
    Barlow, M. J., Krause, O., Swinyard, B. M., et al. 
    2010, \aap, 518, L138
\bibitem[Binchi \& Schneider(2007)]{bia07}
    Bianchi, S., \& Schneider, R. 2007, \mnras, 378, 973
\bibitem[Blander \& Katz(1972)]{bla72}
    Blander, M., \& Katz, J., 1972, J. Stat. Phys., 4, 55
\bibitem[Cherchneff \& Dwek (2009)]{che09} 
    Cherchneff, I., \& Dwek, E. 2009, \apj, 703, 642
\bibitem[Chigai et al.(1999)]{chi99} 
    Chigai, T., Yamamoto, T., \& Kozasa, T. 1999, \apj, 510, 999
\bibitem[Clayton et al.(2013)]{cla13} 
    Clayton, D. D. 2013, \apj, 762, 5
\bibitem[Clayton et al.(2001)]{cla01} 
    Clayton, D. D., Deneault, E. A.-N., \& Meyer, B. S.
    2001, \apj, 562, 480
\bibitem[Clayton et al.(1999)]{cla99} 
    Clayton, D. D., Liu, W., \& Dalgarno, A.
    1999, Science, 283, 1290
\bibitem[Deneault et al.(2003)]{Den03} 
    Deneault, E. A.-N., Clayton, D. D., \& Heger, A.
    2003, \apj, 594, 312
\bibitem[Dwek et al.(2007)]{Dwe07} 
    Dwek, E., Galliano, F., \& Jones, A. P. 2007, \apj, 662, 927
\bibitem[Donn \& Nuth(1985)]{Don85} 
    Donn, B., \& Nuth, J. A. 1985, \apj, 288, 187
\bibitem[Fallest et al.(2011)]{fal11} 
    Fallest, D. W., Nozawa, T., Nomoto, K., Umeda, H., Maeda, K.,
    Kozasa, T., \& Lazzati, D. 2011, \mnras, 418, 571
\bibitem[Ferrarotti \& Gail(2001)]{far01} 
    Ferrarotti, A. S. \& Gail, H.-P. 2001, \aap, 371, 133
\bibitem[Gall et al.(2011)]{gol11} 
    Gall, C., Hjorth, J., \& Andersen, A. C. 2011, \araa, 19, 43
\bibitem[Gomez et al.(2012)]{gom12} 
    Gomez, H. L., Krause, O., Barlow, M. J., et al. 
    2012, \apj, 760, 96
\bibitem[Goumans \& Bromley(2012)]{gou12} 
    Goumans, T. P. M., \& Bomley, S. T. 2012, \mnras, 420, 3344
\bibitem[Hasegawa \& Kozasa (1988)]{has88} 
    Hasegawa, H., \& Kozasa, T. 1988, Prog. Theor. Phys. Suppl., 96, 107
\bibitem[Keith \& Lazzati(2011)]{kei11} 
    Keith, A. C., \& Lazzati, D. 2011, \mnras, 410, 685
\bibitem[Kotak et al.(2009)]{kot09} 
    Kotak, R., Meikle, W. P. S., Farrah, D., et al. 
    2009, \apj, 704, 306
\bibitem[Kozasa et al.(1996)]{koz96} 
    Kozasa, T., Dorschner, J., Henning, Th., \& Stognienko, R. 
    1996, \aap, 307, 551
\bibitem[Kozasa \& Hasegawa(1987)]{koz87} 
    Kozasa, T., \& Hasegawa, H. 1987, Prog. Theor. Phys., 77, 1402 
\bibitem[Kozasa \& Hasegawa(1988)]{koz88} 
    Kozasa, T., \& Hasegawa, H. 1988, Icarus, 73, 180
\bibitem[Kozasa et al.(1989)]{koz89} 
    Kozasa, T., Hasegawa, H., \& Nomoto, K. 1989, \apj, 344, 325
\bibitem[Kozasa et al.(1991)]{koz91} 
    Kozasa, T., Hasegawa, H., \& Nomoto, K. 1991, \aap, 249, 474
\bibitem[Kozasa \& Sogawa(1997)]{koz97} 
    Kozasa, T., \& Sogawa, H. 1997 \apss, 251, 165
\bibitem[Kozasa \& Sogawa(1998)]{koz98} 
    Kozasa, T., \& Sogawa, H. 1998 \apss, 255, 437
\bibitem[Laki\'{c}evi\'{c} et al.(2012)]{lak12} 
    Laki\'{c}evi\'{c}, M., van Loon, J. Th., Stanke, T., De Breck, C.,
    \& Patat, F. 2012, \aap, 541, L1
\bibitem[Landau \& Lifshitz(1976)]{lan76} 
    Landau, L. D., \& Lifshitz, E. M. 1976, Statistical Physics,
    3rd Edition, Part 1, Pergamon Press, Oxford
\bibitem[Liu \& Dalgarno(1994)]{liu94} 
    Liu, W., \& Dalgarno, A. 1994, \apj, 428, 769
\bibitem[Liu \& Dalgarno(1996)]{liu96} 
    Liu, W., \& Dalgarno, A. 1996, \apj, 471, 780
\bibitem[Maeda et al.(2013)]{mae13} 
    Maeda, K., Nozawa, T., Sahu, D. K., et al. 
    2013, ApJ, accepted
\bibitem[Matsuura et al.(2011)]{mat11} 
    Matsuura, M., Dwek, E., Meixner, M., et al. 
    2011, Science, 333, 1258
\bibitem[Nozawa et al.(2008)]{noz08} 
    Nozawa, T., Kozasa, T., Tominaga, N., et al. 
    2008, \apj, 684, 1343
\bibitem[Nozawa et al.(2006)]{noz06} 
    Nozawa, T., Kozasa, T., \& Habe, A. 2006, \apj, 648, 435
\bibitem[Nozawa et al.(2007)]{noz07} 
    Nozawa, T., Kozasa, T., Habe, A., Dwek, E., Umeda, H., Tominaga, N., 
    Maeda, K., \& Nomoto, K. 2007, \apj, 666, 955
\bibitem[Nozawa et al.(2010)]{noz10} 
    Nozawa, T., Kozasa, T., Tominaga, N., Maeda, K., Umeda, H., 
    Nomoto, K., \& Krause, O. 2010, \apj, 713, 356
\bibitem[Nozawa et al.(2003)]{noz03} 
    Nozawa, T., Kozasa, T., Umeda, H., Maeda, K., \& Nomoto, K.
    2003, \apj, 598, 785
\bibitem[Nozawa et al.(2011)]{noz11} 
    Nozawa, T., Maeda, K., Kozasa, T., Tanaka, M., Nomoto, K.
    \& Umeda, H. 2011, \apj, 736, 45
\bibitem[Paquette \& Nuth.(2011)]{paq11} 
    Paquette, J. A., \& Nuth, J. A. 2011, \apj, 737, L6
\bibitem[Sibthorpe et al.(2010)]{sib10} 
    Sibthorpe, B., Ade, P. A. R., Bock, J. J., et al. 
    2010, \apj, 719, 1553
\bibitem[Silvia et al.(2010)]{sil10} 
    Silvia, D. W., Smith, B. D., \& Shull, J. M. 2010, \apj, 715, 1575
\bibitem[Silvia et al.(2012)]{sil12} 
    Silvia, D. W., Smith, B. D., \& Shull, J. M. 2012, \apj, 748, 12
\bibitem[Todini \& Ferrara(2001)]{tod01} 
    Todini, P., \& Ferrara, A. 2001, \mnras, 325, 726
\bibitem[Yamamoto et al.(2001)]{yam01} 
    Yamamoto, T., Chigai, T., Watanabe, S., \& Kozasa, T. 
    2001, \aap, 380, 373
\bibitem[Yasuda \& Kozasa(2012)]{yas12} 
    Yasuda, Y., \& Kozasa, T. 2012, \apj, 745, 159
\bibitem[Zhukovska et al.(2008)]{zhu08} 
    Zhukovska, S., Gail, H.-P., \& Trieloff, M. 2008, \aap, 479, 453

\end{thebibliography}
\end{document}